\documentclass[apj, numberedappendix, appendixfloats, twocolappendix]{emulateapj}
\shorttitle{The Nuclei of Early-Type Galaxies}

\shortauthors{Turner et~al.}

\begin{document}

\title{The ACS Fornax Cluster Survey. VI. The Nuclei of Early-Type Galaxies in the Fornax Cluster\altaffilmark{1}}

\author{
Monica L. Turner\altaffilmark{2,3,\textasteriskcentered},
Patrick C\^ot\'e\altaffilmark{4},
Laura Ferrarese\altaffilmark{4},
Andr\'es Jord\'an\altaffilmark{5},
John P. Blakeslee\altaffilmark{4},
Simona Mei\altaffilmark{6,7},
Eric W. Peng\altaffilmark{8,9},
Michael J. West\altaffilmark{10}
}

\altaffiltext{1}{Based on observations with the NASA/ESA {\it Hubble
Space Telescope} obtained at the Space Telescope Science Institute,
which is operated by the association of Universities for Research in
Astronomy, Inc., under NASA contract NAS 5-26555.}
\altaffiltext{2}{Department of Physics and Astronomy, University of Victoria, 
Victoria, BC, V8W 3P6, Canada}
\altaffiltext{3}{Leiden Observatory, Leiden University, Postbus 9513, 
2300 RA, Leiden, The Netherlands}
\altaffiltext{4}{Herzberg Institute of Astrophysics, National Research Council 
of Canada, Victoria, BC, V9E 2E7, Canada}
\altaffiltext{5}{Departamento de Astronom\'ia y Astrof\'isica, Pontificia Universidad 
Cat\'olica de Chile, Av. Vicu\~na Mackenna 4860, Macul 7820436, Santiago, Chile}
\altaffiltext{6}{University of Paris Denis Diderot,  75205 Paris Cedex 13, France}
\altaffiltext{7}{GEPI, Observatoire de Paris, Section de Meudon, 5 Place J. Janssen, 
92195 Meudon Cedex, France}
\altaffiltext{8}{Department of Astronomy, Peking University, Beijing 100871, China}
\altaffiltext{9}{Kavli Institute for Astronomy and Astrophysics, Peking University, 
Beijing 100871, China}
\altaffiltext{10}{European Southern Observatory, Alonso de Cordova 3107, Vitacura, 
Santiago, Chile}
\altaffiltext{\textasteriskcentered}{E-mail: turnerm@strw.leidenuniv.nl}

\slugcomment{Accepted for publication in the {\it Astrophysical Journal Supplement Series}}

\begin{abstract}
The {\it Advanced Camera for Surveys}  (ACS) Fornax Cluster Survey is a Hubble Space Telescope program to 
image 43 early-type galaxies in the Fornax cluster, using the F475W and F850LP bandpasses of the ACS. We 
employ both 1D and 2D techniques to characterize the properties of the {\it stellar nuclei} in these galaxies, 
defined as the central ``luminosity excesses'' relative to a S\'ersic model fitted to the underlying host. We find 
$72\pm13$\% of our sample (31 galaxies) to be nucleated, with only three of the nuclei offset by more than 
0\farcs5 from their galaxy photocenter, and with the majority of nuclei having colors bluer than their hosts. 
The nuclei are observed to be larger, and brighter, than typical Fornax globular clusters, and to follow different
 structural scaling relations. A comparison of our results to those from the ACS Virgo Cluster Survey reveals striking 
similarities in the properties of the nuclei belonging to these different environments. We briefly review a variety of 
proposed formation models and conclude that, for the low-mass galaxies in our sample, the most important mechanism for 
nucleus growth is probably infall of star clusters through dynamical friction, while for higher mass galaxies, gas 
accretion triggered by mergers, accretions and tidal torques is likely to dominate, with the relative importance of 
these two processes varying smoothly as a function of galaxy mass. Some intermediate-mass galaxies in our sample show 
a complexity in their inner structure that may be the signature of ``hybrid nuclei'' that arose through parallel 
formation channels.
\end{abstract}


%
% INTRO
%

\section{Introduction} \label{sec:introduction}

Once viewed as relatively simple objects that formed in a single, ``monolithic" collapse, early-type galaxies are
now widely believed to have been assembled hierarchically through repeated mergers and accretions
 \citep[e.g.,][]{white78,searle78,white91,kauffman00,cole00,sp05,bower06}. 
A property of most {\it luminous} (e.g., $M_r \lesssim -22.5$) early-type galaxies is that they appear to have formed 
the majority of their stars at high redshift ($z\gtrsim1$, corresponding to ages of 
$\tau \gtrsim 7$-$8$~Gyr) and on short timescales ($\Delta\tau \lesssim1$~Gyr) 
\citep[e.g.,][]{bower92, franx93, thomas99b, trager00, wake06}. 
These features may be related to feedback from active galactic nuclei (AGN), which can generate 
jets and outflows that blow away gas and suppress star formation 
\citep[e.g.,][]{silk98, king03, murray05, fabian06, robertson06}. The general trends in the star formation 
histories of low- and intermediate-luminosity early-type galaxies are not as well understood, but they are known
to show considerable diversity and to depend sensitively on environment  \citep[see, e.g.,][]{tolstoy09}.

The discovery of the $\mathcal{M}_{\rm BH}$--$\sigma$ relation \citep{ferrarese00,gebhardt00} 
points to a fundamental connection between the central black holes powering
these AGN, and the dynamical properties of their host galaxies.  There are several other galaxy properties 
that have also been found to scale with black hole mass, including
luminosity \citep[e.g.,][]{kormendy95, ferrarese00},
light concentration \citep[e.g.,][]{graham01},
global velocity dispersion \citep[e.g.,][]{ferrarese00, gebhardt00, gultekin09},
bulge mass \citep[e.g.,][]{magorrian98, marconi03, haring04},
and total gravitational mass of the host \citep{bandara09}.  
Thus, it has become clear that an understanding of the central regions of galaxies, including black holes and AGN, is essential if we are to
make sense of the formation and evolution of galaxies themselves.

However, the direct detection of black holes remains very challenging: see, e.g., Chapter~11 of 
\citet{ferrarese05} for an overview of the observational difficulties. 
For kinematic measurements, a high central surface brightness
is needed to obtain spectra of adequate S/N, and this requirement 
can pose problems for massive early-type galaxies with shallow surface brightness profiles
in their cores. At the distances of the Virgo and Fornax clusters, the small angular
size of the black hole ``sphere of influence" in most galaxies introduces a further complication. 
For example, at 20~Mpc, the distance of Fornax, a black hole in a 
galaxy with $\sigma=200$~km~s$^{-1}$ has a sphere of influence of only $0\farcs2$ in radius
(assuming the  $\mathcal{M}$--$\sigma$ relation from \citealt{lauraa06}). 
It is therefore not surprising that  a dynamical black hole mass measurement exists for only a single 
early-type galaxy in the Fornax cluster \citep[FCC~213;][]{houghton06, gebhardt07}. 
 
\defcitealias{cote06}{C06}

On the other hand, the correlation between a galaxy's mass and 
that of its black hole was recently shown to extend down to the central
nuclear star clusters found in low-mass galaxies \citep{laurab06, wehner06}. 
Other studies have reported similar relationships between black hole or nucleus
mass and the host bulge luminosity, mass, and S\'ersic index 
\citep{rossa06, balcells07, graham07}. 
These results are suggestive of a global 
relationship between galaxies and {\it both} types of central massive object 
(CMO; \citealt[hereafter C06]{cote06}): however, it is still an open question as to whether
black holes and nuclei form via the same mechanisms, or whether nuclei form first and 
serve as seeds for black hole formation.

The hydrodynamical simulations of \citet{li07} of a shared formation mechanism 
for both nuclei and black holes via the gravitational collapse of gas in bulgeless disks were
able to reproduce a CMO and host mass correlation even without imposing
an  {\em a priori} $M$-$\sigma$ relation, and were observed to be in agreement with \citet{laurab06}.
Alternatively, \citet{laurab06} noted that nuclei could, in principle, form in all galaxies, but 
in massive galaxies they might either collapse or be destroyed (or otherwise altered)
by binary black holes. Using semi-analytic models, it was the demonstrated by 
\citet{devecchi09} and \citet{devecchi10} that nuclei could form at high redshifts 
and act as possible black hole seeds.  

If nuclei and black holes form simultaneously, then it is possible that 
momentum feedback determines which object will eventually dominate the CMO mass.
\citet{mclaughlin06} noted that the same momentum flux that drives
out gas from black holes \citep{king03, king05} could also regulate 
the growth of nuclear star clusters. \citet{nayakshin09} used this finding to explain why 
nuclei, not black holes, appear more likely to form in less massive hosts.
Both objects can form simultaneously as gas is driven to the center of a galaxy through
an event such as a merger, but it is the mass of the host bulge that sets the individual
formation rates. Some evidence for such a scenario comes from observations of 
intermediate-luminosity galaxies \citep{filippenko03, gonzalez08, seth08, graham09}, 
as well as a number of dwarfs \citep{barth04, reines11}, that have been found to contain 
both a central stellar nucleus and a black hole. 
Indeed, using observations in the Virgo cluster, \citet{gallo10} estimated that
hybrid nuclei could occur in 0.3--7\% of galaxies with stellar masses below $10^{11}M_\odot$,
and in less than 32\% of hosts above this stellar mass.   
In short, the study of nuclei presents us 
with a new opportunity to deepen our understanding of how galaxies and black holes co-evolve.

Like black holes, nuclei pose some observational challenges of their own.
Although their existence in some dwarf galaxies has been known for decades, comprehensive surveys of 
galaxy clusters --- in which the frequency of nucleation within complete galaxy samples could be robustly 
measured --- did not appear until \citet{binggeli87} published their Virgo Cluster Catalog (VCC). 
This program observed 1277 members and 574 probable members of the Virgo cluster using the 2.5~m Las 
Campanas telescope;  about 26\% of all dwarf galaxies in the VCC sample were found to be nucleated.  
Shortly thereafter, a similar survey of the Fornax cluster by \citet{ferguson89} ---
the Fornax Cluster Catalog (FCC) --- found nuclei in 103/249  $\approx$ 41\% of their dwarf galaxies.  
In the above studies, dwarf galaxies were identified primarily morphologically by their flat surface brightness
profiles, although in general they were found to be fainter than $M_B\simeq-18$~mag \citep{sandage84}. 

Given the 
low luminosities and small sizes of most of these nuclei, the frequencies of nucleation estimated from ground-based 
photographic studies are certainly lower limits. For instance,
 \citet{lotz04} used WFPC2 on the Hubble Space Telescope (HST) to observe 69 dwarf elliptical galaxies in both Virgo and Fornax, finding
nuclei in six galaxies that were previously classified as non-nucleated
in the VCC and FCC.  Based on wide-field imaging of Virgo dwarfs from the Isaac Newton Telescope, \citet{grant05} was
able to identify many faint nuclei that were missed in the earlier photographic survey. 
In fact,  imaging of {\it late-type} galaxies with HST commonly revealed ``nuclear clusters" 
that had gone unnoticed in earlier studies, with an overall
frequency of nucleation of $\approx70\%$ \citep[e.g.,][]{carollo98, matthews99, 
boker04, walcher05, seth06}.

The first study to find a comparable frequency of nucleation among early-type galaxies was carried out by 
\citetalias{cote06} with the Advanced Camera for Surveys (ACS) on HST: i.e., 
the ACS Virgo Cluster Survey \citep[ACSVCS;][]{cote04}.\footnote{Related papers from the ACSVCS on the central
structure of early-type galaxies include \citet{lauraa06, laurab06, cote07, glass11}.}  In addition to establishing a 
high frequency of nucleation for early-type galaxies (at least 66\% for galaxies brighter
than $M_B\approx-15$), the high-resolution imaging made it possible to characterize the detailed properties of the
nuclei for the first time, including their luminosity function, structural properties,
color-magnitude relation, and nucleus-to-galaxy luminosity ratio.
We note here that although in \citetalias{cote06} and this work, we call the central excess of light
rising above a galaxy's extrapolated outer surface brightness profile a ''nucleus``, these objects
are not limited to being nuclear star clusters; certainly, some could be described as 
disks, bars, or other large scale structures, which have been observed by previous studies of early-type 
galaxy centers \citep[e.g.][]{lauraa06, balcells07, morelli10}.
In this paper, which is part of the
ACS Fornax Cluster Survey (ACSFCS), we examine the properties of nuclei belonging to galaxies in
the Fornax Cluster, which is located at a distance of $D = 20\pm0.3\pm1.4$~Mpc (statistical $+$
systematic error) \citep{blake09}.  
This cluster is smaller, denser, more dynamically evolved, and more regular in shape 
than the Virgo cluster, and therefore allows us to study the properties of the nuclei
of galaxies residing in a new and different environment. 

\defcitealias{jordan07a}{Paper~I}
\defcitealias{cote07}{Paper~II}
\defcitealias{glass11}{Paper~IV}
\defcitealias{blake09}{Paper~V}
%\defcitealias{tbd12}{Paper~III}
\defcitealias{masters10}{Paper~VII}
\defcitealias{villegas10}{Paper~VIII}
\defcitealias{mieske10}{Paper~IX}
\defcitealias{liu11}{Paper~X}

Other papers in the ACSFCS series have described the data reduction procedures used in the survey
\citep[hereafter Paper~I]{jordan07a}, systematic variations in the central structure of galaxies
\citep[hereafter Paper~II]{cote07}, the logarithmic slope of the galaxy central surface 
brightness profiles \citep[hereafter Paper~IV]{glass11}, and the use of surface brightness 
fluctuations as a distance indicator \citep[hereafter Paper~V]{blake09}. Paper~III (2012, in prep.) 
of the ACSFCS will present a detailed isophotal
analysis of the ACSFCS galaxies, including their dust properties, axial ratios, 2D structure, total
magnitudes, colors, and surface brightness and color profiles.
Papers studying the properties of globular clusters (GCs) in ACSFCS galaxies have examined their half-light 
radii \citep[hereafter Paper~VII]{masters10}, luminosity function \citep[hereafter Paper~VIII]{villegas10}, 
color-magnitude relation  \citep[hereafter Paper~IX]{mieske10}, and color gradients 
\citep[hereafter Paper~X]{liu11}. 

The outline of this paper is as follows. In \S\ref{sec:observations} we describe
the observations and methodologies used to measure photometric and structural
parameters for the nuclei; in \S\ref{sec:results} we examine the nucleus properties, including their frequency of 
nucleation, luminosity function, sizes, surface brightness parameters, and colors; in \S\ref{sec:discussion}
we put our results into the context of current formation scenarios; and in 
\S\ref{sec:summary} we summarize our main results. An appendix presents a comparison
of 1D and 2D methods for measuring photometric and structural parameters of nuclei and
their host galaxies. 

%
% OBSERVATIONS AND ANALYSIS
%

\section{Observations and Analysis}\label{sec:observations}

The ACSFCS sample was constructed by selecting all galaxies from the FCC with: (1)
blue magnitudes $B_T \leq 15.5$; and (2)  early-type morphologies: i.e., E, S0, SB0, dE, dE,N or dS0,N. 
These morphological types were taken directly from Ferguson (1989) which are in turn
based on the classification scheme of \citet{sandage84}.
In addition to the 42 FCC galaxies that met these criteria, two ellipticals that lie just beyond the FCC survey region
(NGC~1340 and IC~2006) were added, giving a total of 44  targets.
Unfortunately, due to a shutter failure during execution, no images were obtained 
for FCC~161 (NGC~1379). Our final sample therefore consists of 43 early-type galaxies, which is
complete (apart from FCC~161) down to a limiting magnitude of  $B_T \approx15.5$ mag 
($M_B \approx -16.0$ mag).  For all galaxies in this survey, membership in the cluster has been 
confirmed through radial velocity measurements. More details on the sample can be found in Papers~I and~III.

In \S\ref{sec:compare}, we will compare our results to a sample of galaxies and nuclei
from the ACSVCS, which consists of 100 early-type members of the Virgo Cluster.
That survey was magnitude limited down
to $B_T \approx12$ mag ($M_B\approx-19$ mag) and 44\% complete down to its limiting
magnitude of $B_T \approx16$ mag ($M_B\approx-15$ mag).
Both the Fornax and Virgo galaxies were observed with the ACS using
Wide Field Channel (WFC) mode with the F475W and F850LP filters, which
correspond closely to the $g$- and $z$-band filters in the Sloan
Digital Sky Survey (SDSS) system \citep[see, e.g.,][]{fukugita96,york00,sirianni05}.

Basic data for the ACSFCS galaxies are presented in Table~\ref{tab:datagal}.
 The ACSFCS identification number, the FCC number from
\citet{ferguson89}, and any alternate names are reported in the first three columns.  
The table is ordered by increasing FCC blue magnitude, $B_T$, which is given in 
column 4.  In calculating absolute magnitudes, we used the individual
surface brightness fluctuation (SBF) distances measured in Paper V. 
Beginning in \S3, all reported magnitudes are extinction corrected, 
using dust maps from \citet{schlegel98}, with the ratios of total-to-selective absorption 
in the WFC filters taken from  \citet{sirianni05}; 
the adopted $B$-band extinctions are shown in column 5.  
The galaxy $g$- and $z$- band surface brightness at a geometric mean radius of $1\arcsec$,
measured by spline interpolation, are recorded in columns 6 and 7. 
Note that all HST/ACS magnitudes quoted in this paper are AB magnitudes. 

% Nucleus data

\tabletypesize{\scriptsize}	
\begin{deluxetable*}{rllrccccc}
%\rotate
\tablecolumns{9} 
\tablewidth{0pt}
\tablecaption{Basic Data for ACSFCS Galaxies\label{tab:datagal}}
\tablehead{
\colhead{ID}         & 
\colhead{Name}        & 
\colhead{Other}      & 
\colhead{$B_T$}      & 
\colhead{$A_B$}   & 
\colhead{$\mu_g(1\arcsec$)}       & 
\colhead{$\mu_z(1\arcsec$)}       &
\colhead{Class}    &
\colhead{Class}    \\
\colhead{} & 
\colhead{} & 
\colhead{} & 
\colhead{(mag)} & 
\colhead{(mag)} & 
\colhead{(mag/$\Box\arcsec$)} &
\colhead{(mag/$\Box\arcsec$)} &
\colhead{(FCC)}&
\colhead{(ACS)} \\
\colhead{(1)} &
\colhead{(2)} &
\colhead{(3)} &
\colhead{(4)} &
\colhead{(5)} &
\colhead{(6)} &
\colhead{(7)} &
\colhead{(8)} &
\colhead{(9)}
}
\startdata
 1  &    21  &      N1316  &   9.06  &  0.090  &  15.61  &  14.12  &  N  &  cS   \\
 2  &   213  &      N1399  &  10.04  &  0.056  &  16.78  &  15.17  &  N  &  cS   \\
 3  &   219  &      N1404  &  10.96  &  0.049  &  16.45  &  14.88  &  N  &  cS   \\
 4  &  1340  &  E418-G005  &  11.23  &  0.077  &  17.00  &  15.56  &  N  &  cS   \\
 5  &   167  &      N1380  &  10.84  &  0.075  &  16.88  &  15.32  &  N  &  S1   \\
 6  &   276  &      N1427  &  11.79  &  0.048  &  17.07  &  15.59  &  N  &  S1   \\
 7  &   147  &      N1374  &  11.95  &  0.060  &  17.14  &  15.58  &  N  &  S1   \\
 8  &  2006  &  E359-G007  &  12.59  &  0.048  &  17.72  &  16.18  &  N  &  S2   \\
 9  &    83  &      N1351  &  12.33  &  0.061  &  17.35  &  15.83  &  N  &  S1   \\
10  &   184  &      N1387  &  11.77  &  0.055  &  16.70  &  15.05  &  N  &  S1   \\
11  &    63  &      N1339  &  12.77  &  0.057  &  17.13  &  15.56  &  N  &  S2   \\
12  &   193  &      N1389  &  12.59  &  0.046  &  17.34  &  15.88  &  N  &  S2   \\
13  &   170  &      N1381  &  12.91  &  0.058  &  17.12  &  15.62  &  N  &  S2   \\
14  &   153  &      I1963  &  13.55  &  0.062  &  18.32  &  16.91  &  N  &  S2   \\
15  &   177  &     N1380A  &  13.60  &  0.063  &  18.83  &  17.58  &  N  &  S2   \\
16  &    47  &      N1336  &  13.34  &  0.049  &  18.50  &  17.11  &  N  &  S2   \\
17  &    43  &      I1919  &  13.82  &  0.062  &  19.99  &  18.83  &  Y  &  S2   \\
18  &   190  &     N1380B  &  13.79  &  0.074  &  19.32  &  17.89  &  N  &  S2   \\
19  &   310  &      N1460  &  13.68  &  0.047  &  19.32  &  17.96  &  N  &  S2   \\
20  &   249  &      N1419  &  13.61  &  0.056  &  17.68  &  16.25  &  N  &  S2   \\
21  &   148  &      N1375  &  13.39  &  0.063  &  18.20  &  17.02  &  N  &  S2   \\
22  &   255  &   E358-G50  &  13.99  &  0.025  &  19.50  &  18.26  &  Y  &  S2   \\
23  &   277  &      N1428  &  14.01  &  0.044  &  18.84  &  17.45  &  N  &  S2   \\
24  &    55  &   E358-G06  &  14.23  &  0.043  &  19.68  &  18.41  &  Y  &  S2   \\
25  &   152  &   E358-G25  &  14.13  &  0.044  &  20.44  &  19.25  &  N  &  S1   \\
26  &   301  &   E358-G59  &  14.22  &  0.039  &  18.61  &  17.31  &  N  &  S2   \\
27  &   335  &   E359-G02  &  14.90  &  0.063  &  20.40  &  19.27  &  N  &  S2   \\
28  &   143  &      N1373  &  14.19  &  0.061  &  18.39  &  16.96  &  N  &  S1   \\
29  &    95  &        G87  &  15.01  &  0.064  &  20.16  &  18.83  &  N  &  S2   \\
30  &   136  &        G99  &  15.00  &  0.069  &  20.73  &  19.39  &  Y  &  S2   \\
31  &   182  &        G79  &  15.01  &  0.057  &  19.61  &  18.18  &  N  &  S2   \\
32  &   204  &   E358-G43  &  15.33  &  0.045  &  20.50  &  19.23  &  Y  &  S2   \\
33  &   119  &        G26  &  15.44  &  0.060  &  21.35  &  20.10  &  N  &  S1\tablenotemark{a}   \\
34  &    90  &       G118  &  15.10  &  0.052  &  19.55  &  18.76  &  N  &  S2   \\
35  &    26  &   E357-G25  &  15.26  &  0.067  &  19.80  &  19.39  &  N  &  S1   \\
36  &   106  &        G47  &  15.34  &  0.046  &  19.89  &  18.62  &  Y  &  S2   \\
37  &    19  &   E301-G08  &  15.81  &  0.085  &  21.56  &  20.49  &  Y  &  S2   \\
38  &   202  &      N1396  &  15.50  &  0.057  &  20.71  &  19.41  &  Y  &  S2   \\
39  &   324  &   E358-G66  &  15.83  &  0.042  &  22.16  &  21.01  &  N  &  S2   \\
40  &   288  &   E358-G56  &  15.82  &  0.025  &  21.03  &  19.85  &  Y  &  S2   \\
41  &   303  &       NG47  &  15.74  &  0.046  &  21.63  &  20.49  &  Y  &  S2   \\
42  &   203  &   E358-G42  &  15.82  &  0.051  &  21.50  &  20.28  &  Y  &  S2   \\
43  &   100  &        G86  &  15.75  &  0.062  &  22.18  &  21.08  &  Y  &  S2   \\
\enddata
\tablenotetext{~}{Column key:\\
(1) ACSFCS Identification number;\\
(2) Galaxy name, mainly from the Fornax Cluster Catalog (FCC) of \citet{ferguson89};\\
(3) Alternative names in the NGC, ESO or IC catalogs;\\
(4) Total blue magnitude from ACSFCS (Paper~III);\\
(5) $A_B$ from \citet{schlegel98};\\ 
(6)--(7) $g$- and $z$-band surface brightness measured at a geometric radius of 1\arcsec;\\
(8) Nuclear classification in the FCC: Y = nucleated, N = non-nucleated;\\
(9) Nuclear classification in ACSFCS:  {\tt cS}=core-S\'ersic (non-nucleated), 
{\tt S1}=S\'ersic (non-nucleated), {\tt S2}=double-S\'ersic (nucleated)}
\tablenotetext{a}{Due to the offset of the nucleus and the amount of central dust,
the nucleus parameters for FCC 119 were derived using a King profile fit to the ACS image.}
\end{deluxetable*}

  The final two columns in Table~\ref{tab:datagal} give the classifications
of the galaxies as nucleated from \citet{ferguson89}, and the ones
  derived from our surface brightness profile analysis.
The parameterization of these profiles is discussed in \S\ref{sec:sbp}, and
the fitting methods used are outlined in \S\ref{sec:fitProc}.
Finally, the nucleus properties obtained from the above procedure are described in 
\S\ref{sec:idNuc}. Additional information about our program galaxies, such as 
coordinates and morphological classifications, can be found in Papers I and III.

% Schematic showing systematic trends along the LF

\begin{figure*}
	\figurenum{1}
	\plotone{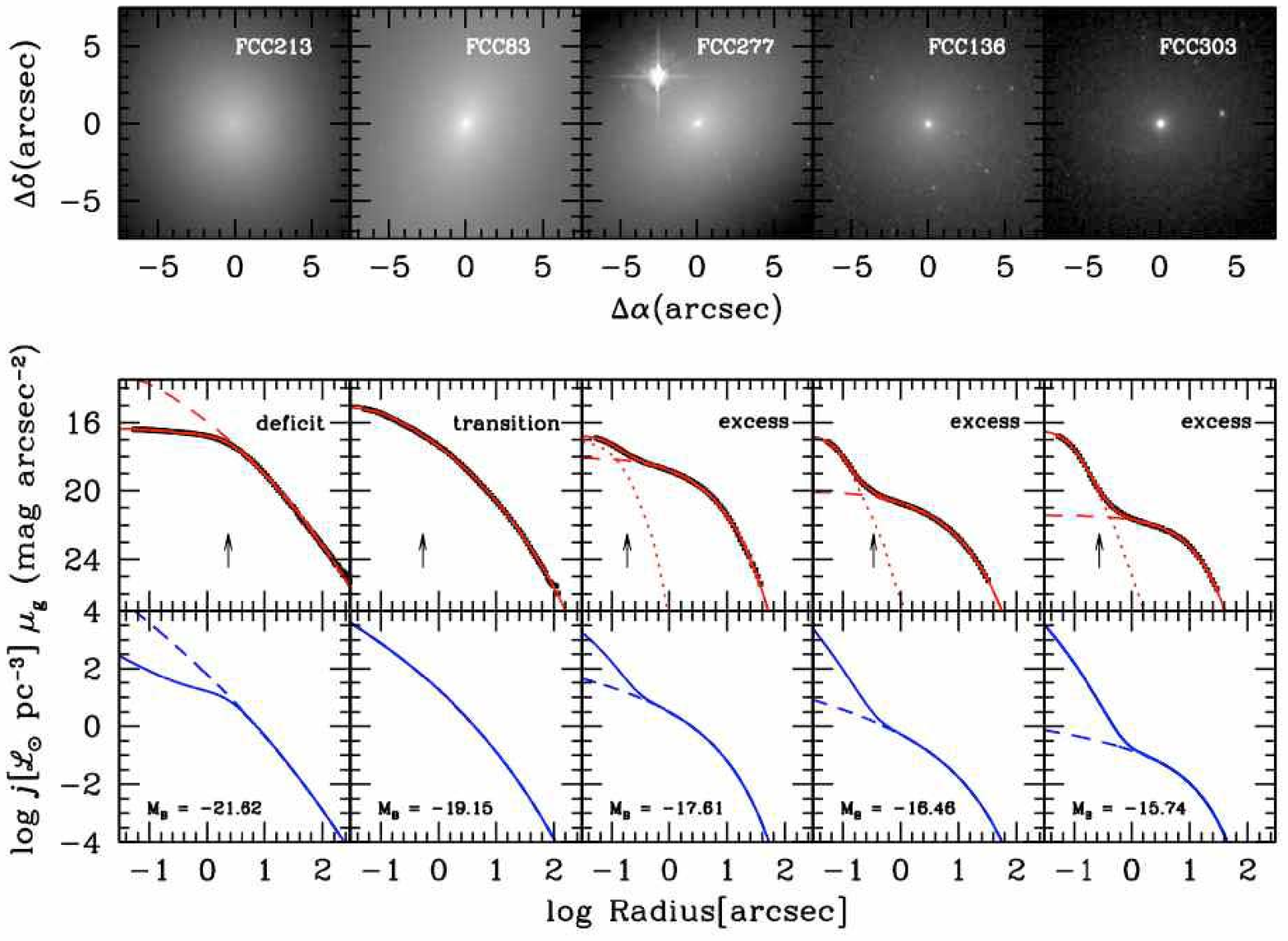}
	\caption{Five galaxies from the ACSFCS chosen to illustrate systematic trends in central and global structure 
	along the luminosity function.
	{\it Top Row:} ACS/F475W images showing the inner 15$\arcsec\times$15$\arcsec$ ($\approx 1.5$ kpc $\times$$1.5$ kpc) for each galaxy.
	{\it Middle Row:} Azimuthally-averaged surface brightness profiles and the best-fit PSF-convolved models. For
	FCC~213, the best-fit ``core-S\'ersic" model is shown. For FCC~83, the solid curve shows a fitted S\'ersic model. 
	The three remaining galaxies --- which show central nuclei, or luminosity ``excesses" relative to an underlying S\'ersic model --- 
	are fitted with double-S\'ersic models, with the dotted curves
	showing the separate nuclear and global components.   The arrow in each panel is drawn at 2\% of the effective radius of the galaxy 
	\citepalias{cote07}.
	{\it Bottom Row:} Deprojected luminosity profiles (which represent the true 3-dimensional density distribution without any PSF convolution) 
	for the same five galaxies from \citetalias{glass11}. The solid curves show the deprojected profiles corresponding to the 
	solid curves shown in the middle row. Dashed curves show the profiles corresponding to the inward extrapolation of the S\'ersic models that 
	best fit the outer (galaxy) profile.}
	\label{fig:schematic}
\end{figure*}

% Galaxy images
\begin{figure*}
	\figurenum{2}
	\plotone{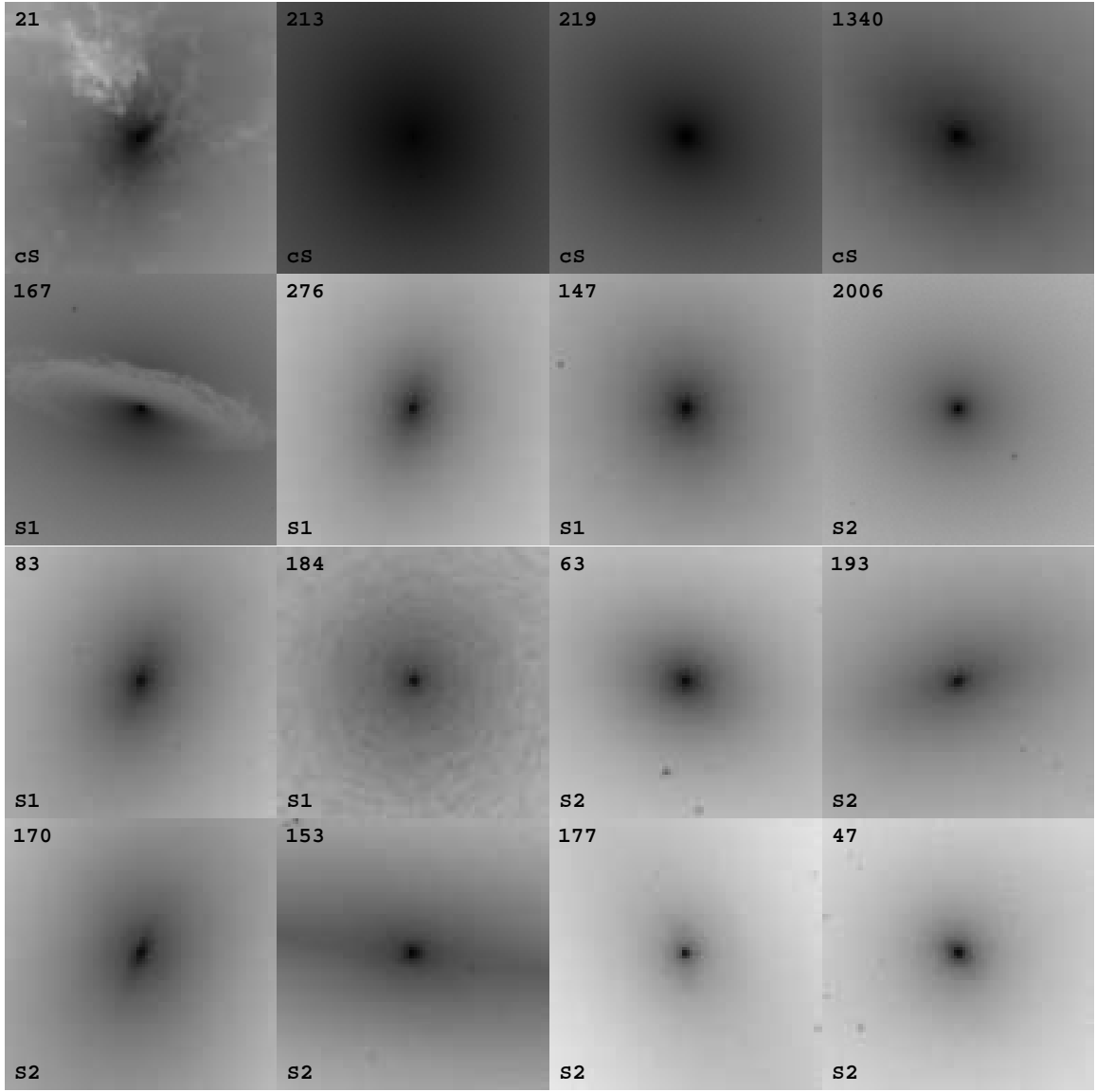}
	\caption{F475W ($g$-band) images of the inner $10\arcsec\times10\arcsec$ 
	($\sim1\rm{kpc}\times1\rm{kpc}$) regions of the ACSFCS galaxies.  
	The galaxies are arranged in order of increasing blue magnitude 
	(i.e., decreasing luminosity) from left to right, and from top to bottom.
	Each galaxy's FCC number is displayed in the top left, and the bottom
	left denotes the model used to fit the galaxy, either
	{\tt S1} (S\'ersic), {\tt cS} (core-S\'ersic), or {\tt S2} (double-S\'ersic).}
	\label{fig:acs_images}
\end{figure*}
\begin{figure*}
	\figurenum{2}
	\plotone{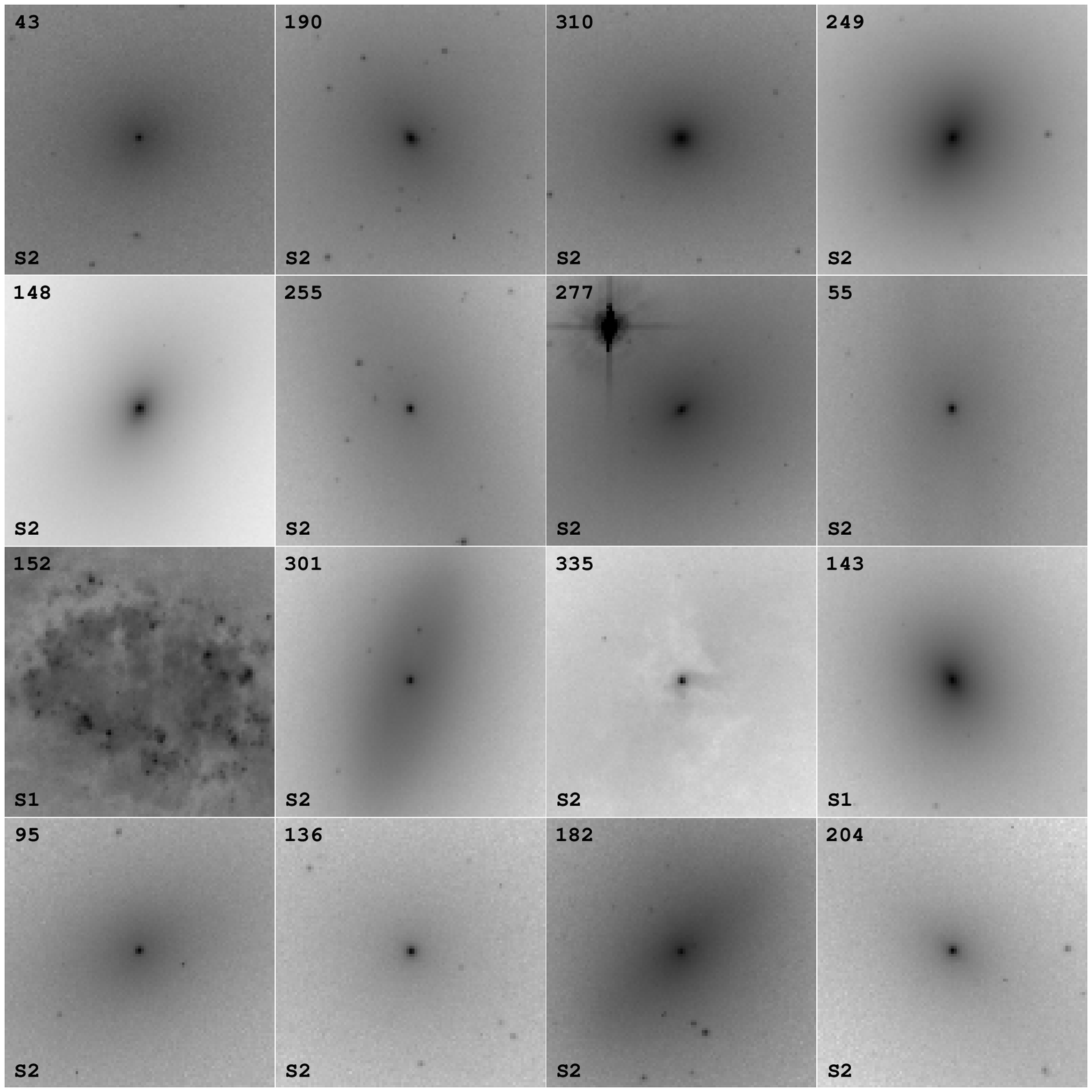}
	\caption{\emph{Continued}}
\end{figure*}
\begin{figure*}
	\figurenum{2}
	\plotone{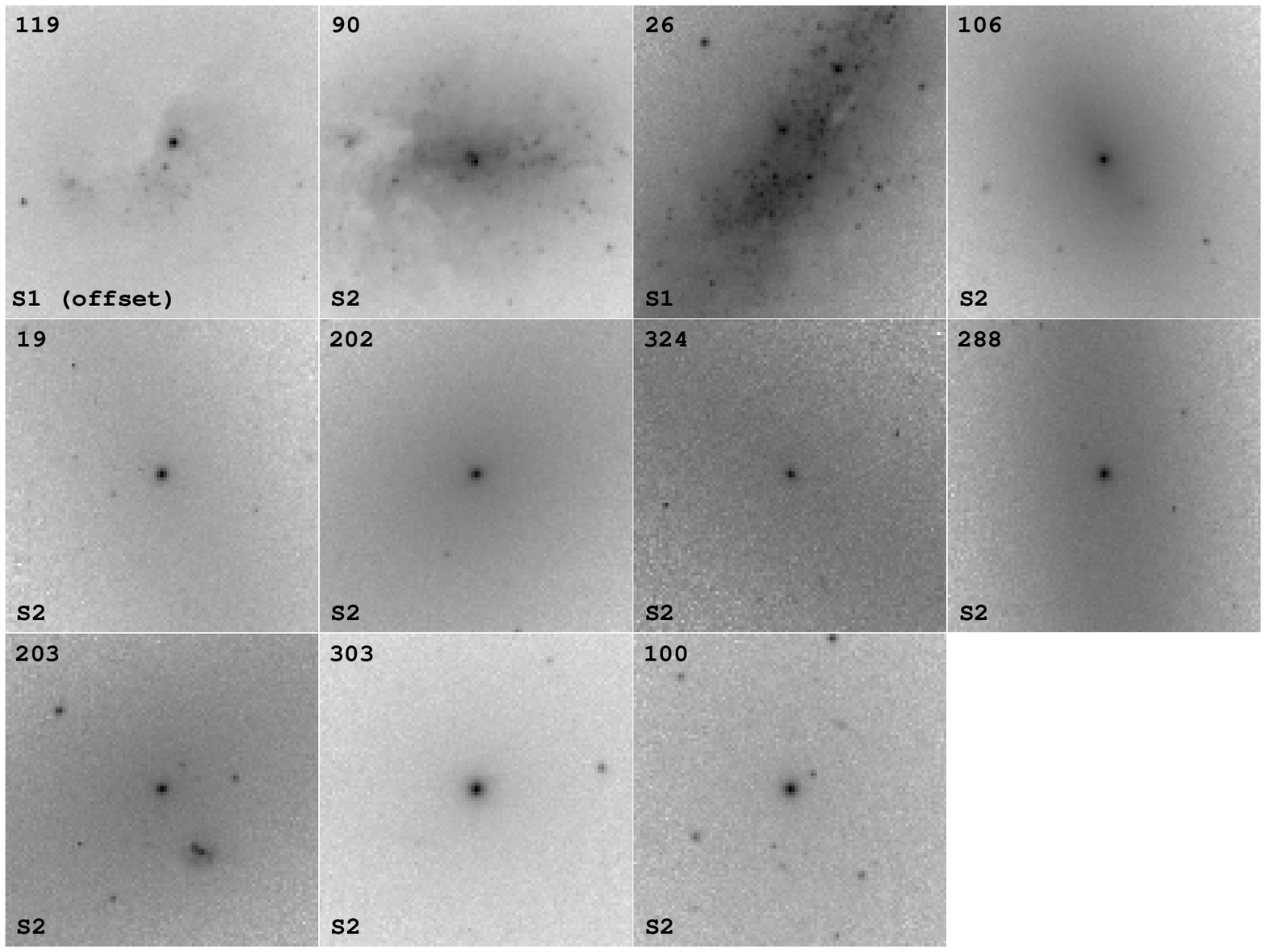}
	\caption{\emph{Continued}}
\end{figure*}

% Galaxy profiles
\begin{figure*}
	\figurenum{3}
	\plotone{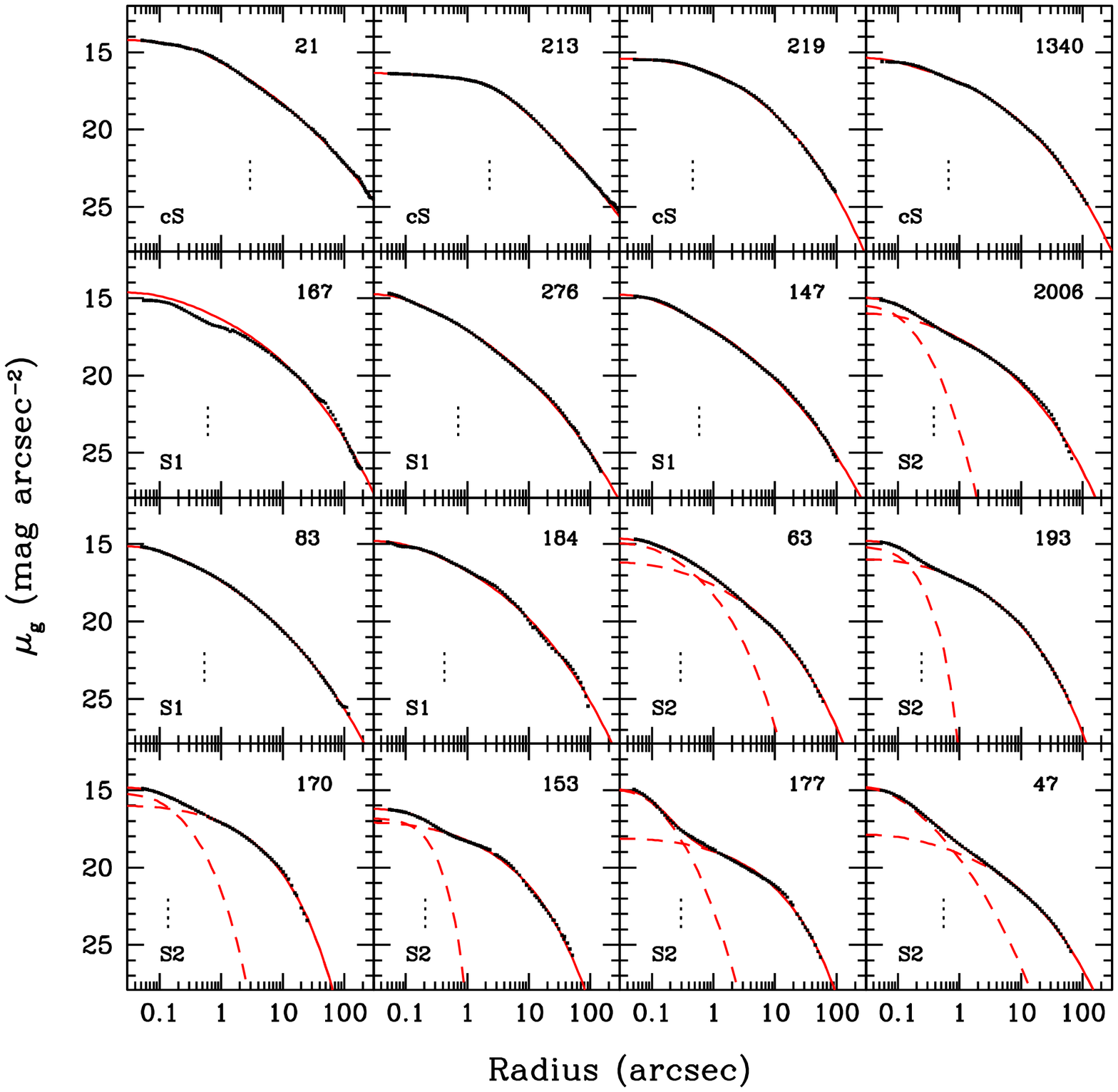}
	\caption{Azimuthally-averaged surface brightness profiles
	for the ACSFCS galaxies. As in Figure~\ref{fig:acs_images}, galaxies are ordered by blue magnitude,  with
	luminosity decreasing from left to right, and from top to bottom.  The black points plot the
	measured $g$-band profiles. The red curves show
	the fitted models with the two separate components (nucleus and galaxy) indicated by the dashed curves; their
	sum is shown as the solid curve.  The dotted vertical lines are drawn at a radius of $0.02R_e$ in all cases.
	The top right label denotes the galaxy FCC number, and
	the three types of fitted models are denoted in the bottom left,
	by {\tt S1} (S\'ersic), {\tt cS} (core-S\'ersic),  or {\tt S2} (double-S\'ersic). Note that FCC 167 contains a prominent
	central dust disk (Figure~\ref{fig:acs_images}, Paper~III), so the models were fitted outside $R = 5\arcsec$.}
	\label{fig:sb_profiles}
\end{figure*}
\begin{figure*}
	\figurenum{3}
	\plotone{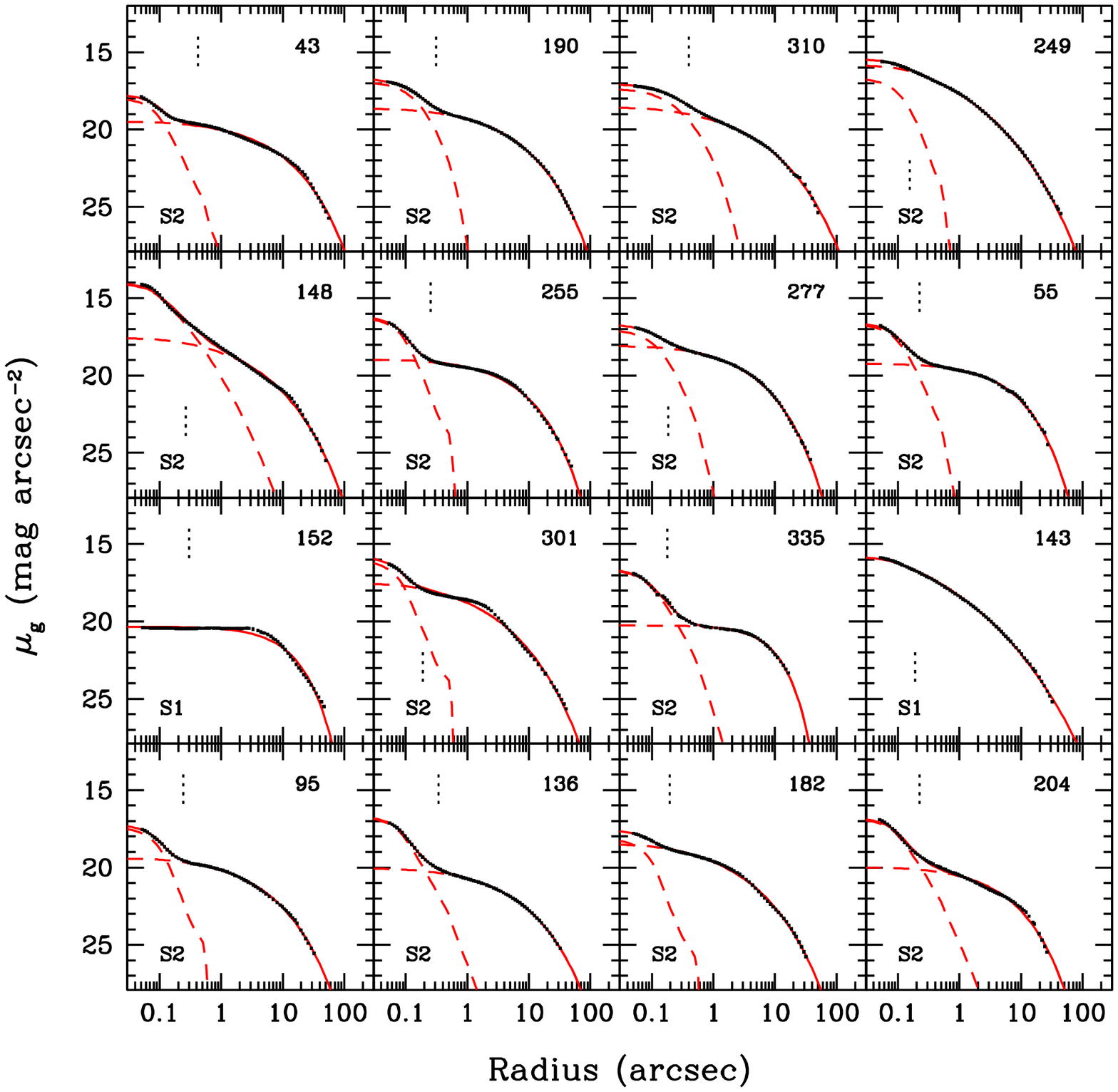}
	\caption{\emph{Continued}}
\end{figure*}
\begin{figure*}
	\figurenum{3}
	%\plotone{f3c.eps}
	\centering 
	\leavevmode 
	\includegraphics[width = 0.85\linewidth, trim =0cm 4cm 0cm 0cm, clip=true]{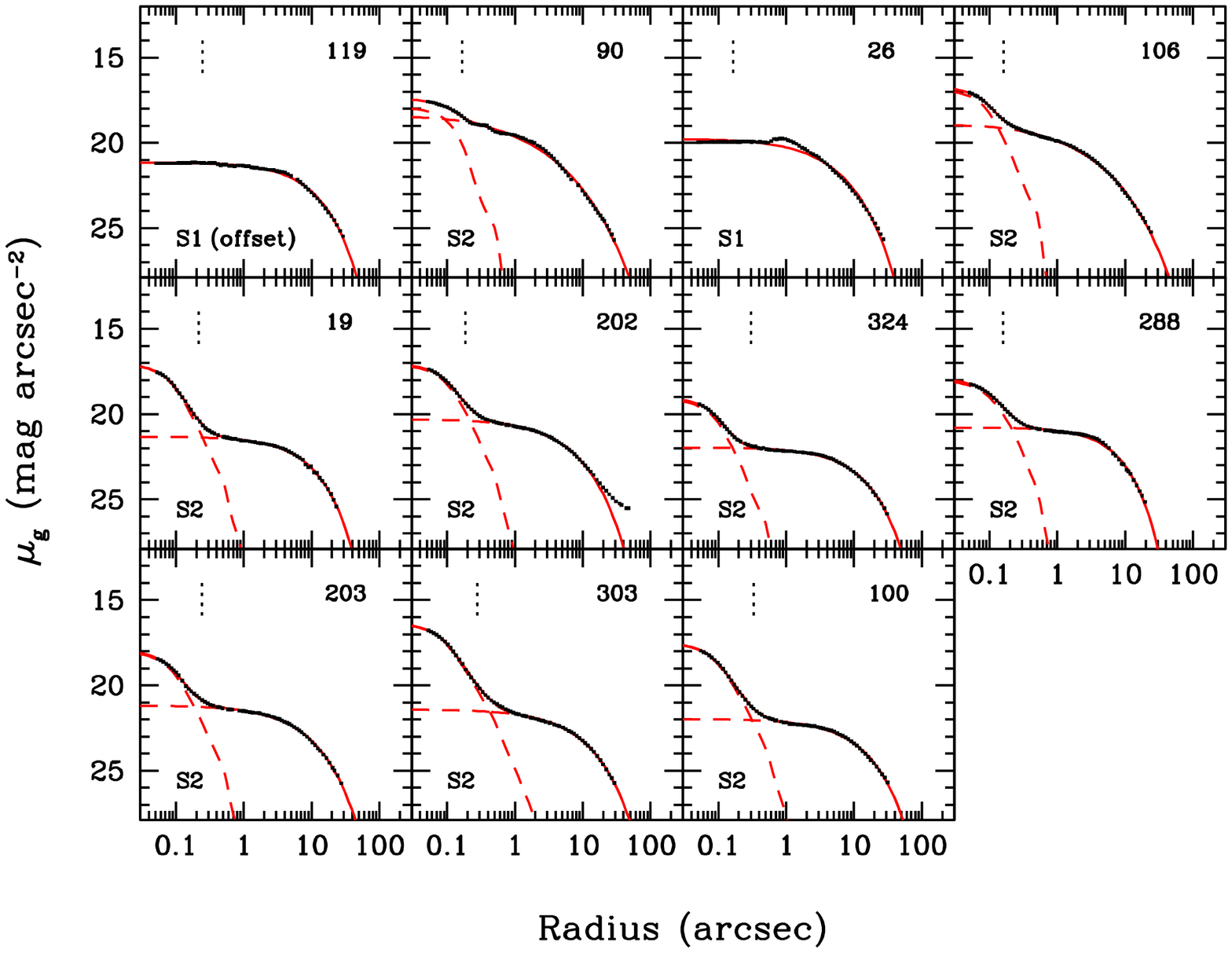}
	\caption{\emph{Continued}}
\end{figure*}

\subsection{Parameterization of the Surface Brightness Profiles}\label{sec:sbp}

As stellar nuclei, which are the focus of this study, are found in the luminous central regions of their
host galaxies, accurately modeling the underlying galaxy surface brightness is necessary 
to measure their photometric and structural parameters.  Indeed, for the faintest nuclei, or for some nuclei embedded in
 high surface brightness galaxies with steeply rising brightness profiles, this can
be important for even {\it identifying} a central nuclear component (see Appendix~A of \citetalias{cote06}).
Using the IRAF  task {\tt ellipse}, which is based on the algorithm of \citet{jed87},
elliptical isophotes with logarithmically increasing semi-major axis length 
were fitted to the galaxies. In most cases, all ellipse parameters (center,
ellipticity, and position angle) were allowed to vary. 
However, to achieve convergence, the
galaxies with large amounts of central dust required the ellipse centers to be held
fixed throughout the fit (FCC 335, FCC 119, FCC 90), as well as the position angles
and ellipticities while fitting the innermost areas (FCC 119 and FCC 90), where the
fixed parameter values were determined by ellipse fits to the outer regions
($R_{e}\gtrsim5\arcsec$). For more details on the fitting procedures, see
\S3.2 of \citet{lauraa06} and Paper~III.

The results from the {\tt ellipse} isophotal analysis were used to derive azimuthally-averaged
radial surface brightness profiles, which were then fitted using one of  three 
different parameterizations for the global surface brightness profile.
The first parameterization is the well known S\'ersic profile \citep{sersic68}, a three parameter
model which has the form
\begin{equation}
	I_{\rm S}(R)=I_e \exp \left\{ -b_n \left[ \left(\frac{R}{R_e} 
	\right)^{1/n}-1 \right] \right\},
	\label{eq:S}
\end{equation}
where $I_e$ is the intensity at the effective radius, $R_e$, and  the S\'ersic index, $n$, characterizes 
the overall shape of the light profile. The
constant $b_n$ is defined such that $\Gamma(2n)=2\gamma(2n,b_n)$, where
$\Gamma$ and $\gamma$ are the complete and incomplete gamma functions,
respectively \citep{ciotti91}.  For lower values of $n$, the S\'ersic
profile is shallow in the inner regions and steep in the outer regions;
$n=1$ produces a pure exponential profile, which  generally provides a reasonable
fit to dwarf galaxies. 
Higher values of $n$ yield functions which are steep in the inner regions and extended at large radii, 
with a less pronounced radial dependence on slope; these profiles generally fit bright
ellipticals quite well (i.e., $n=4$ reduces to a classical de Vaucouleurs profile).  

Historically, these two types of profiles have been used to separately
parameterize dwarfs and giants. However, more complete studies of galaxies 
have found that $n$ actually varies steadily with galaxy luminosity 
(e.g., \citealt{graham03b, gavazzi05, lauraa06, kormendy09, misgeld11}; 
McLaughlin et~al. 2012, in prep.). In what follows, we will refer
to these single-component parameterizations as {\tt S1} models.

Although the S\'ersic profile describes the outer 
component (typically beyond a few percent of the the effective radius) 
of galaxies remarkably well --- a consequence of the wide range in
concentration, spatial scale and  surface brightness that is possible by
varying $n$, $R_e$ and $I_e$, respectively --- there can be variations in 
the {\it central} structure that cannot be accounted for in this simple model (see, e.g.,
Figures~1 and 2 of \citetalias{cote07}).  
Specifically, the brightest ellipticals tend to show a {\it luminosity deficit} in their
central regions; for these objects, the six-parameter ``core-S\'ersic" model 
\citep{graham03a} provides a good description of their surface brightness profiles.
The core-S\'ersic model, referred to hereafter as a {\tt cS} profile, can be written as
\begin{equation}
	I_{\rm cS}(R) = I^{\prime}\left[I+\left(\frac{R_b}{R}
	\right)^{\alpha}\right]^{\gamma/\alpha} \exp\left[-b_n\left(
	\frac{R^{\alpha}+R^{\alpha}_b}{R^{\alpha}_b}
	\right)^{1/\alpha n}\right],
	\label{eq:cSa}
\end{equation}
where
\begin{equation}
	I^{\prime}=I_b2^{-\gamma/\alpha}\exp\left[b_n\left(2^{1/\alpha}
	R_b/R_e\right)^{1/n}\right].
	\label{eq:cSb}
\end{equation}
This parameterization consists of the usual S\'ersic profile, with effective 
radius $R_e$ and S\'ersic index $n$, outside of a ``break'' radius $R_b$ (where
the intensity is $I_b$).  At $R_b$, the outer profile transitions to an inner power-law component 
with slope $\gamma$, according to the ``sharpness'' parameter $\alpha$ (where smaller values translate to
smoother transitions).

By contrast, most of the low- and intermediate-luminosity galaxies in our sample show evidence for a
{\it luminosity excess} in their cores which is, by definition, the signature of a central nucleus 
\citep[see Appendix~A of][]{cote06}.\footnote{The ACSVCS finding of ``luminosity excesses" in Virgo cluster galaxies relative to the inward
extrapolation of S\'ersic models fitted to the outer profiles
was subsequently confirmed by \citet{kormendy09} who reanalyzed a subset of the ACSVCS sample.}
A central excess in the surface brightness profile
can then be modeled by adding a second S\'ersic component. This double-S\'ersic profile (which we denote
hereafter as an {\tt S2} profile) has the form
\begin{equation}
\begin{array}{rcl}
	I_{\rm S2}(R) &=&  I_{e,1}\exp\left\{- b_{n,1} 
	\left[\left(\frac{R}{R_{e,1}} 
	\right)^{1/n_1}-1 \right] \right\} \\
	 &+& I_{e,2}\exp\left\{- b_{n,2} \left[\left(\frac{R}{R_{e,2}} 
	\right)^{1/n_2}-1 \right] \right\}
	\label{eq:dS}
\end{array}
\end{equation}
where the enumerated subscripts indicate the S\'ersic parameters for the outer and inner
components.  

It should be noted that, in \citetalias{cote06}, double-S\'ersic profiles were not used to fit the nucleated
galaxies.  Instead, the central nuclei were represented by King profiles \citep{michie63, king66}, 
while the outer component was represented by either a core-S\'ersic or
S\'ersic profile.  Our decision to use a double-S\'ersic parameterization in the ACSFCS analysis is
motivated by two considerations.
Firstly, modeling the inner component with the S\'ersic profiles allows for a  
diversity of possible physical systems, due to the range of the S\'ersic parameter (see above).
For $n\sim1$, the profile is a pure exponential and is thus suitable for embedded disks, 
whereas $n\sim2$ represents Galactic
GCs quite accurately, and presumably, nuclear star clusters as well. This is supported
by the findings of \citet{graham09}, who measured S\'ersic indices of of $n=3.0$, $2.3$, and $1.6$
for the nuclear star clusters of the Milky Way, M32, and NGC~205, respectively; and by 
\citet{seth10} who observed a S\'ersic index of $n\sim2$ for NGC~404. Secondly, the use of 
S\'ersic profile for both the inner and outer components allows straightforward and convenient 
comparisons of their respective structural properties.  

The overall trends described here are illustrated in Figure~\ref{fig:schematic}, which shows the systematic variations
in the core and global structure of early-type galaxies along the luminosity function 
(see also \citetalias{cote06}; \citealt{lauraa06, laurab06}; \citetalias{cote07, glass11}).
The upper panels in this  figure show $15\arcsec\times15\arcsec$ images centered on five representative galaxies
from the ACSFCS, arranged in order of decreasing luminosity. The middle rows show model fits to the $g$-band surface brightness
profiles as described above: i.e., FCC~213 ({\tt cS}), FCC~83 ({\tt S1}), FCC~277 ({\tt S2}), FCC~136 ({\tt S2}) and FCC~303 ({\tt S2}). Note
the systematic decline in galaxy surface brightness from left to right,  and the emergence of an increasingly prominent
central nuclear component  as galaxy luminosity decreases. At low and intermediate luminosities, these luminosity ``excesses" (i.e., nuclei) relative to the 
underlying galaxy model correspond to a steady 
steepening of the three-dimensional luminosity density on small scales, as shown in the lower panel of Figure~\ref{fig:schematic} (from \citetalias{glass11}).
Images and brightness profiles for the full sample of ACSFCS galaxies will be discussed below.

\subsection{Fitting Procedure} \label{sec:fitProc}

As described in \citetalias{jordan07a}, the ACSFCS uses the {\it Lanczos3} kernel for
drizzling rather than the {\it Gaussian} kernel which was selected for the ACSVCS.  Due
to the slightly larger distance of the Fornax cluster --- 20.0 vs. 16.5 Mpc (\citealt{mei07}; \citetalias{blake09}) 
--- and the fact that some of the 
Virgo nuclei were only marginally resolved in the ACSVCS (\citetalias{cote06}; \citealt{lauraa06, laurab06}),
the sharper point spread function (PSF) possible with the {\it Lanczos3} kernel was deemed to be more important for 
the ACSFCS galaxies than the {\it Gaussian} kernel's ability to repair bad pixels.
	 	
New PSFs for the ACSFCS were constructed in an identical manner
using more than a thousand stars from the GO-10048 
and GO-10375 programs to obtain photometric calibrations of the Galactic GC 
47 Tucanae (PI = J. Mack).  Using 
multiple observations allowed PSFs to be extracted from data
that were acquired no more than two months away from the ACSFCS observation times;  this proved to
be important since on 2004 December 20, the HST
secondary mirror was moved by $4.6$~$\mu$m.  

After running KINGPHOT \citep{jordan05} on the GC candidates identified in the 
ACSFCS images,\footnote{KINGPHOT fits two-dimensional,
PSF-convolved King models to candidate GCs in the ACS images.} 
it was found that, for a subset of galaxies (FCC~213, IC~2006, 
FCC~193, FCC~249, FCC~277, FCC~19, and FCC~202), the mean half-light radius for GC candidates
was  significantly larger in the $g$-band than in the $z$-band: i.e., by roughly 0.5 pixels in F475W, 
which is much larger than  the $\lesssim0.1$ pixel differences
found in the ACSVCS.  \citet{anderson06} showed that the WFC PSF
exhibits unpredictable variations on orbital timescales, particularly in the bluer 
filters, with differences in flux values of up to $\sim10$\% in the central regions.
To correct the seven galaxies whose imaging suffered from this variability, stellar sources in the individual
images were used to adjust empirically the 47 Tucanae PSFs.  Full details on 
this procedure are given in \citetalias{jordan07a}.

The azimuthally-averaged, one-dimensional (1D) surface brightness profiles were fitted
using a $\chi^2$ minimization scheme to determine if a S\'ersic
or core-S\'ersic model was most appropriate.  If visual inspection of the images and/or surface brightness
profiles revealed a nucleus, then an {\tt S2} parameterization was 
adopted.
At each iteration of the fitting procedure, the models used were convolved 
with the PSF in two dimensions (assuming spherical symmetry),
and both the models and PSF were oversampled by a factor of 10 with respect
to the ACS pixel size (i.e. they were sampled every 0\farcs005).

All profile parameters, except for intensity, were first fitted to both bandpasses
simultaneously.  These preliminary values were then used as initial guesses for the
independent $g$- and $z$-band fits for  most of the galaxies, with the exception of those with high 
central surface brightness that appear to be nucleated.  In these galaxies, the nuclei 
are often quite extended, and difficult to differentiate from the underlying galaxy light; 
thus, only the intensity parameters were allowed to vary between the two bands.
 As many previous investigators have noted, it is possible to measure reliably the total magnitudes
and effective radii of marginally resolved stellar systems (i.e., star clusters, nuclei) using HST imaging,
whereas the concentrations can usually be constrained with considerably lower precision 
\citep{kundu98, larsen99, carlson01, jordan05}. This is understandable given that the measurement
of concentration  (or S\'ersic index) for a stellar system requires the {\it curvature} of the profile to be measured on scales
smaller than the PSF.
Fortunately, the derived radii and magnitudes 
are quite insensitive to S\'ersic index, at least insofar as the adopted model is an accurate
representation of the actual nuclear profile.

A conservative resolution limit of $0\farcs025$ was estimated in \citetalias{cote06} based on the 
half-light radii of King models fit to stars classified as unresolved by
KINGPHOT, and from the size of the central non-thermal point source 
found in VCC~1316 (M87).  
\citetalias{cote06} further showed that most of their detected nuclei were more extended
than point sources, by fitting point source profiles in addition 
to King profiles, and comparing the residuals.  Four of the nucleated
galaxies (FCC~301, FCC~249, FCC~255, FCC~95) in Table~\ref{tab:datanuc} have best-fit effective radii that are measured to be
smaller than our resolution limit in one, or both, photometric bands; these nuclei
are thus unresolved --- or nearly so --- in our HST imaging.

% Nuclei
\tabletypesize{\tiny}	
\begin{deluxetable*}{rlcccccccc}
%\rotate
\tablecolumns{10} 
\tablewidth{0pt}
\tablecaption{Data for ACSFCS Nuclei\label{tab:datanuc}}
\tablehead{
\colhead{ID}         & 
\colhead{Name}        & 
\colhead{$g$}    &
\colhead{$z$}    &
\colhead{$(g-z)$}    &
\colhead{$(g-z)^a$}    &
\colhead{$R_{e,g}$} &
\colhead{$R_{e,z}$} &
\colhead{$L_{>R,g}/L_g$} & 
\colhead{$L_{>R,z}/L_z$} \\
\colhead{} & 
\colhead{} & 
\colhead{(mag)} &
\colhead{(mag)} &
\colhead{(mag)} &
\colhead{(mag)} &
\colhead{($\arcsec$)} &
\colhead{($\arcsec$)} &
\colhead{($R=0\farcs5$)} & 
\colhead{($R=0\farcs5$)} \\
\colhead{(1)} &
\colhead{(2)} &
\colhead{(3)} &
\colhead{(4)} &
\colhead{(5)} &
\colhead{(6)} &
\colhead{(7)} &
\colhead{(8)} &
\colhead{(9)} &
\colhead{(10)}\\
}
\startdata
 8  &  2006  &     $18.17\pm0.12$  &     $16.35\pm0.10$  &     $1.82\pm0.15$  &     $1.66\pm0.06$  &     $0.132\pm0.013$  &     $0.139\pm0.012$  &     0.07  &     0.08  \\
11  &    63  &     $15.22\pm0.07$  &     $13.53\pm0.07$  &     $1.70\pm0.09$  &     $1.49\pm0.04$  &     $0.889\pm0.047$  &     $0.927\pm0.046$  &     0.69  &     0.71  \\
12  &   193  &     $17.97\pm0.07$  &     $16.54\pm0.06$  &     $1.43\pm0.09$  &     $1.41\pm0.05$  &     $0.100\pm0.006$  &     $0.097\pm0.004$  &     0.00  &     0.00  \\
13  &   170  &     $17.26\pm0.04$  &     $15.82\pm0.03$  &     $1.45\pm0.05$  &     $1.43\pm0.03$  &     $0.228\pm0.008$  &     $0.207\pm0.004$  &     0.19  &     0.15  \\
14  &   153  &     $19.06\pm0.05$  &     $18.29\pm0.03$  &     $0.77\pm0.07$  &     $0.64\pm0.04$  &     $0.153\pm0.004$  &     $0.153\pm0.003$  &     0.01  &     0.00  \\
15  &   177  &     $17.76\pm0.09$  &     $16.95\pm0.06$  &     $0.82\pm0.10$  &     $0.84\pm0.03$  &     $0.130\pm0.020$  &     $0.099\pm0.010$  &     0.10  &     0.06  \\
16  &    47  &     $16.09\pm0.19$  &     $14.86\pm0.20$  &     $1.24\pm0.30$  &     $1.33\pm0.05$  &     $0.750\pm0.125$  &     $0.612\pm0.119$  &     0.61  &     0.56  \\
17  &    43  &     $21.57\pm0.21$  &     $20.05\pm0.55$  &     $1.52\pm0.64$  &     $0.88\pm0.06$  &     $0.039\pm0.028$  &     $0.127\pm0.173$  &     0.02  &     0.16  \\
18  &   190  &     $19.67\pm0.17$  &     $18.64\pm0.07$  &     $1.03\pm0.18$  &     $0.96\pm0.05$  &     $0.129\pm0.022$  &     $0.121\pm0.007$  &     0.02  &     0.01  \\
19  &   310  &     $18.64\pm0.22$  &     $17.29\pm0.11$  &     $1.35\pm0.27$  &     $1.37\pm0.08$  &     $0.359\pm0.061$  &     $0.328\pm0.026$  &     0.35  &     0.30  \\
20  &   249  &     $20.08\pm0.12$  &     $19.22\pm0.06$  &     $0.85\pm0.12$  &     $0.99\pm0.08$  &     $0.038\pm0.007$  &     $0.018\pm0.004$  &     0.00  &     0.00  \\
21  &   148  &     $16.38\pm0.16$  &     $15.69\pm0.17$  &     $0.70\pm0.25$  &     $0.70\pm0.03$  &     $0.270\pm0.082$  &     $0.233\pm0.057$  &     0.33  &     0.29  \\
22  &   255  &     $20.22\pm0.03$  &     $19.14\pm0.02$  &     $1.08\pm0.04$  &     $0.95\pm0.03$  &     $0.028\pm0.002$  &     $0.023\pm0.002$  &     0.00  &     0.00  \\
23  &   277  &     $20.08\pm0.16$  &     $18.75\pm0.07$  &     $1.33\pm0.18$  &     $1.32\pm0.06$  &     $0.089\pm0.017$  &     $0.082\pm0.005$  &     0.02  &     0.01  \\
24  &    55  &     $20.16\pm0.02$  &     $18.98\pm0.02$  &     $1.18\pm0.03$  &     $1.05\pm0.03$  &     $0.064\pm0.002$  &     $0.057\pm0.002$  &     0.01  &     0.00  \\
26  &   301  &     $20.32\pm0.03$  &     $19.29\pm0.02$  &     $1.03\pm0.04$  &     $0.87\pm0.03$  &     $0.016\pm0.003$  &     $0.015\pm0.002$  &     0.00  &     0.00  \\
27  &   335  &     $19.95\pm0.02$  &     $18.81\pm0.02$  &     $1.14\pm0.03$  &     $1.19\pm0.03$  &     $0.094\pm0.003$  &     $0.066\pm0.002$  &     0.05  &     0.02  \\
29  &    95  &     $21.25\pm0.04$  &     $20.10\pm0.04$  &     $1.15\pm0.06$  &     $1.12\pm0.04$  &     $0.035\pm0.004$  &     $0.013\pm0.005$  &     0.00  &     0.00  \\
30  &   136  &     $20.38\pm0.03$  &     $19.31\pm0.03$  &     $1.07\pm0.04$  &     $0.95\pm0.03$  &     $0.055\pm0.003$  &     $0.042\pm0.003$  &     0.05  &     0.03  \\
31  &   182  &     $22.15\pm0.07$  &     $21.62\pm0.13$  &     $0.53\pm0.15$  &     $0.38\pm0.15$  &     $0.038\pm0.002$  &     $0.038\pm0.002$  &     0.00  &     0.00  \\
32  &   204  &     $20.00\pm0.10$  &     $18.86\pm0.09$  &     $1.13\pm0.14$  &     $1.00\pm0.03$  &     $0.092\pm0.020$  &     $0.093\pm0.018$  &     0.10  &     0.10  \\
33  &  119\tablenotemark{a}  &     $20.20\pm0.02$  &     $19.56\pm0.12$  &     $0.63\pm0.12$  &     $0.58\pm0.05$  &     $0.025\pm0.003$  &     $0.030\pm0.010$  &  \nodata  &  \nodata  \\
34  &    90  &     $21.28\pm0.08$  &     $20.31\pm0.07$  &     $0.97\pm0.10$  &     $0.84\pm0.07$  &     $0.073\pm0.004$  &     $0.066\pm0.004$  &     0.00  &     0.00  \\
36  &   106  &     $20.69\pm0.04$  &     $19.54\pm0.05$  &     $1.15\pm0.07$  &     $1.08\pm0.04$  &     $0.042\pm0.003$  &     $0.036\pm0.003$  &     0.00  &     0.00  \\
37  &    19  &     $20.86\pm0.04$  &     $20.02\pm0.03$  &     $0.85\pm0.05$  &     $0.78\pm0.03$  &     $0.042\pm0.002$  &     $0.032\pm0.003$  &     0.01  &     0.01  \\
38  &   202  &     $20.57\pm0.02$  &     $19.64\pm0.03$  &     $0.94\pm0.04$  &     $0.88\pm0.03$  &     $0.053\pm0.003$  &     $0.047\pm0.003$  &     0.01  &     0.01  \\
39  &   324  &     $22.92\pm0.04$  &     $22.13\pm0.03$  &     $0.79\pm0.05$  &     $0.70\pm0.04$  &     $0.040\pm0.004$  &     $0.028\pm0.004$  &     0.00  &     0.00  \\
40  &   288  &     $21.32\pm0.03$  &     $20.41\pm0.03$  &     $0.91\pm0.04$  &     $0.82\pm0.03$  &     $0.081\pm0.003$  &     $0.075\pm0.003$  &     0.00  &     0.00  \\
41  &   303  &     $19.72\pm0.03$  &     $18.77\pm0.03$  &     $0.96\pm0.04$  &     $0.83\pm0.03$  &     $0.079\pm0.004$  &     $0.078\pm0.005$  &     0.08  &     0.08  \\
42  &   203  &     $21.78\pm0.08$  &     $20.92\pm0.06$  &     $0.86\pm0.10$  &     $0.78\pm0.04$  &     $0.051\pm0.005$  &     $0.040\pm0.004$  &     0.01  &     0.01  \\
43  &   100  &     $21.01\pm0.04$  &     $20.10\pm0.03$  &     $0.91\pm0.05$  &     $0.75\pm0.03$  &     $0.072\pm0.003$  &     $0.071\pm0.003$  &     0.03  &     0.02  \\
\enddata
\tablenotetext{~}{Column key:\\
(1) ACSFCS Identification number;\\
(2) Galaxy name, mainly from the Fornax Cluster Catalog (FCC) of \citet{ferguson89};\\
(3)--(4) $g$- and $z$-band magnitudes for the nuclei;\\
(5) Integrated color of nuclei;\\
(6) Nuclei color within a 4-pixel radius aperture;\\
(7)--(8) {\tt S2} model effective (half-light) radius in the $g$- and $z$-bands;\\
(9)--(10) $g$- and $z$-band luminosity fraction residing beyond $0\farcs5$.}
\tablenotetext{a}{Due to the offset of the nucleus and the amount of central dust,
the nucleus parameters for FCC 119 were derived using a King profile fit to the ACS image.}
\end{deluxetable*}

After some experimentation, we have estimated uncertainties on the fitted parameters for the nuclei (and their host galaxies) 
using a Monte Carlo approach in which the $g$- and $z$-band
surface brightness profiles for each galaxy are independently simulated 200 times. We included an amount of noise at each
data point in the profile assuming a Gaussian distribution of errors and using the uncertainty on the intensity at each point 
computed by {\tt ELLIPSE}. An additional source of error for the profiles comes from the determination of the background level,
which we have also included by assuming a 10\% error in the adopted background for each galaxy (estimated roughly by
the galaxy-to-galaxy scatter in the measured background levels; see Figure~5 of \citetalias{jordan07a}). Errors on the magnitudes, colors
and effective radii estimated from these Monte Carlo simulations are given in Table~1. We hasten to point out that these errors do not include possible sources of 
systematic errors, such as errors in the PSF, and that they are therefore best viewed in a relative sense, and as lower limits 
on the true errors.

Finally, as a check on the (1D) method, we also fitted surface brightness 
profiles to our galaxies using 2-dimensional (2D) techniques. The full results of 
this analysis are described in \S\ref{app:2D}. In brief, the 
structural parameters obtained from the two procedures are largely in agreement, but
due to the increased difficulty of characterizing complex structures using 2D fitting, 
we proceed with results from the 1D method, which we consider most appropriate for this study.

\subsection{Identification of the Nuclei} \label{sec:idNuc}

The classification of a galaxy as nucleated or non-nucleated 
was performed in the following way.
The program galaxies were all fitted with  pure S\'ersic profiles
outside of a geometric mean radius of 0\farcs5 ($\sim 50$~pc).  
The geometric mean radius was
derived from the fitted elliptical isophotes, and is thus defined as 
$R\equiv a(1-\epsilon)^{1/2}$, where $a$ is the semi-major axis, and $\epsilon$ 
is the ellipticity.  If an inward extrapolation
of this profile revealed an excess of light in the center, then the
full profile was refitted by adding a second S\'ersic component, and 
the galaxy was thus considered nucleated and classified as {\tt S2}.  
In general, the level of nucleation 
was slightly greater in the $g$-band, as the nuclei 
are often found to be somewhat bluer than their hosts (see \S\ref{sec:colors}). 

One of our program galaxies, FCC~119, appears to have a distinct 
nucleus offset that is from its photocenter by $\sim0\farcs7$. 
Due to the presence of dust in the inner regions of the galaxy, the ellipse 
centers were held fixed to the photocenter throughout the fit; 
thus, the nucleus is not apparent in the 1D surface brightness profile (discussed below).
We therefore use parameters derived from a KINGPHOT fit to this object, 
and consider this galaxy to be nucleated for the remainder of our analysis.

Galaxy classifications as nucleated or non-nucleated in the FCC, and our 
revised classification, are presented in columns 8 and 9 of Table~\ref{tab:datagal}.
In Table~\ref{tab:datanuc}, we record the parameters of the S\'ersic profile fit 
to the nucleus of all {\tt S2} galaxies, as well as the KINGPHOT fit to FCC~119.
Specifically, we have measured the $g$- and $z$-band integrated 
nucleus magnitudes (columns 3 and 4), integrated and 4-pixel radius aperture nucleus colors 
(columns 5 and 6), and $g$- and $z$-band nucleus half-light radii (columns 7 and 8). 
Error estimates for each of these parameters are also included in this table, derived using the Monte 
Carlo approach described in \S\ref{sec:fitProc}.
We have also calculated, by integrating the S\'ersic profiles, the fraction of luminosity
occurring outward of $R>0\farcs5$ (columns 7 and 8).

Although the nuclei half-light radii in a few galaxies were measured to be somewhat
larger in the $g$-band than in the $z$-band (i.e. FCC~310, FCC 177, FCC 95), we note that 
these are not the same galaxies that suffered from the variable PSF 
discussed in \S\ref{sec:fitProc}; for the most part, these  differences reflect the fact that 
size measurements are particularly challenging for under-luminous or extended nuclei in galaxies with
steeply rising surface brightness profiles.  For the reasons discussed in \S\ref{sec:fitProc},
we do not report the best-fit S\'ersic indices in Table~\ref{tab:datanuc}, although we note
that the indices for all nuclei in our sample have $0.5 \lesssim n \lesssim 4$, with a median
of $n=2.0\pm0.7$.

F475W images for the central $10\arcsec\times10\arcsec$ region of the program 
galaxies, where a distinct nuclear component is 
often discernible, are shown in Figure~\ref{fig:acs_images}.
The FCC number of the galaxy  is
labeled in each of the panels, along with the type of profile fitted; {\tt  S2} therefore
indicates that the galaxy was considered to
be nucleated (that is, fitted with a double-S\'ersic profile).
Individual fits to the azimuthally-averaged $g$-band surface brightness 
profiles are shown in Figure~\ref{fig:sb_profiles}. These images illustrate the systematic
trend noted in \citetalias{cote07}, in which the central regions of early-type galaxies 
transition from shallow ``cores" in the brightest systems \citep{ferrarese94, lauer95, faber97, rest01, ravindranath01, lauraa06} to a
two-component structure (nucleus+galaxy) as one moves down the luminosity function:
i.e., toward fainter, and lower surface brightness, galaxies.

% Nucleus luminosity histogram
\begin{figure}
	\figurenum{4}
	\centering 
	\leavevmode 
	\includegraphics[width = 0.85\linewidth, trim =0cm 2cm 0cm 0cm, clip=true]{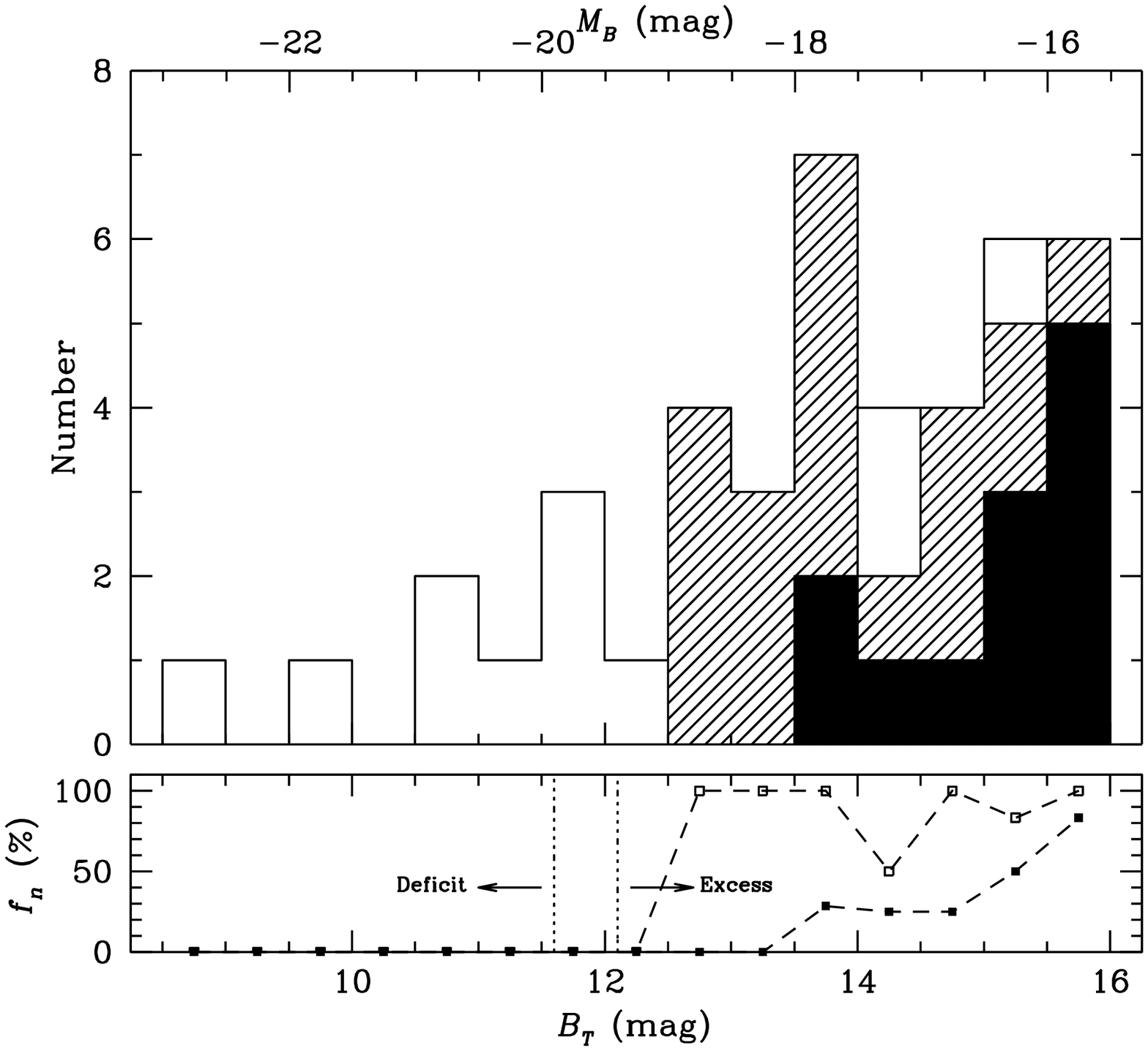}
	%\plotone{f4.eps}
	\caption{\emph{Top}: Luminosity distribution of the 43 
	ACSFCS program galaxies (open histogram).  The overlaid 
	hatched histogram shows the distribution 
	of the 31 galaxies classified as nucleated in this study, while the solid 
	histogram shows the distribution of the 12 nucleated galaxies according 
	to the FCC.
	\emph{Bottom}: The percentage of nucleated galaxies ($f_n$) 
	in this study (open squares) and in the FCC (solid squares). The approximate luminosity regimes
	where galaxies show central surface brightness ``deficits" and ``excesses" (i.e., nuclei) 
	are indicated \citepalias[see][]{cote07, glass11}.}
	\label{fig:lum_hist}
\end{figure}

% Nucleus SB vs nucleus mag
\begin{figure}
	\figurenum{5}
	\plotone{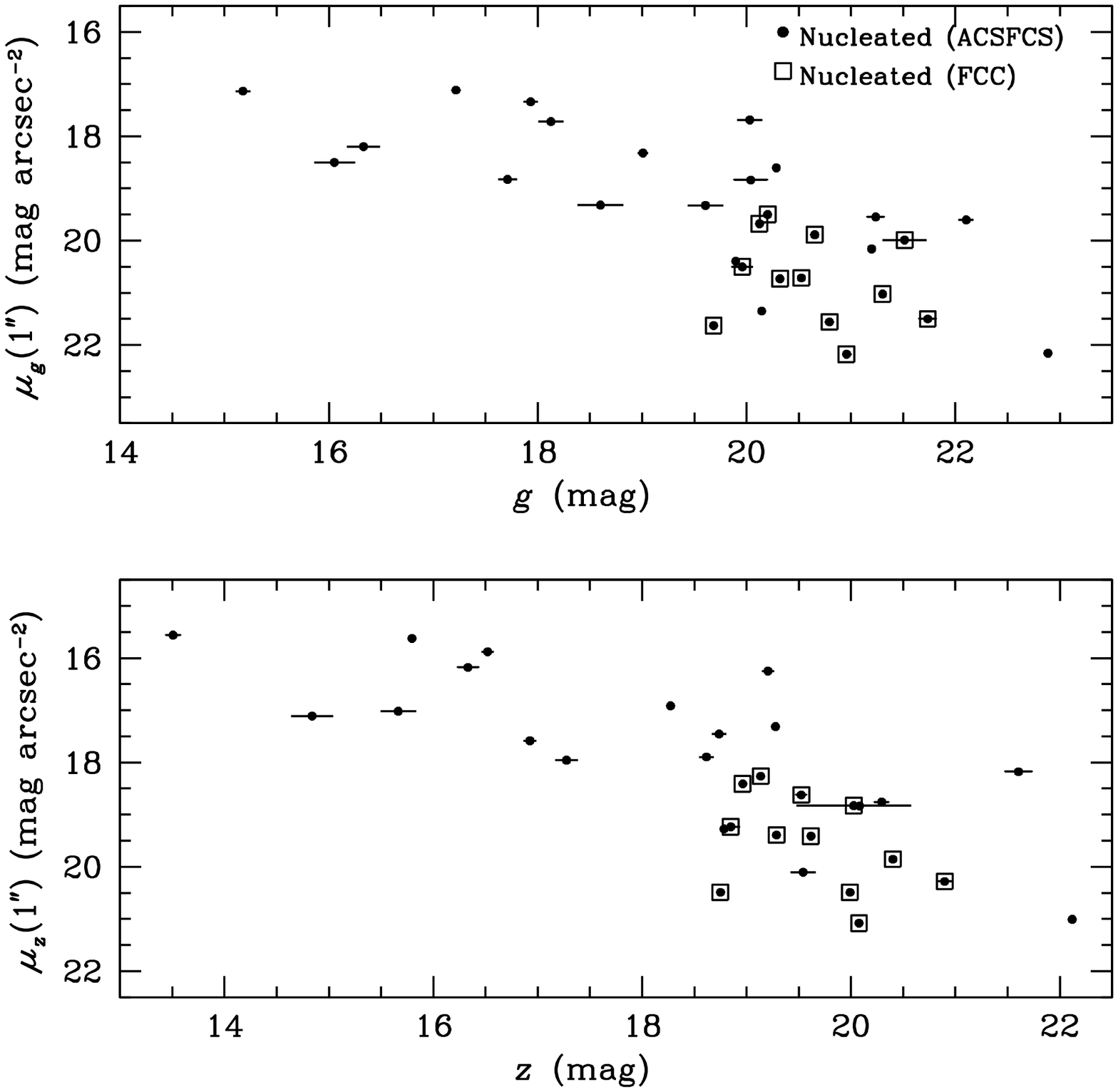}
	\caption{Galaxy surface brightness in the $g$-band (\emph{top}) and 
	$z$-band (\emph{bottom}) measured at a mean radius of $1\arcsec$, 
	plotted against the magnitude of the nucleus.  The 
	filled circles show 
	the 31 galaxies found to be nucleated by this study, while the open 
	squares show the 12 galaxies classified as nucleated in the FCC.}
	\label{fig:galsb_vs_nucmag}
\end{figure}

%
% RESULTS
%

\section{Results} \label{sec:results}

In the following section, we analyze the 
properties of the ACSFCS nuclei derived from the above
parameterization. Specifically, we examine the frequency
of nucleation (\S~\ref{sec:freq}), offset of the nuclei
from their hosts (\S~\ref{sec:offset}), nucleus-to-galaxy
luminosity ratio (\S~\ref{sec:eta}), and nucleus
luminosity function (\S~\ref{sec:lf}). Additionally, 
we compare their structural
properties and scaling relations (\S~\ref{sec:scaling}),
as well as colors (\S~\ref{sec:colors}), with those of 
their host galaxies and GCs. 

\subsection{Frequency of Nucleation}\label{sec:freq}

Only 12 out of our 43 program galaxies were classified
as nucleated in the FCC, which sets the frequency of nucleation at $f_n\approx28\%$.\footnote{We include
NGC~1340 and IC~2006 in this calculation; although they do not appear in the catalog of Ferguson (1989),
both have ``E" classifications in NED (i.e., non-nucleated ellipticals).} 
Column 8 of Table~\ref{tab:datagal} shows the classification as nucleated or
non-nucleated in the FCC. These can be compared to our classification 
in the ACSFCS, where the use of the double-S\'ersic ({\tt S2}) model
indicates that we consider the galaxy to be nucleated.  
We find all galaxies previously classified as nucleated in the FCC to be
nucleated in our sample, as well as an additional 19 objects, for a total of 
31/43 galaxies, or $f_n\approx72\%$. 

The cause of this sharp rise in frequency of nucleation
can be attributed to both observational selection effects and the definition in our analysis
of a nucleus as a central luminosity excess relative to a fitted galaxy model.
In the top panel of Figure~\ref{fig:lum_hist}, the open histogram shows the luminosity
of all of the program galaxies, while the hatched and solid histograms
denote those found to be nucleated in the ACSFCS and the FCC, respectively.
The bottom panel of Figure~\ref{fig:lum_hist} plots $f_n$ as a function of luminosity 
for the two surveys.  The ACSFCS uncovers many more nuclei in more luminous host 
galaxies, as the high resolution of the WFC allows us to  
resolve nuclei in their high surface brightness cores.  
This selection effect is explored further in 
Figure~\ref{fig:galsb_vs_nucmag}.

Galaxy surface brightness at a geometric mean radius of $R=1\arcsec$ ($\approx97$ pc) 
was calculated using linear spline interpolation, in the $g$- and $z$-bands. 
By measuring surface brightness at a constant radius 
(rather than at some function of the effective radius), the result is a model-independent 
measure of central surface brightness, at a distance large enough to avoid the contribution 
from a typical nucleus, if present.  Figure~\ref{fig:galsb_vs_nucmag} plots the integrated 
nucleus magnitude derived from the S2 fit against galaxy surface brightness measured at a 
distance of 1\arcsec. The filled circles and open squares show the measurements for galaxies 
classified as nucleated in the ACSFCS and FCC, respectively.  Clearly, the nuclei that went 
undetected in the earlier (photographic) survey come in two forms: bright nuclei that are 
embedded at the centers of galaxies with intermediate luminosity (which also have steeply 
rising profiles, see Figure~\ref{fig:sb_profiles}) and faint nuclei belonging to the lowest 
luminosity galaxies. Needless to say, it is possible that we too may be missing some nuclei, so
we take $f_n\approx72\%$ as a lower limit on the true frequency of nucleation in the ACSFCS sample.

% OFfset 
\begin{figure}
	\figurenum{6}
	\plotone{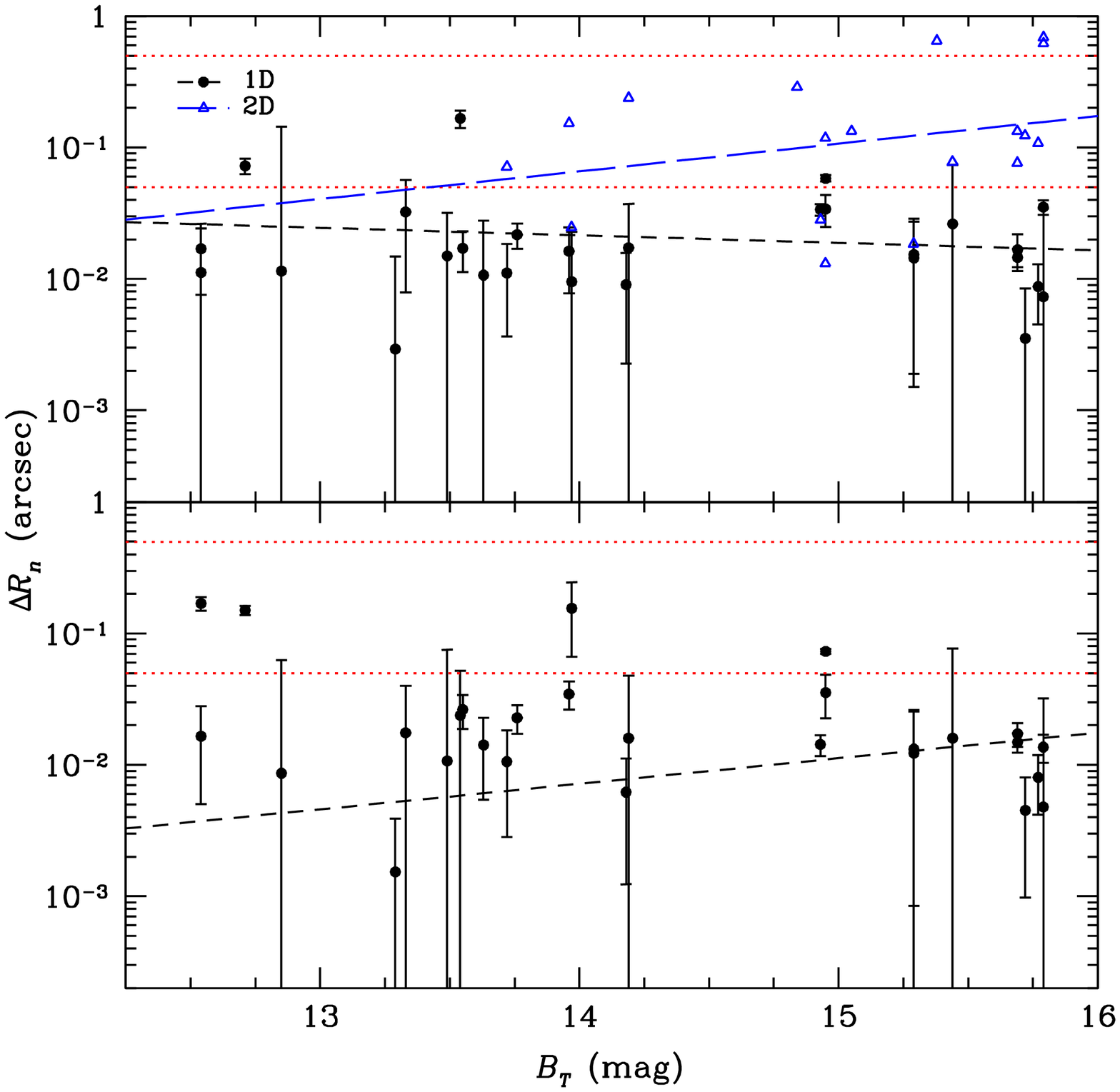}
	\caption{\emph{Top}: Projected offset between the nucleus and the galaxy
		photocenter in the $g$-band, plotted against host galaxy magnitude. 
		Offsets were calculated using our 1D (black circles) and 2D (blue 
		triangles) analyses. The 
		two dotted red lines show offsets one and ten ACS/WFC pixels 
		($0\farcs05$ and $0\farcs5$). The black short-dashed line
		and blue long-dashed line represent the best-fit relation for 1D 
		and 2D offsets, respectively.
		\emph{Bottom}: Same as above, but for the $z$-band. }
	\label{fig:offsetA}
\end{figure}

\subsection{Offset Nuclei}\label{sec:offset}

The offset of each nucleus from its host galaxy photocenter was measured for 28 
of our 31 nucleated galaxies.  For FCC 335, FCC 119, and FCC 90, the elliptical isophote fitting 
was performed with the ellipse centers held 
fixed, as convergence could not be otherwise achieved due to dust in their 
central regions. Thus, offsets for these nuclei could not be 
measured using the technique described below, although we did examine their offsets 
using our 2D (GALFIT) analysis, and they are also included in this section.

For the remaining 28 galaxies in question, an analysis slightly different
to that used in \citetalias{cote06} was performed. In \citetalias{cote06}, the galaxy photocenter was determined 
by taking the mean of the positions of all fitted isophotes satisfying 
$1\arcsec \leq R \leq R_e$. However, because of the possibility that isophotes
might drift from the center due to bright sources in the field of view (causing an 
artificial offset in the photocenter calculated using the above method), we have adopted a different
procedure in this work, where the photocenter and its error was determined by running {\tt ellipse} 
to fit a single isophote with a semi-major axis length of approximately $R_e / 2$. 
As in \citetalias{cote06}, the position and error of the centroid of the nucleus were taken as the smallest 
fitted ellipse from the full {\tt ellipse} run. We note that the geometrical parameter errors 
output by {\tt ellipse} are calculated from the errors of the harmonic fit, with the first
and second harmonics removed. The errors from the photocenter and centroid were then added 
in quadrature to obtain the total error on the offset.
The results of this procedure are plotted as the black filled circles in Figure~\ref{fig:offsetA}.
We find that almost all of the galaxies in our sample have an offset of less than $0\farcs1$.
The four that do have a larger offset (FCC~63, FCC~193, FCC~177, and FCC~277) have
$\Delta R_n > 0\farcs1$ in only one of the two bands. 

We generally observe the offsets from our 2D analysis to be larger than those determined using
our 1D method. This is due to the fact that the 2D fitting procedure does not allow the ellipse
parameters to vary with radius, and returns the model that best fits the average parameters of the entire
galaxy, giving more weight to the outer regions in the determination of the photocenter. 
Thus, for our 2D analysis, we are not concerned with offsets larger than $\sim 0\farcs5$, and
only three galaxies are found to have offsets larger than this --- FCC 119 ($0\farcs65$), 
FCC 324 ($0\farcs70$), and FCC 288 ($0\farcs62$). FCC 119 is fairly irregular in structure, with a significant amount 
of dust in its core. FCC 324 and FCC 288 are both low surface brightness, highly flattened galaxies,
with no obvious clusters near the photocenter that may have caused 
source confusion with what we consider to be the nucleus. We conclude that in
our sample, at most $10$\% of the nuclei are offset at the level of $0\farcs5$ or more, 
consistent with the findings of \citetalias{cote06} for the Virgo cluster. 

To measure any trend between offset and galaxy luminosity, we perform a weighted least-squares fit
to the data from Figure~\ref{fig:offsetA}. 
Using the offsets from our 1D analysis, we obtain
\begin{equation}
%\begin{split}
\begin{array}{rcl}
\log \Delta R_{g} & = & (-0.057\pm0.070)\, B_{T} - (0.87\pm1.06) \\
\log \Delta R_{z} & = & (0.19\pm0.09)\, B_{T} - (4.9\pm1.3) ,\\
%\end{split}
\end{array}
\end{equation}
and from our 2D analysis,
\begin{equation}
%\begin{split}
\begin{array}{rcl}
\log \Delta R_{\rm 2D} & = & (0.21\pm0.18)\, B_{T} - (4.1\pm2.7). 
%\end{split}
\end{array}
\end{equation}
The slopes of these relations do not indicate any significant trend between offset and
galaxy luminosity. The errors on the fitted parameters are the standard errors.

Finally, some of the galaxies in our sample that we do
not find to be nucleated may, in fact, be ``dIrr/dE transition" objects, where a nucleus
could be in the process of formation.\footnote{A prototype for this class is VCC1512 in
the Virgo cluster which contains a prominent central excess that is composed
of blue, densely packed star clusters.}
In particular, FCC 152 and FCC 26 are irregular in shape and contain many star clusters 
and significant amounts of dust in their central regions. It is possible that one or more 
of these clusters could be nucleus progenitors that will migrate inwards through dynamical
friction (see \S\ref{sec:gcinfall}).

% Nucleus-to-galaxy luminosity ratio
\begin{figure}
	\figurenum{7}
	%\plotone{f7.eps}
	\centering 
	\leavevmode 
	\includegraphics[width = 0.85\linewidth, trim =0cm 3.5cm 2cm 0cm, clip=true]{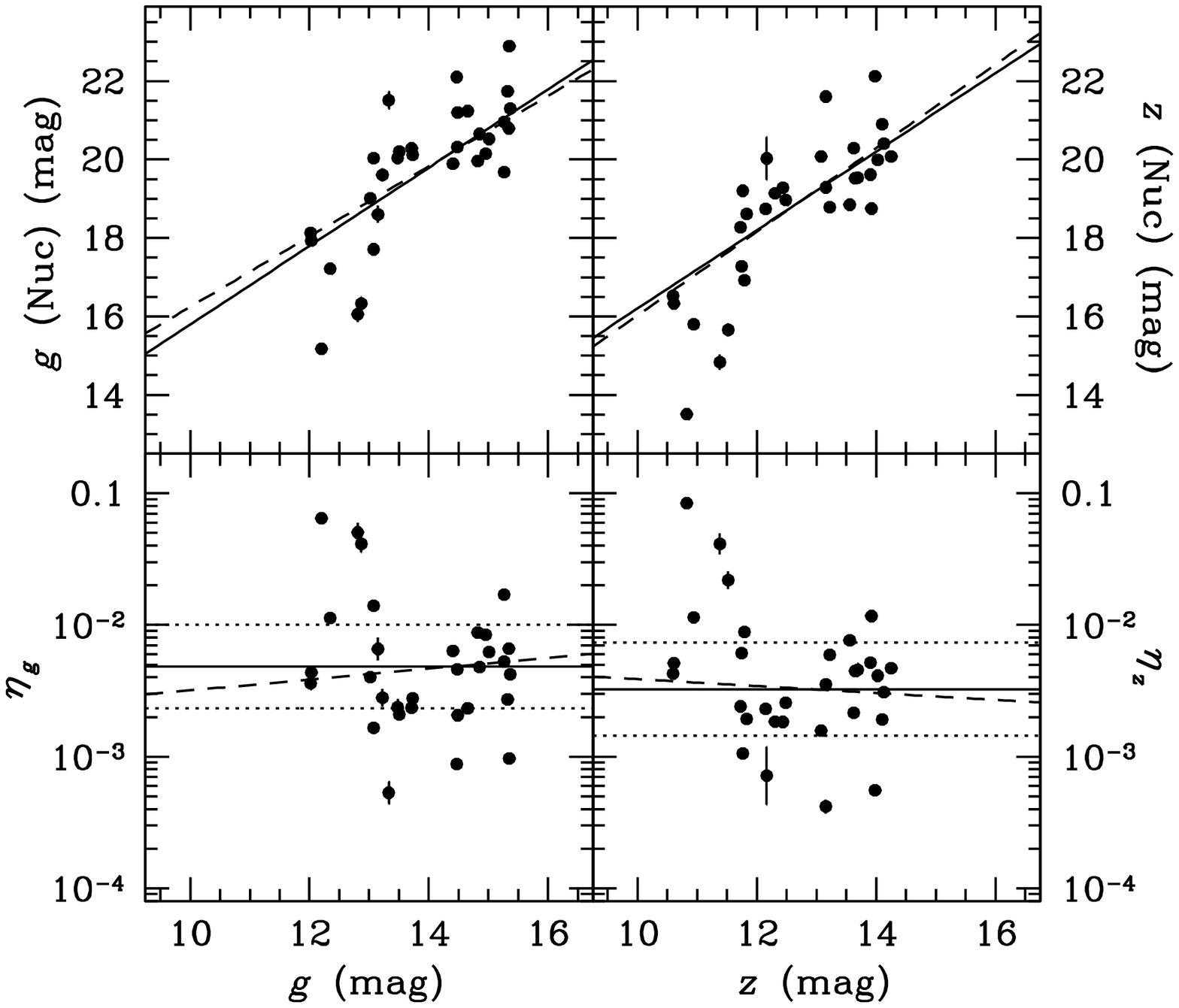}
	\caption{\emph{Top}: Nucleus magnitude plotted against host galaxy magnitude
	for nucleated galaxies in the ACSFCS; results for the $g$ and $z$ bands are 
	shown on the left and right, respectively.  The lines show the best-fit relations, with the slope
	held fixed at unity (\emph{solid}) and allowed to vary (\emph{dashed}). 
	\emph{Bottom}: Nucleus-to-galaxy luminosity ratio, $\eta$, against host galaxy
	magnitude for the $g$-band (\emph{left}) and $z$-band (\emph{right}). The solid 
	and dotted lines show the mean value of $\eta$ and its $\pm$1$\sigma$ limit, while the
	dashed line shows the best fit relation given by the dashed line in the upper panel,
	recast in terms of $\log(\eta)$ and host magnitude.}
	\label{fig:eta}
\end{figure}

\subsection{The Nucleus-to-Galaxy Luminosity Ratio}
\label{sec:eta}

Previous studies of early-type dwarfs \citep{lotz04, grant05, graham03b}, including \citetalias{cote06}, found that nucleus brightness increases 
with host galaxy brightness. Similar relations are known to exist for the nuclear clusters in late-type galaxies
\citep[see, e.g.,][]{carollo98,boker04}.  A plot of nucleus versus host galaxy magnitude,  the latter calculated by 
integrating the S\'ersic profile best-fitting the main galaxy component over all radii, is shown at 
the top of Figure~\ref{fig:eta}.  Weighted best-fit linear relations of the form:
\begin{equation}
%\begin{split}
\begin{array}{rcl}
g_{\rm nuc} & = & {\alpha}g_{\rm gal} + \beta \\
z_{\rm nuc} & = & {\alpha}z_{\rm gal} + \beta \\
\label{eq:eta1}
%\end{split}
\end{array}
\end{equation}
%\begin{equation}
%\begin{split}
%g^{\prime}_{\rm n} &= g^{\prime}_{\rm g}+b_2\\
%z^{\prime}_{\rm n} &= z^{\prime}_{\rm g}+b_2
%\end{split}
%\end{equation}
were fitted to the data, where $g_{\rm n}$ and $z_{\rm n}$ are nuclei magnitudes
and $g_{\rm g}$ and $z_{\rm g}$ are the galaxy magnitudes. The best-fit parameters
($\alpha_1$, $\beta_1$, $\beta_2$) are given in Table~\ref{tab:eta}, where the quoted errors
are the standard errors. Results are given for the two cases of fixing the
slope at $\alpha_2 \equiv 1$, and allowing it to vary freely (shown respectively as the solid and dashed lines in the
upper panel of Figure~\ref{fig:eta}).

% Eta 
\begin{deluxetable}{lcccccccc}
%\tabletypesize{\footnotesize}	
\tabletypesize{\small}
\tablecaption{Nucleus-to-Galaxy Luminosity Ratios\label{tab:eta}}
\tablewidth{0pt}
\tablehead
{
	\colhead{Sample} & 
	\colhead{Band} & 
	\colhead{$\alpha_1$} & 
	\colhead{$\beta_1$} & 
	\colhead{$\beta_2$} & 
	\colhead{$\langle{\log{\eta}}\rangle$} & 
	\colhead{$\sigma$} \\
	\colhead{} & 
	\colhead{} & 
	\colhead{} & 
	\colhead{(mag)} & 
	\colhead{(mag)} & 
	\colhead{(dex)} & 
	\colhead{(dex)} 
}
\startdata
ACSFCS & $g$ & $0.90\pm0.17$ & $7.27\pm2.45$ & $5.79\pm0.15$ & $-2.31$ & $0.32$  \\
ACSFCS & $z$ & $1.07\pm0.16$ & $5.37\pm2.12$ & $6.20\pm0.16$ & $-2.49$ & $0.35$ 
\enddata
%\tablecomments{}
%\tablenotetext{a}{}
\end{deluxetable}

Since the best-fit slope of the nucleus-galaxy luminosity relation is very nearly one,
we consider the possibility of a constant nucleus-to-galaxy luminosity ratio,  
$\eta=\mathcal{L}_{\rm n}/\mathcal{L}_{\rm g}$,
%defined as:
%\begin{equation}
%\eta=\mathcal{L}_{\rm n}/\mathcal{L}_{\rm g}
%\end{equation}
where $\mathcal{L}_{\rm n}$ and $\mathcal{L}_{\rm g}$ are nucleus and galaxy luminosity, 
respectively.   In the bottom of Figure~\ref{fig:eta},  $\eta$ is plotted as a function 
of host galaxy magnitude in the same band.  The values of the weighted means and standard deviations 
are given in Table~\ref{tab:eta}, while the weighted mean ratio and standard error on the mean from both bands is 
\begin{equation}
	\langle\eta\rangle = 0.41\%\pm0.04\%.
\end{equation}
This is $0.11\pm0.06\%$ larger than the value of $\langle\eta\rangle = 0.30\pm0.04$\% found in \citetalias{cote06} (a 1.9$\sigma$ discrepancy). At
 first glance, this might suggest that, at a given luminosity, early-type galaxies in Fornax were slightly more efficient in
 assembling their nuclei than those in Virgo; however, the difference is due to the use
of S\'ersic rather than King models in fitting the ACSFCS nuclei.  As S\'ersic
profiles with even moderate $n$ have somewhat extended wings, they increase the inferred luminosity of the 
nuclei relative to the King models (whose defining characteristic is a tidal truncation radius).
Re-fitting the Virgo data with {\tt S2} profiles confirms this conclusion --- in \S~\ref{sec:virgoEta}, where
the new fits to Virgo are presented, we find agreement between $\eta$ for both clusters. 

Finally, due to the definition of $\eta$, the best-fit relation from Equation~\ref{eq:eta1}
can be recast in terms of $\log(\eta)$ and galaxy magnitude, where $\alpha_\eta = -0.4 \left(\alpha_1 -1\right)$ 
and $\beta_\eta = -0.4 \beta_1$. This relation is plotted as the dashed line in the bottom panels of 
Figure~\ref{fig:eta}. Although we do not see any trend between $\eta$ and galaxy luminosity,
 we note that in a study of galaxies containing
both a nucleus and a black hole by \citet{graham09}, it was found that the ratio 
of total CMO mass to spheroid mass tended to decrease in more massive galaxies.

\subsection{Luminosity Function}\label{sec:lf}

\begin{figure}
	\figurenum{8}
	\plotone{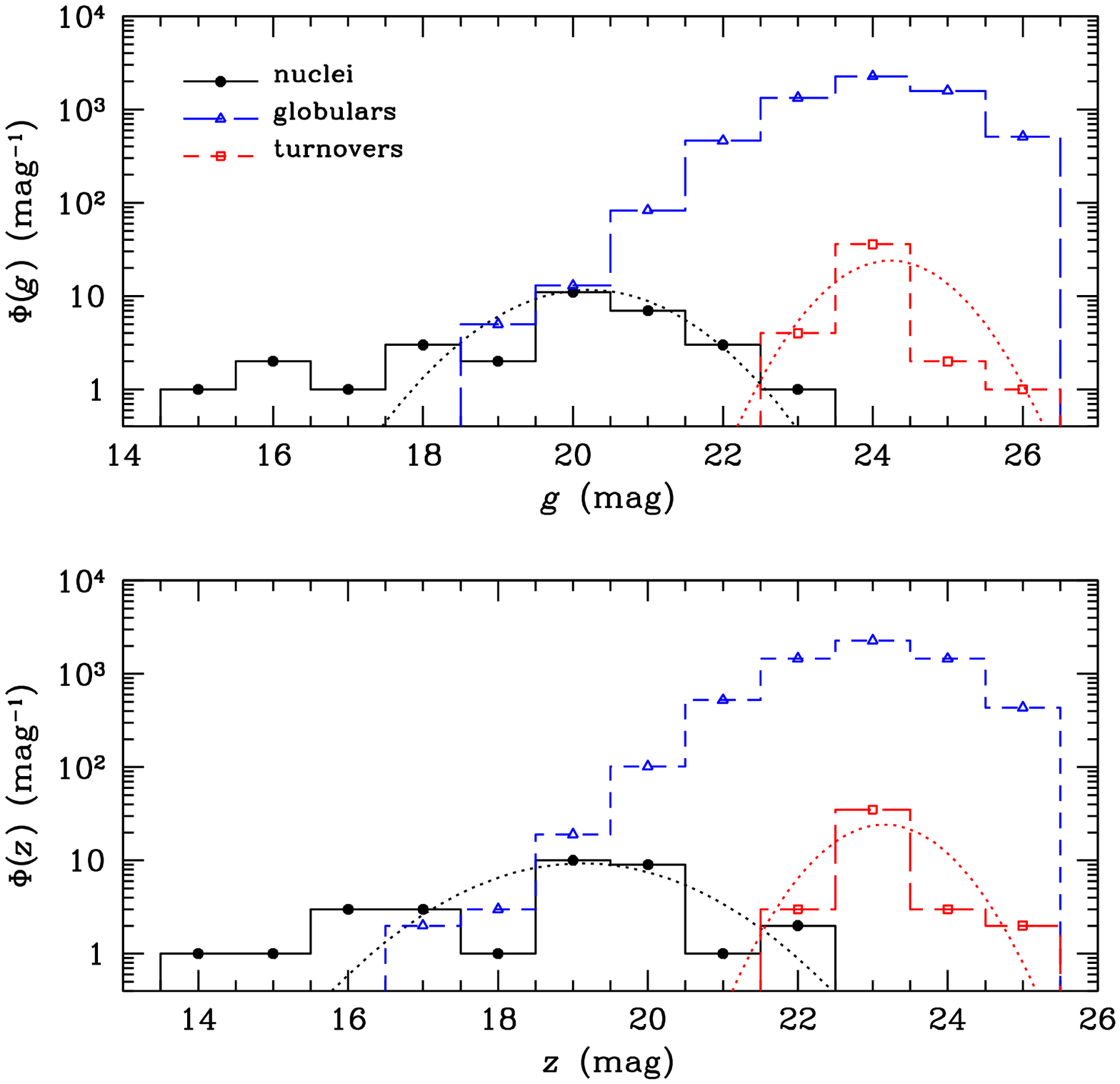}
	\caption{Luminosity function for the nuclei (closed black circles) 
	in the $g$ (\emph{top}) and $z$ bands (\emph{bottom}). The distribution
	of the ACSFCS GC candidates(open blue triangles) , as well as the luminosity function of their
	turnovers (open red squares) from \citetalias{villegas10} plotted for comparison. 
	Both luminosity functions are derived by fitting normalized Gaussians.
	For the GC turnovers, each turnover magnitude was weighted by the number
	of GCs in the galaxy.}
	\label{fig:lum_fun}
\end{figure}

One mechanism for the formation of galaxy nuclei is through multiple mergers
of GCs that sink to the galaxy center by dynamical friction 
\citep[e.g.,][]{tremaine76, capuzzo93, capuzzo99, lotz01, bekki04}.
A comparison of the luminosity function of our nuclei with that of the 
GCs identified in the ACSFCS can offer some insight into this process.  
In Figure~\ref{fig:lum_fun}, we present the results of a weighted maximum-likelihood fit to the 
luminosity functions of the nuclei, using a normalized Gaussian:
\begin{equation}
	\Phi(m_{\rm nuc}^0) \propto \exp\left[-(m_{\rm nuc}^0-\bar{m}_{\rm nuc}^0)
	/2\sigma_{\rm nuc}^2\right].
	\label{eq:gaussian}
\end{equation}
While this choice of parameterization is commonly used for GCs, there
is no physical reason that the nuclei should have a Gaussian distribution. 
It is, nevertheless, a useful departure point for the purpose of comparison with the GCs.
To parameterize the GC luminosity function, we also performed a maximum-likelihood
fit of a normalized Gaussian, using the GC turnover magnitudes for each galaxy 
(which have been corrected for completeness), taken from \citetalias{villegas10}. Each 
turnover magnitude was weighted by the number of GCs in the galaxy.

 % Nucleus and glob luminosity function

% Convolved Schechter function
\begin{figure}
	\figurenum{9}
	\plotone{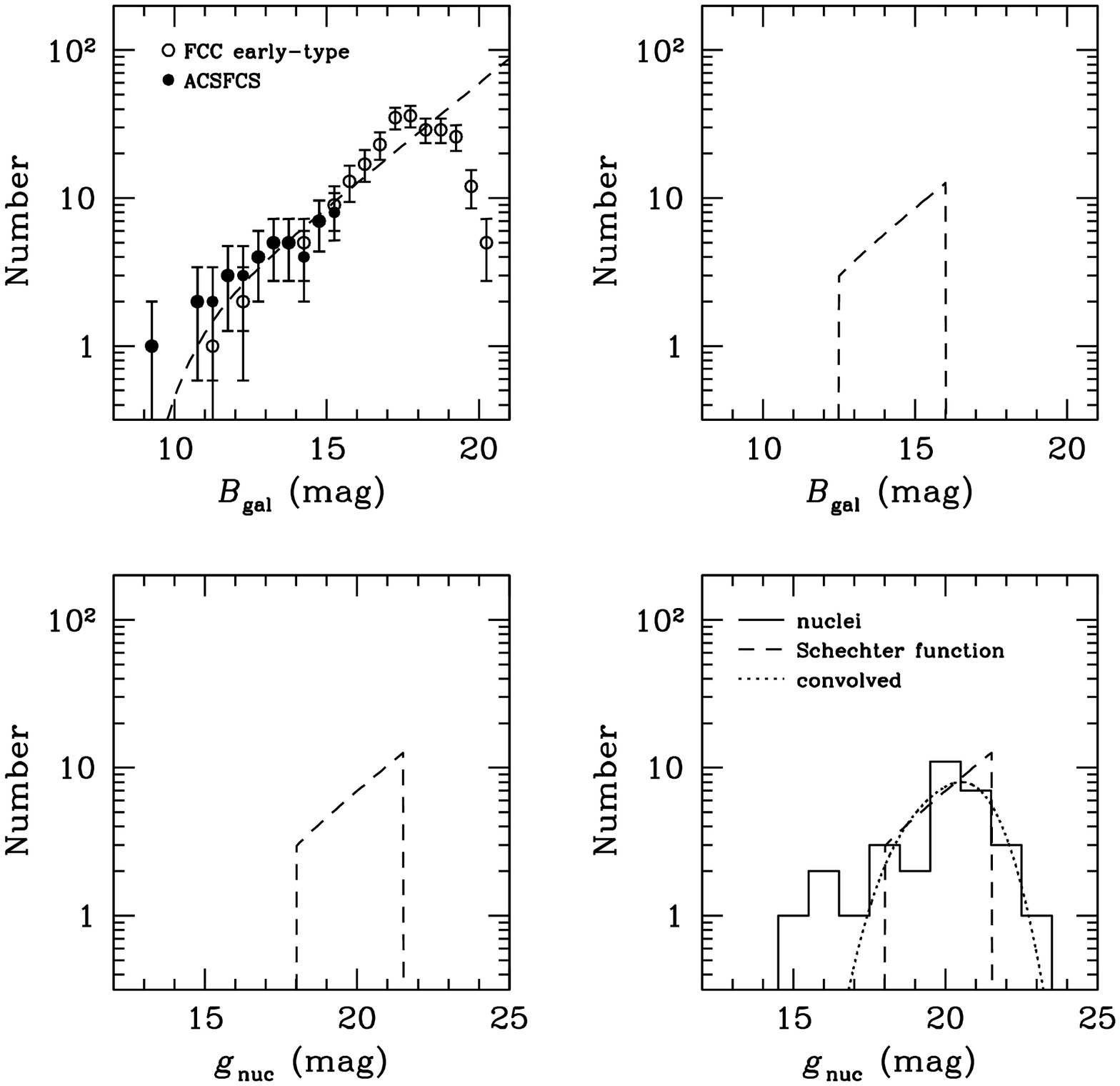}
	\caption{\emph{Top left}: A Schechter function plotted over the $B$-band
	galaxy luminosity distribution for all early-type galaxies in the 
	FCC (open circles) and the ACSFCS sample (filled circles). 
	\emph{Top right}: The previous Schechter function truncated at $B_{T}=12.5$
	and $16$, and reduced by 91\%, so that it represents the
	nucleated galaxies in our sample.
	\emph{Bottom left}: The previous Schechter function shifted by $+6.0$~mag, 
	corresponding to a constant $\langle\eta\rangle$ of 0.41\%, and $-0.4$~mag
	to convert from $B$ to $g$. It should now be roughly correspond
	to the nucleus luminosity distribution, although without taking
	into account the scatter. 
	\emph{Bottom right}: The luminosity distribution of the nuclei
	(solid line), the previous Schechter function (dashed line), and
	the same Schechter function convolved with a
	Gaussian of $\sigma=0.87$~mag, the unweighted standard deviation of $\eta_{g}$
	(dotted line). }
	\label{fig:schechter}
\end{figure}

Our GC sample consists of $\approx2000$ candidates with probability index 
$\mathcal{P}_{\rm gc} \geq 0.5$ (see \citetalias{villegas10} for more details on the GC probability 
index, and a detailed study of the GC luminosity functions).
The best-fit parameters for both nuclei and GCs are given in Table~\ref{tab:lum_fun}, 
where the errors on the fitted parameters are the standard errors.
We find the luminosity function of the nuclei to be both brighter, and have a  greater
spread, than that of the GCs.  
The difference in the means is $\Delta = 4.03$ mag in each band;
that is, on average the nuclei are $\sim40\times$ brighter than a typical GC. 

In reality, since we find the nucleus-to-galaxy luminosity ratio to be roughly constant,
the nucleus luminosity function should reflect that of the host galaxies 
(albeit with more scatter), and
 is most likely parameterized by a Schechter function truncated on both ends --- on the bright end because we find no
bright galaxies that are nucleated, and on the faint end because our sample is 
magnitude limited.
To illustrate this, we show in Figure~\ref{fig:schechter} a Schechter function overlaid 
on the $B$-band galaxy luminosity
distribution of the FCC and the ACSFCS. 
We then apply cutoffs at $B_{T}=12.5$ and $16$~mag (so that we are 
left with the magnitude range of the nucleated galaxies in our sample), 
and scale it down by $91\%$ (since we have 34 galaxies in this magnitude 
range, 31 of which are nucleated), and shift the Schechter function over 
by $+6.0$~mag (which corresponds to an
$\langle\eta\rangle$ of 0.41\%) and $-0.4$~mag (to convert  roughly from
$B$ to $g$). Finally, we convolve it with a Gaussian with $\sigma=0.87$~mag, 
the unweighted standard deviation of $\eta$, to take the scatter around 
$\langle\eta\rangle$ into account.\footnote{We use the unweighted standard
deviation in this case, because we are looking to reproduce the observed 
rather than intrinsic scatter.} The resulting function plotted over
the nucleus luminosities shows good agreement, apart from a few bright outliers;
specifically, the nuclei from FCC~63, FCC~47, and FCC~148 have $B_T \lesssim 17$.
Each of these galaxies have a complex central structure that may be contributing to 
their brightness, either by causing accurate nucleus parameterization to be 
more difficult, or due to the fact that, in these cases, the 
nucleus themselves may be more complex.

% Luminosity function
\begin{deluxetable}{lccccc}
%\tabletypesize{\scriptsize}
\tabletypesize{\small}
%\rotate
\tablecaption{Nucleus and Globular Cluster Luminosity Functions \label{tab:lum_fun}}
\tablewidth{0pt}
\tablehead
{
\colhead{Sample} & 
\colhead{Band} & 
\colhead{$\bar{m}^0_n$} & 
\colhead{$\sigma_n$} &
\colhead{$\bar{m}^0_{gc}$} & 
\colhead{$\sigma_{gc}$} \\
\colhead{} & 
\colhead{} & 
\colhead{(mag)} & 
\colhead{(mag)} &
\colhead{(mag)} & 
\colhead{(mag)}
}
\startdata
ACSFCS & $g$ & $20.21\pm0.01$ & $1.06\pm0.01$ & $24.24\pm0.01$ & $0.71\pm0.01$\\
ACSFCS & $z$ & $19.12\pm0.01$ & $1.33\pm0.01$ & $23.15\pm0.01$ & $0.71\pm0.01$
\enddata
%\tablecomments{}
%\tablenotetext{a}{}
\end{deluxetable}

% Nucleus and GC half-light radii
\begin{figure}
	\figurenum{10}
	\plotone{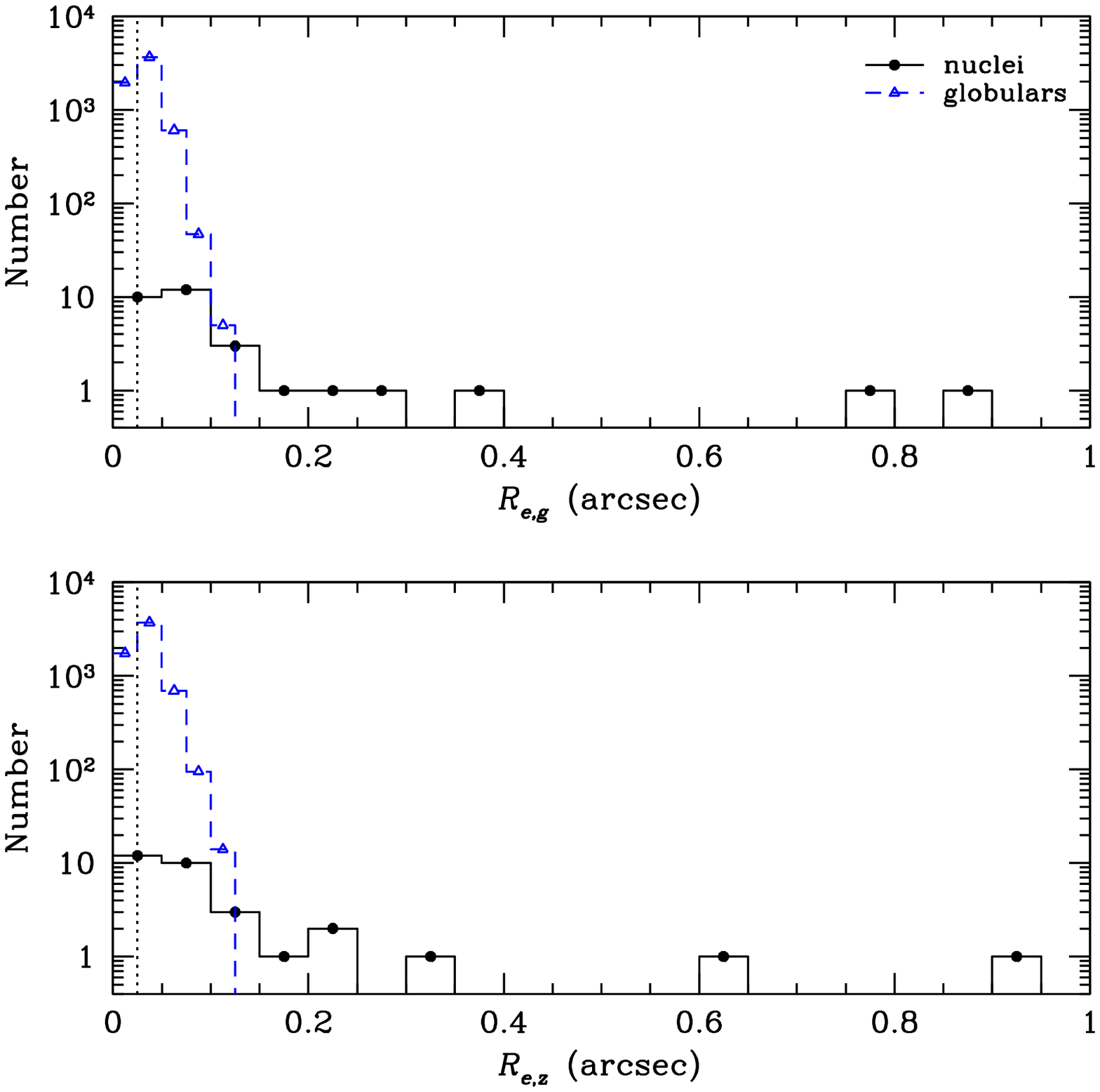}
	\caption{Distribution of half-light radii of the 31 nuclei identified in this study 
	(black filled circles), as well as the candidate ACSFCS GCs (blue open triangles), measured in the
	the $g$- ({\it top}) and $z$-bands ({\it bottom}). 
	The vertical dotted lines indicate the adopted resolution limit of $\sim0\farcs025$.}
	\label{fig:size_hist}
\end{figure}

% Nucleus scaling relations
\begin{figure}
	\figurenum{11}
	\plotone{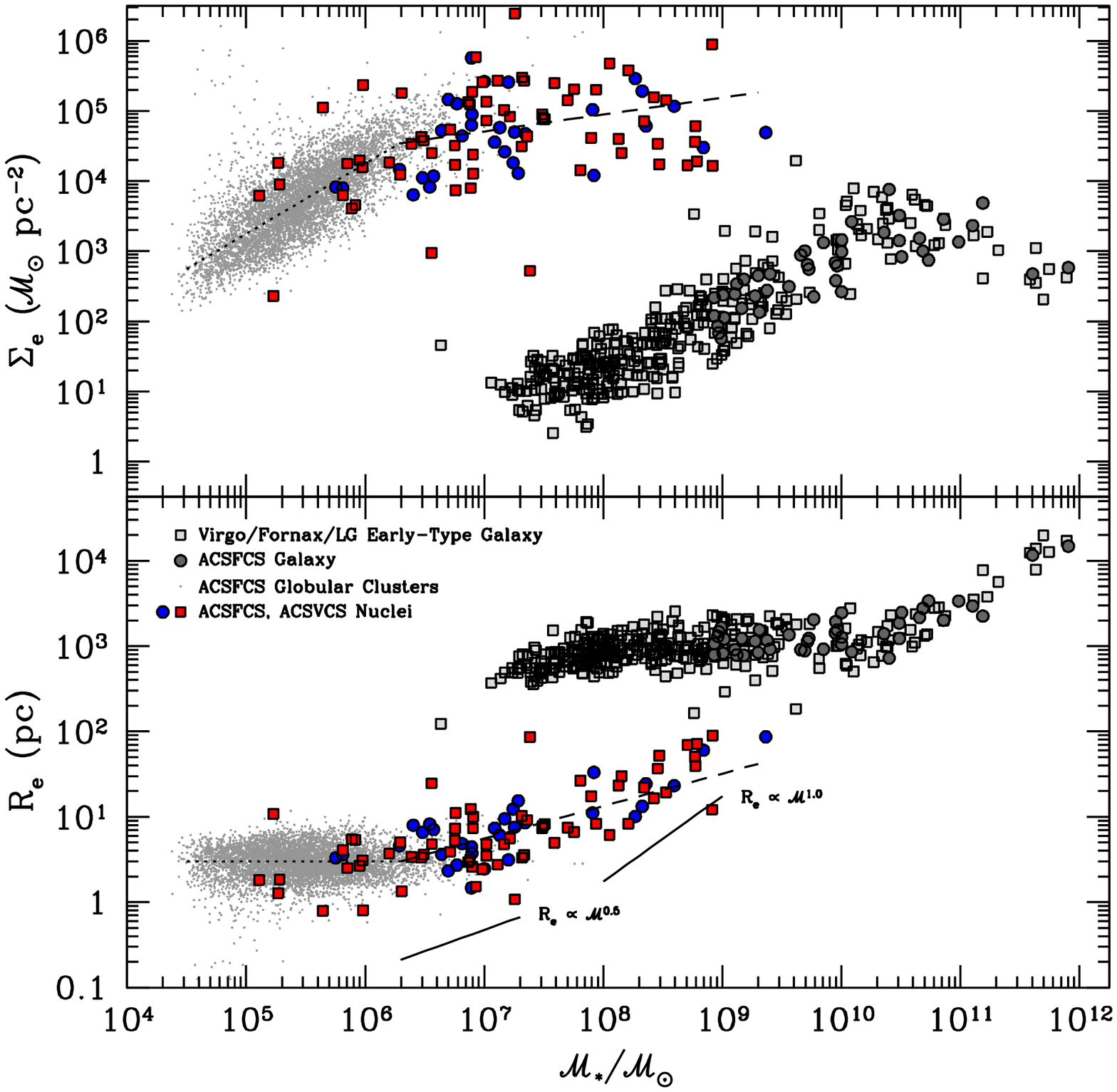}
	\caption{Scaling relations of nuclei compared to galaxies and globular clusters from the ACSFCS and ACSVCS surveys
	(with data for galaxies from McLaughlin et~al. 2012, in prep.).
	\emph{Top}: Stellar mass surface density plotted against stellar mass. The dotted line shows the relation for GCs, which
	have $R_e \simeq 3$~pc (\citealt{jordan05}; \citetalias{masters10}). The dashed line shows the relation calculated from 
	the \citet{bekki04} finding that
	nuclei assembled from repeated GC mergers have $R_e \propto L^{0.38}$.
	\emph{Bottom}: Effective radius plotted against stellar mass for the same stellar systems. The curves are the same as
	in the previous panel. The solid lines show scaling relations of the form $R_e \propto M^{0.5}$ and $R_e \propto M$.
	See text for details.}
	\label{fig:scaling}
\end{figure}
 
\subsection{Structural Properties and Scaling Relations}\label{sec:scaling}

Nuclei at the distance of the Fornax cluster are almost never resolved in ground-based
imaging, as 1\arcsec~corresponds to $\approx$100~pc
at a distance of 20~Mpc. However, with ACS resolution it is possible to measure 
sizes for nuclei as small as $R_e \sim 0\farcs025$ \citepalias[see][]{cote06}.
In Figure~\ref{fig:size_hist}, we present a comparison of the effective radii of the ACSFCS nuclei and 
GC candidates.  On average, the nuclei are larger in size, and have a much
greater spread, than the GCs, although the considerable overlap 
between the two distributions shows that the most compact nuclei are 
very nearly the same size as typical GCs \citepalias{masters10}.

The two most prominent outliers are the nuclei of FCC~63 and FCC~47; both of these nuclei
were also found to be the brightest in our sample (see \S~\ref{sec:lf}).
Regardless, Figure~\ref{fig:size_hist} clearly demonstrates that the nuclei have a size distribution that 
peaks at compact sizes, and an extended tail populated by larger nuclei. The
median sizes of the full sample are found to be 
 $0\farcs073$ (7.2~pc) and $0\farcs071$ (7.0~pc) in the $g$- and $z$-bands, respectively.

Figure~\ref{fig:scaling} shows scaling relations for the nuclei from the ACSFCS and
ACSVCS. In the upper panel, we plot the effective mass surface density, 
$\Sigma_e \equiv {\cal M}_*/{2{\pi}R_e^2}$, against
total stellar mass, ${\cal M}_*$, calculated from the observed $(g-z)$ colors and the relations of \citet{bell03}.
The lower panel shows effective radius as a function of stellar mass. In both panels,
we also plot ACSFCS GCs, and the  sample of early-type galaxies in Virgo, Fornax
and the Local Group from McLaughlin et~al. (2012, in prep.). As found by \citet{jordan05}, the
GCs have a size of $R_e \simeq 3$~pc that is nearly independent of mass, while the early-type
galaxies show a smoothly varying ${\cal M}_*$--$R_e$ relation,
a reflection of the fact that galaxies form a non-homologous family (McLaughlin et~al. 2012, in prep.).

This figure highlights several other interesting properties of the nuclei. First, there is an obvious 
similarity in the scaling relations of the Fornax and Virgo nuclei; we shall return to
this point and its implications for nucleus formation models in \S\ref{sec:compare} and \S\ref{sec:models}.
The addition of the ACSFCS nuclei also reaffirms the trend noted by \citetalias{cote06} that the nuclei, unlike GCs, obey a  
size-mass relation that merges with the GC sequence at low mass. For reference, the 
dashed line in the lower panel of Figure~\ref{fig:scaling} shows the predicted scaling relation for nuclei
that are assembled from repeated GCs mergers, $R_e \propto {\cal M}_*^{0.38}$ \citep{bekki04}. 
The corresponding $\Sigma_e$--${\cal M}_*$ relation is shown in the upper panel.
Based on structural parameters alone, we conclude that the GC merger model is broadly
consistent with the data (although the extremely red colors of the brightest nuclei pose
a challenge to this model in its simplest form). The two solid lines in the lower panel show
relations of the form  $R_e \propto {\cal M}_*^{0.5}$ and $R_e \propto {\cal M}_*$, which
will be discussed in \S\ref{sec:gcinfall}.

\subsection{Nuclei Colors}
\label{sec:colors}

\begin{figure}
	\figurenum{12}
	\plotone{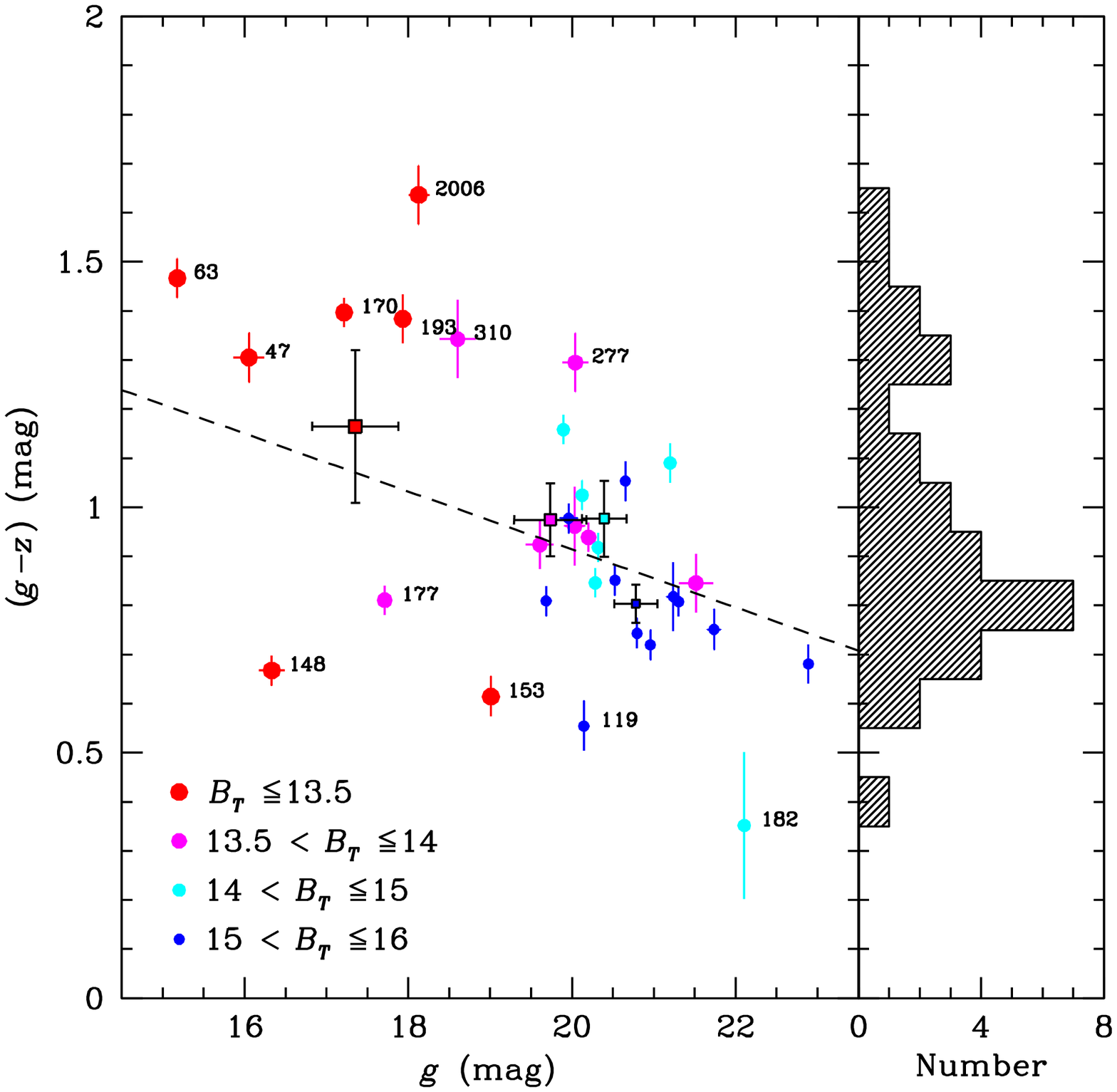}
	\caption{Color-magnitude diagram 
	for the 31 nuclei identified in this
	study, with colors derived using 4-pixel apertures. Point size is scaled with magnitude
	as indicated. The dashed line is the weighted best fit for galaxies fainter
	than $B_T=13.5$.  Galaxies with $B_T\leq13.5$ or with unusually red or blue
	nuclei are labeled. The mean and standard error of the mean for each luminosity 
	bin are indicated by the outlined squares.  \emph{Right}: Histogram of the nuclei colors showing
	a possible bimodal, or skewed, distribution.}
	\label{fig:nuc_col_ap}
\end{figure}

As in \citetalias{cote06}, we find a relationship between nuclei colors and magnitudes
with the brighter nuclei having redder colors and residing
in more luminous hosts. This is shown in Figure~\ref{fig:nuc_col_ap},
where we plot the nucleus 4-pixel aperture colors against $g$-band magnitudes.  
Significant scatter is seen for the brighter galaxies, which are labeled with
their FCC number. This scatter was also seen in \citetalias{cote06}, although in Virgo galaxies 
bright nuclei appeared to be preferentially red, while in the case of Fornax, 
bright nuclei are seen to scatter to both red and blue colors. The increased 
scatter in the color of the bright nuclei may simply reflect the more complex 
formation and enrichment histories in their inner regions of brighter, more 
massive galaxies: mergers, gas inflow, star formation, and GC accretion would naturally 
lead to a greater degree of scatter in the general color-magnitude trend. However, we 
caution that firm conclusions are difficult to draw, since at least part of the scatter 
is likely the result of larger observational errors, given the  difficulty of measuring accurate
photometric parameters for nuclei residing in luminous, high surface brightness galaxies. 

% Nucleus aperture-magnitude

% Nucleus vs. host color
\begin{figure}
	\figurenum{13}
	\plotone{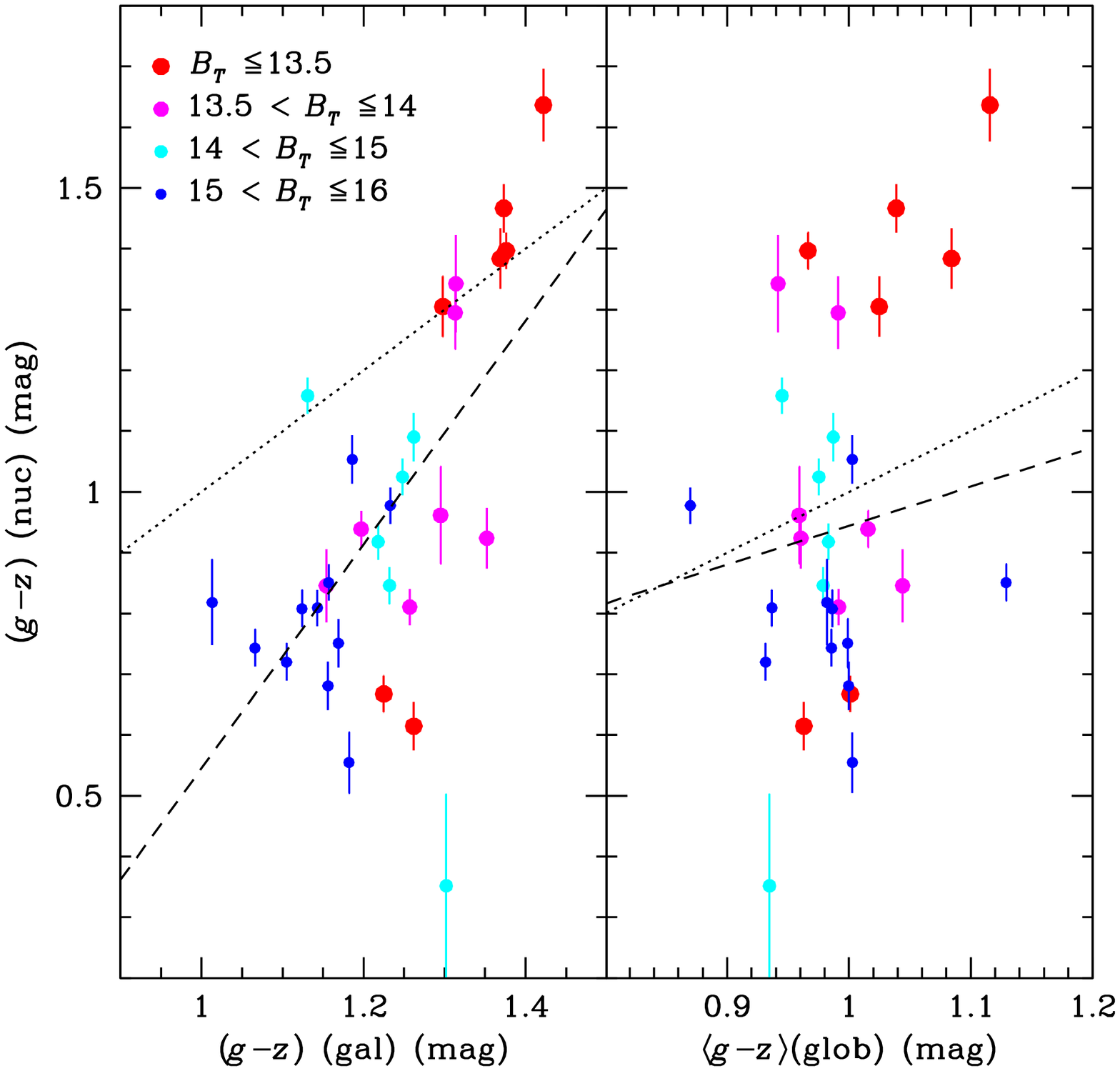}
	\caption{Nucleus aperture colors plotted against host galaxy colors 
	from \citetalias{blake09} (\emph{left}) and mean colors of the GC sample
	from \citetalias{villegas10} (\emph{right}).
	The sizes of the circles are proportional to the magnitude of the host galaxy.
	The dotted lines indicate equal colors, while the dashed lines 
	show the weighted best fit relation for the plotted points. }
	\label{fig:col_col}
\end{figure}

% Galaxy and nucleus colors against host magnitude
\begin{figure}
	\figurenum{14}
	\plotone{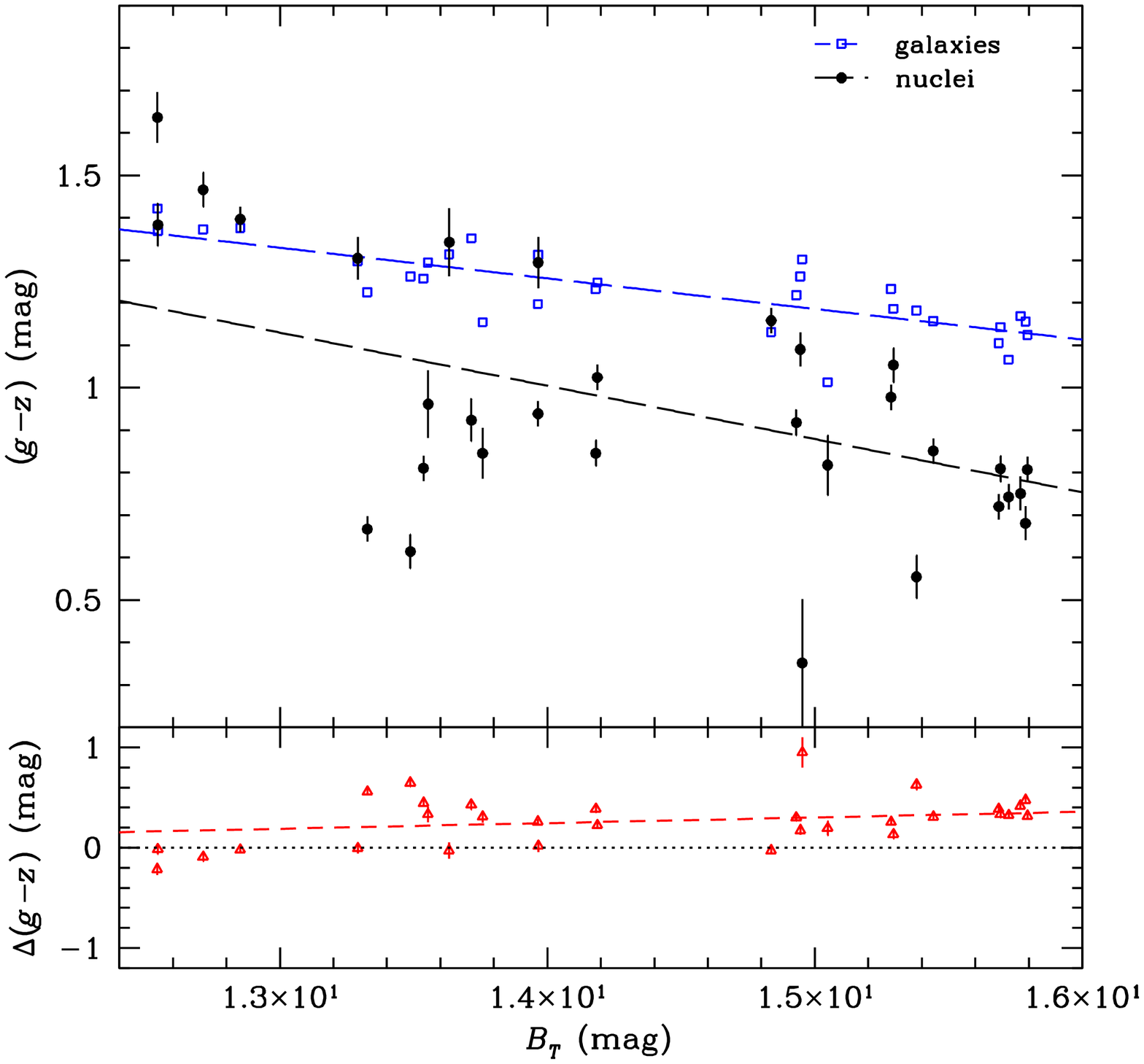}
	\caption{\emph{Top}: Galaxy colors (blue open squares, from \citetalias{blake09}) and nucleus
	aperture colors (black closed circles)
	plotted against host galaxy magnitude.
	Only galaxies that we find to be nucleated are shown. The black short-dashed
	and blue long-dashed lines are the best fit to the galaxies and nuclei, 
	respectively. 
	\emph{Bottom}: Difference between galaxy and nucleus color, as a function
	of host galaxy magnitude. The black dotted line marks a 
	difference of zero,
	while the red short-dashed line shows the best fit to all points.
	On average, nuclei are $\approx0.3$~mag
	bluer than their host galaxies.}
	\label{fig:col_mag}
\end{figure}

Figure~\ref{fig:nuc_col_ap} also shows the weighted line of best fit
for nuclei in host galaxies fainter than $B_T = 13.5$:
\begin{equation}
	(g-z)_{\rm nuc} = -(0.059\pm0.034)\,g_{\rm nuc} + (2.1\pm0.7).
	\label{eq:col}
\end{equation}
Such color-magnitude (or possibly metallictiy-mass) relations are generally 
thought to be a sign of self-enrichment in low-mass stellar systems 
\citep[e.g.,][]{dopita86, morgan89, brown91, recchi05, strader08, bailin09}.
It would not be surprising to observe the same self-enrichment in nuclei,
given the location of the nuclei at the centers of their host galaxies, 
where compressive tidal forces would aid in the retention of chemically enriched gas. 

The colors of the nuclei compared to the mean color of their host galaxy's GCs 
(calculated using the GC sample from \citetalias{villegas10}) are examined in the right-hand panel

of Figure~\ref{fig:col_col}. 
We find only a very weak trend that redder nuclei also have redder GCs, where a 
weighted least-squares fit gives (with standard errors) 
\begin{equation}
(g-z)_{\rm nuc} = (0.64\pm0.84)\,(g-z)_{\rm glob} + (0.30\pm0.83).
\end{equation}
Since mean GC color has been found to correlate with that of the host galaxy
\citep[e.g.,][]{larsen01, peng06}, we might expect to find a relation between 
the colors of nuclei and their GCs, given
that we also find a correlation between nuclei and galaxy colors, plotted in the 
left-hand panel of Figure~\ref{fig:col_col}. The weighted best-fit line with standard
errors is given by
\begin{equation}
(g-z)_{\rm nuc} = (1.84\pm0.39)\,(g-z)_{\rm gal} + (1.29\pm0.47),
\end{equation}
which indicates that
bluer nuclei tend to lie in bluer host galaxies, and vice-versa. 
The nuclei are also found to have a larger range in colors, and are 
in most cases bluer, than their host galaxies. 

In Figure~\ref{fig:col_mag}, we show galaxy and nuclei colors as a function
of host galaxy luminosity. The colors of both the galaxies and the nuclei
are found to become redder with increasing host luminosity:
\begin{equation}
%\begin{split}
\begin{array}{rcl}
(g-z)_{\rm gal} &= & -(0.072\pm0.010)\, B_{T} + (2.3\pm0.1), \\
(g-z)_{\rm nuc} &= & -(0.13\pm0.04)\, B_{T} + (2.8\pm0.5). \\
%\end{split}
\end{array}
\end{equation}
where the errors on the fitted parameters are the standard errors.
We find the nuclei colors to vary more steeply with host luminosity than 
those of the galaxies, although the trend for the nuclei
is quite weak for galaxies fainter than $B_T\sim13$.
Examining the offset between galaxy and nucleus colors reveals 
that those nuclei that are redder than their hosts lie predominantly 
in high-luminosity galaxies. The weighted least-squares relation and standard
errors for the color difference is given by
\begin{equation}
	\Delta_{(g-z)} = (0.056\pm0.033)\, B_{T} - (0.54\pm0.48).
\end{equation}
On average, we find the nuclei to be  bluer than their hosts by
$\langle\Delta(g-z)\rangle = 0.28\pm0.04$ mag. If we exclude the nuclei in galaxies with $B_{T}<13$
(the regime in which the nuclei are  found to be  redder than their hosts)
we obtain a mean offset of $\langle\Delta(g-z)\rangle = 0.32\pm0.03$ mag, where the errors
are the standard error of the mean.

%
% DISCUSSION
%

\section{Discussion} \label{sec:discussion}

%
% Comparison to ACSVCS
%

\subsection{The Role of Environment: Comparison to the ACSVCS} 
\label{sec:compare}

As described in \S\ref{sec:introduction}, our Fornax survey was preceded by a 
similar study of 100 early-type galaxies in the Virgo cluster (ACSVCS, \citealt{cote04}) 
where an  investigation 
into the properties of the nuclei in ACSVCS galaxies was carried out by \citetalias{cote06}. 
Our prime motivation for a study
of galaxies in the Fornax cluster is to provide a first glimpse into the properties of nuclei in 
two, rather different, clusters, and an assessment of the role played by
environment in nucleus formation and evolution.
The interested reader is referred to \S~1 of \citetalias{jordan07a}, which compares some 
key properties of the two clusters. Briefly, Virgo is overall a much larger 
cluster, with a mass almost 10 times that of Fornax 
(M$_{200} \sim 4.2\times10^{14}$~M$_\sun$ vs. $(1\mbox{--}7)\times10^{13}$ 
\citep{mclaughlin99, tonry00, drinkwater01}), and a velocity dispersion 
twice as large ($\sigma_v\sim760$ vs. $374$~km~s$^{-1}$ \citep{binggeli87, drinkwater01}).
Compared to the Virgo Cluster, Fornax is poorer (Richness Class 0 vs. 1, 
\citep{abell89, girardi95}) and more compact (R$_{200} \sim 0.7$ vs. $1.55$~Mpc).
Its intracluster medium (ICM) has both lower temperature (1.20 vs. 2.58 keV) and metallicity (0.23 vs. 0.34 solar) 
\citep{fukazawa98}, with the Fornax electron density at a given radius being about 1/4 that of Virgo
\citep{nulsen95, paolillo02}.

In this section, we will directly compare the results from both surveys.
While \citetalias{cote06} used King profiles for the nuclei in their paper, 
the ACSVCS results have since been updated with S\'ersic model fits to the nuclei,
which allows a fair comparison  between the two studies.\footnote{See 
{\tt https://www.astrosci.ca/users/VCSFCS/Data$\_$Products.html}}
Distances from \citetalias{blake09} were used
to calculate absolute magnitudes and physical sizes for both Fornax and Virgo galaxies.
We note that 
the two surveys have slightly different cutoff magnitudes ($B_{T}\simeq16$ for Virgo and $\simeq15.5$ for Fornax) and that the distance
modulus of Fornax is $\sim$~0.5~mag larger than that of Virgo \citepalias{blake09}. Therefore,
 the Virgo galaxies can reach absolute magnitudes roughly 1 mag fainter than those in Fornax.

\begin{figure}
	\figurenum{15}
	\plotone{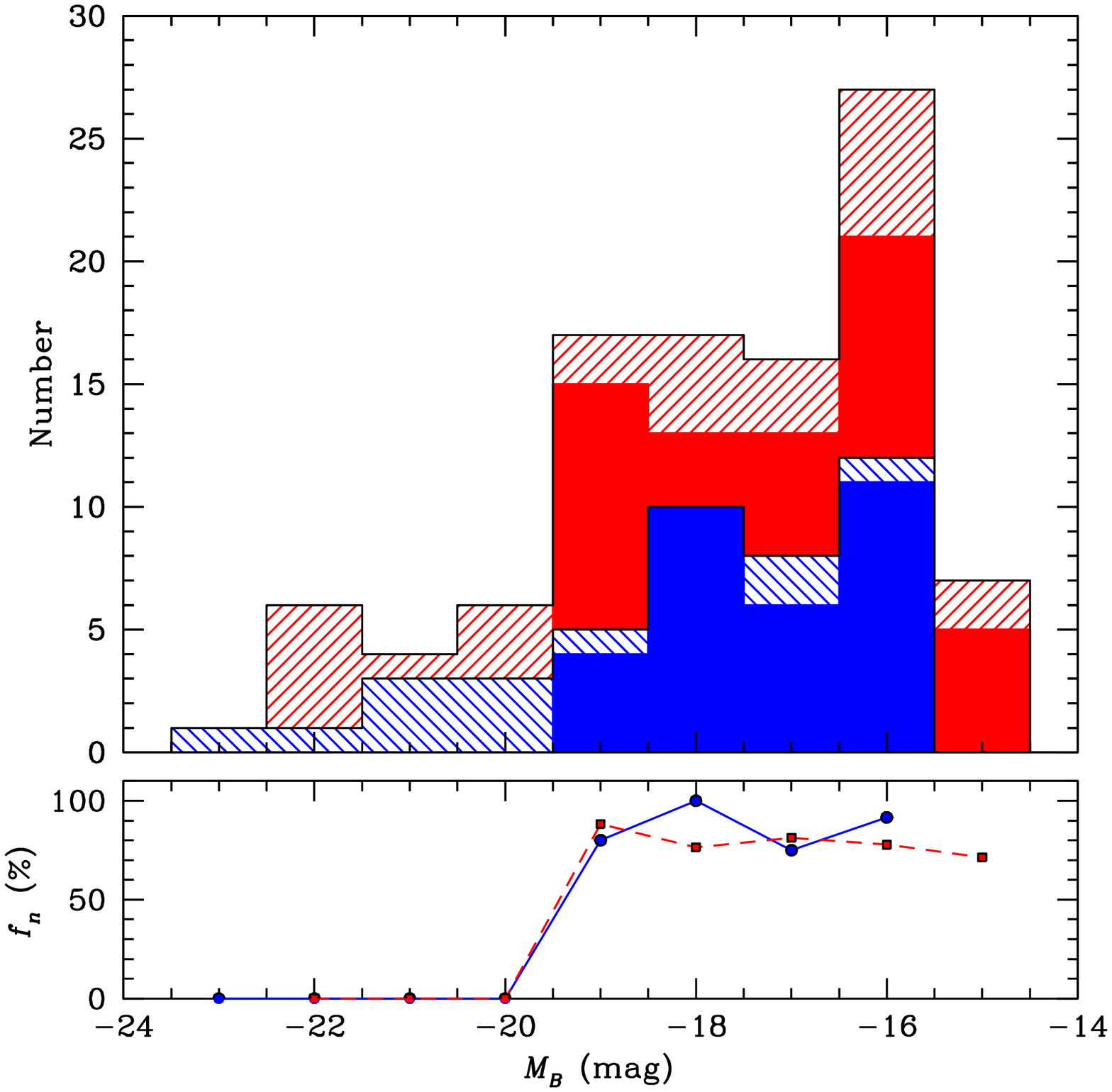}
	\caption{Same as Figure~\ref{fig:lum_hist}, but using absolute magnitudes
	and showing both ACSVCS and ACSFCS program galaxies (143 objects in total).
	\emph{Top}: Luminosity distribution of the
	program galaxies for Virgo (solid and hatched red histograms) 
	and Fornax (solid and hatched blue histograms). The 
	solid histograms show the distribution of the 67 Virgo and 31 Fornax galaxies 
	found to be nucleated by the ACSVCS and ACSFCS.
	\emph{Bottom}: The percentage of galaxies found to be nucleated ($f_n$) 
	for Virgo (red squares) and Fornax (blue circles).}
	\label{fig:virgo1}
\end{figure}

\subsubsection{Frequency of Nucleation}

In Figure~\ref{fig:virgo1}, we plot the frequency of nucleation of the Virgo 
and Fornax program galaxies as a function of their absolute blue magnitude.
The Virgo galaxies appear in red, and the Fornax galaxies are shown in blue.
In the upper panel, we overlay histograms for all galaxies (hatched) and
nucleated galaxies (solid). This figure demonstrates how the Virgo 
galaxy magnitudes extend to $\sim1$ mag below those of Fornax, as explained above. 
Our Virgo sample contains 100 galaxies, 67 of which are found to be nucleated,
so we obtain a total frequency of nucleation, $f_{n}=67\pm8\%$. This is in excellent
agreement with the value of $f_{n}=72\pm13\%$ found for our full Fornax sample. 

The bottom panel shows the frequency of nucleation in each luminosity 
bin. Both clusters exhibit very similar distributions with $f_{n}=0$ 
for the bright galaxies, while fainter than $M_{B}\sim-19.5$,
$f_{n}$ continuously stays above $\sim70\%$. 
Since our Virgo sample has 84 galaxies below $M_{B}=-19.5$, and our Fornax
sample has 35,  we find the total frequency of nucleation for galaxies
fainter than $M_{B}=-19.5$ to be $80\pm10$\% for Virgo and $89\pm16$\% for Fornax. 

Both \citetalias{cote06} and this study have shown that this sharp increase
in frequency of nucleation compared to previous ground based studies
(the VCC and FCC) is due mainly to surface brightness selection
(see Figure~7 and 8 in \citetalias{cote06} and Figure~\ref{fig:galsb_vs_nucmag} 
in this work), which can be attributed to the improved resolution and depth 
offered by the ACS imaging. That is, the excellent angular resolution of  HST 
has allowed us to uncover previously undetected nuclei
in both very high surface brightness galaxies, where the nuclei are difficult to 
distinguish from the main body, and low luminosity galaxies, in which the nuclei 
may lie below the magnitude limit of the older photographic surveys. 

\begin{figure}
	\figurenum{16}
	%\plotone{f16.eps}
	\centering 
	\leavevmode 
	\includegraphics[width = 0.85\linewidth, trim =0cm 3.5cm 2cm 0cm, clip=true]{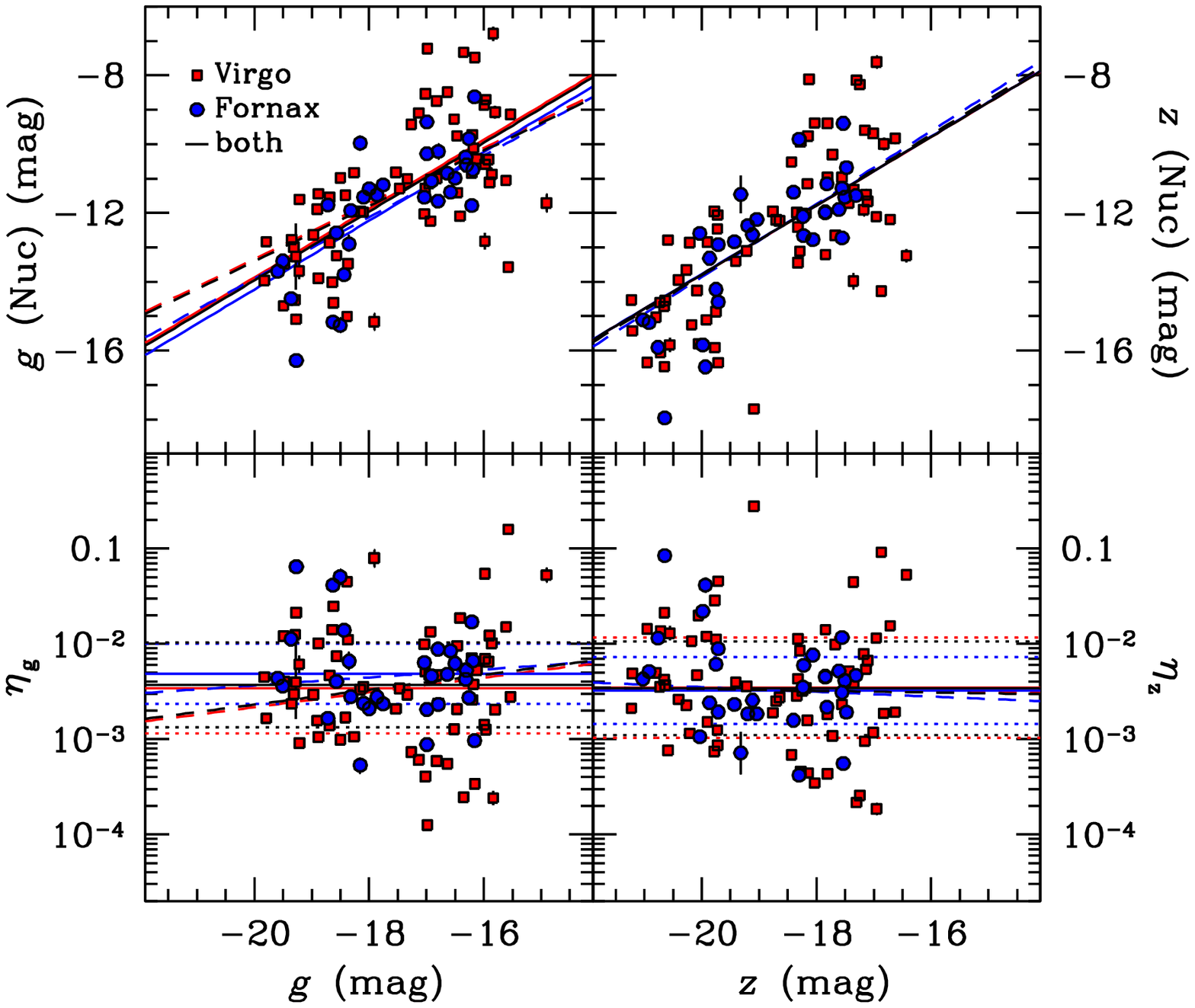}
	\caption{Same as Figure~\ref{fig:eta}, but using absolute magnitudes
	and including 67 ACSVCS and 31 ACSFCS nuclei.
	\emph{Top}: Nucleus magnitude plotted against host galaxy magnitude,
	 for the Virgo (red squares) and Fornax 
	(blue circles) galaxies found to be nucleated, in the $g$- (\emph{left}) and 
	$z$-bands (\emph{right}).  The lines show the weighted best fit relations, with the slope
	held fixed at unity (\emph{solid}) and allowed to vary (\emph{dashed}).
	The red and blue lines correspond to fits to the Virgo and Fornax samples
	respectively, while the black lines show the fits to the combined sample.  
	\emph{Bottom}: Nucleus-to-galaxy luminosity ratio, $\eta$, against host galaxy
	magnitude, for the $g$-band (\emph{left}) and $z$-band (\emph{right}). The solid 
	and dotted lines show the mean and one standard deviation, respectively, while the
	dashed line shows the best fit relation given by the dashed line in the upper panel,
	recast in terms of  $\log(\eta)$ and host magnitude.}
	\label{fig:virgo2}
\end{figure}

\subsubsection{Nucleus-to-Galaxy Luminosity Ratio} \label{sec:virgoEta}

As  in \S\ref{sec:eta} and Figure~\ref{fig:eta}, absolute nucleus magnitude
has been plotted against absolute galaxy magnitude in the top panels of 
Figure~\ref{fig:virgo2}. Relations of the form Equation~\ref{eq:eta1} 
have been fitted using weighted least-squares to the Virgo and Fornax
samples, both separately and combined, and the parameters and standard errors are recorded
in Table~\ref{tab:eta_virgo}, the results of which are in agreement to within
the errors for both galaxy samples.

We also plot nucleus-to-galaxy luminosity ratio $\eta$ 
as a function of absolute galaxy magnitude in the bottom panels of 
Figure~\ref{fig:eta}. The values for the mean and standard deviation
of $\eta$ are given in Table~\ref{tab:eta_virgo}.
Taking the mean nucleus-to-galaxy luminosity ratio of both data
sets combined, we obtain the following values for each band:
\begin{equation}
%\begin{split}
\begin{array}{rcl}
\langle \eta_{g} \rangle & = & 0.37 \% \pm 0.04 \% \\
\langle \eta_{z} \rangle & = & 0.34 \% \pm 0.04 \%, \\
%\end{split}
\end{array}
\end{equation}
which gives a mean value for both bands of
\begin{equation}
\langle \eta \rangle = 0.36 \% \pm 0.03 \%.
\end{equation}
The quoted errors refer the standard error on the mean.

Finally, we note that, due to the definition of $\eta$, the best-fit relation from Equation~\ref{eq:eta1}
can be recast in terms of $\log(\eta)$ and galaxy magnitude, where $\alpha_\eta = -0.4 \left(\alpha_1 -1\right)$ 
and $\beta_\eta = -0.4 \beta_1$. This relation is plotted as the dashed line in the bottom panels of 
Figure~\ref{fig:eta}, and we find that we do not see any significant trend between nucleus-to-galaxy
luminosity ratio and galaxy magnitude. 

% Eta 
\begin{deluxetable}{lccrccccc}
\tabletypesize{\footnotesize}
\tablecaption{Virgo and Fornax Nucleus-to-Galaxy Luminosity Ratios\label{tab:eta_virgo}}
\tablewidth{0pt}
\tablehead
{
	\colhead{Sample} & 
	\colhead{Band} & 
	\colhead{$\alpha_1$} & 
	\colhead{$\beta_1$} & 
	\colhead{$\beta_2$} & 
	\colhead{$\langle\log\eta\rangle$} & 
	\colhead{$\sigma$} \\
	\colhead{} &
	\colhead{} &
	\colhead{} &
	\colhead{(mag)} &
	\colhead{(mag)} &
	\colhead{(dex)} &
	\colhead{(dex)} 
}
\startdata
ACSFCS   & $g$ & $0.90\pm0.17$ & $3.99\pm2.85$ & $5.78\pm0.14$ & $-2.31$ & $0.32$ \\
ACSVCS   & $g$ & $0.81\pm0.11$ & $2.79\pm1.96$ & $6.12\pm0.15$ & $-2.46$ & $0.47$ \\
Combined & $g$ & $0.80\pm0.09$ & $2.68\pm1.60$ & $6.04\pm0.11$ & $-2.43$ & $0.44$ \\
ACSFCS   & $z$ & $1.06\pm0.16$ & $7.38\pm2.97$ & $6.22\pm0.16$ & $-2.49$ & $0.35$ \\
ACSVCS   & $z$ & $1.02\pm0.11$ & $6.51\pm2.07$ & $6.21\pm0.15$ & $-2.46$ & $0.53$ \\
Combined & $z$ & $1.02\pm0.09$ & $6.63\pm1.69$ & $6.21\pm0.11$ & $-2.47$ & $0.49$ 
\enddata
%\tablecomments{}
%\tableotetext{a}{}
\end{deluxetable}

\begin{figure}
	\figurenum{17}
	\plotone{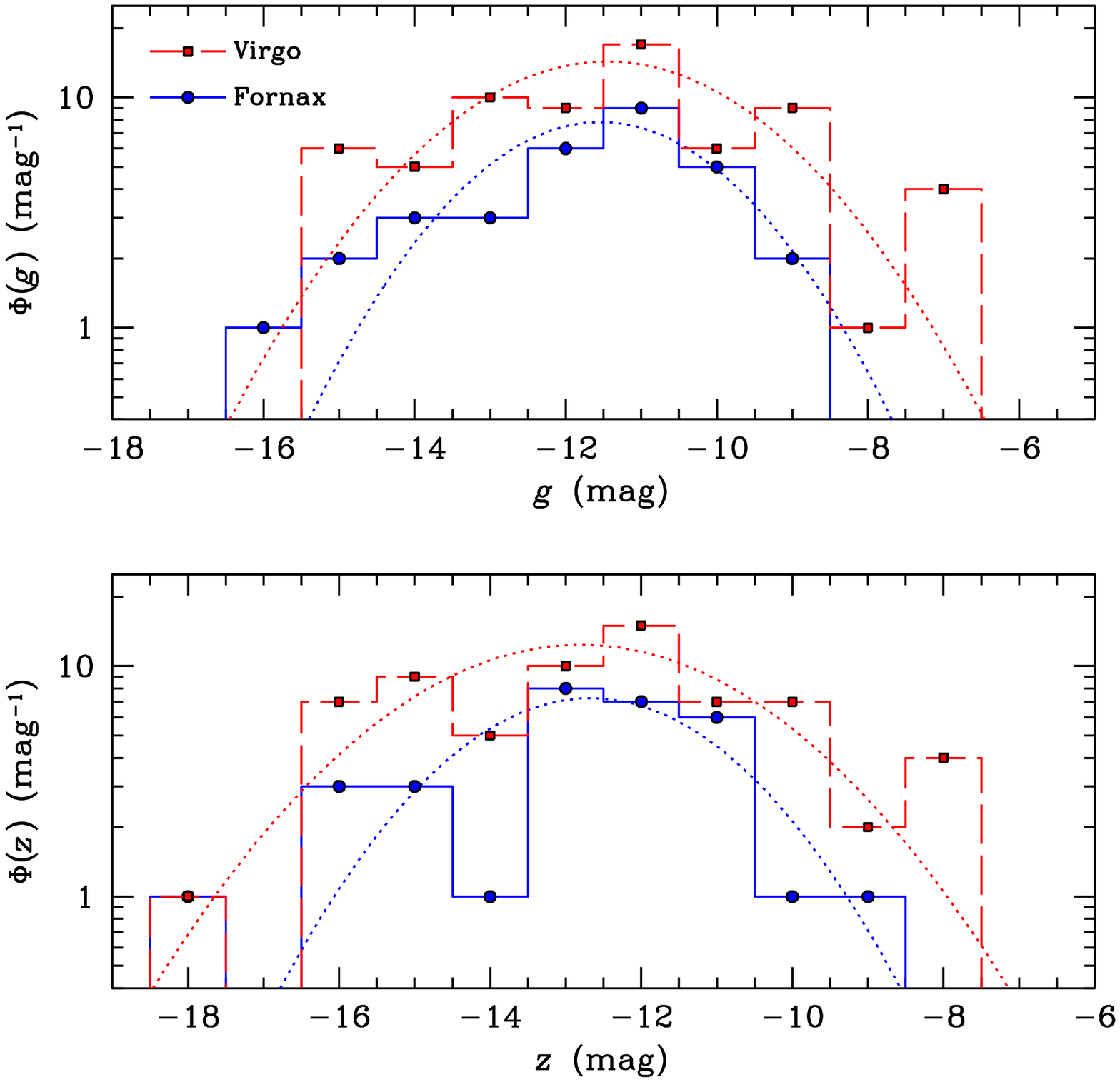}
	\caption{Same as Figure~\ref{fig:lum_fun}, but using absolute magnitudes
	and including 67 ACSVCS and 31 ACSFCS nuclei.
	The luminosity functions for both the Virgo (red squares) and 
	Fornax (blue circles) nuclei are shown, 
	in the $g$-band (\emph{top}) and $z$-band (\emph{bottom}).  
	Both data sets have been fitted with a normalized Gaussian.}
	\label{fig:virgo3}
\end{figure}

\begin{figure}
	\figurenum{18}
	\plotone{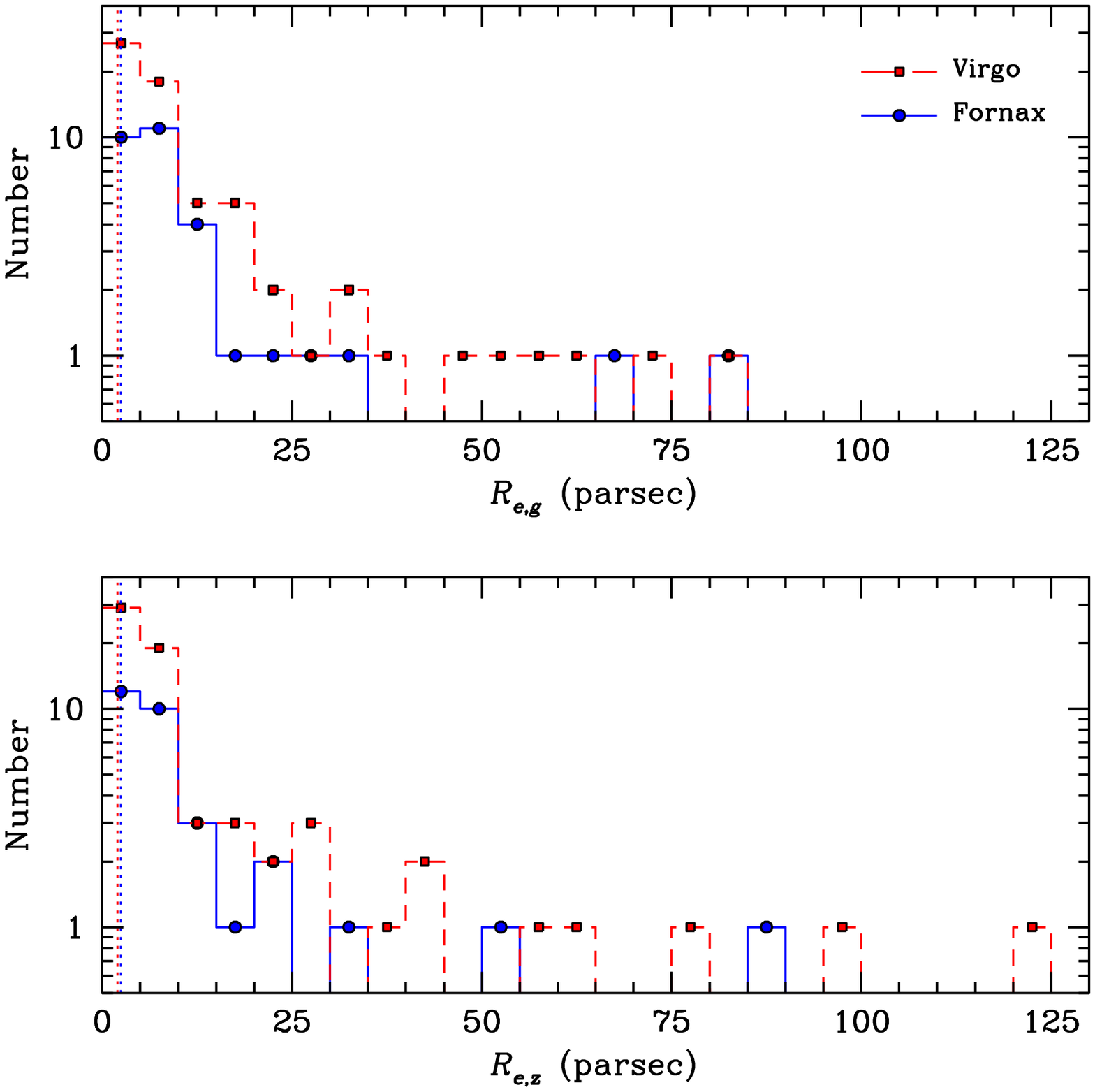}
	\caption{Same as Figure~\ref{fig:size_hist}, but using parsecs and 
	including 67 ACSVCS and 31 ACSFCS nuclei. The distribution 
	of half-light radii for both Virgo (red squares) and Fornax
	(blue circles) nuclei are shown.  
	The red and blue vertical dotted lines indicate the adopted resolution 
	limit of $\sim0\farcs025$, which corresponds to 2.0~pc in the ACSVCS
	and 2.4~pc in the ACSFCS.}	\label{fig:virgo4}
\end{figure}

\begin{deluxetable}{lccc}
%\tabletypesize{\scriptsize}
\tabletypesize{\small}
%\rotate
\tablecaption{Virgo and Fornax Nucleus Luminosity Function\label{tab:lum_fun_virgo}}
\tablewidth{0pt}
\tablehead
{
\colhead{Sample} & 
\colhead{Bandpass} & 
\colhead{$\bar{m}^0_n$} & 
\colhead{$\sigma_n$} \\
\colhead{} & 
\colhead{} & 
\colhead{(mag)} & 
\colhead{(mag)} 
}
\startdata
ACSFCS & $g$ & $-11.54\pm0.03$ & $1.58\pm0.02$ \\
ACSVCS & $g$ & $-11.45\pm0.02$ & $1.87\pm0.02$ \\
ACSFCS & $z$ & $-12.67\pm0.03$ & $1.70\pm0.02$ \\
ACSVCS & $z$ & $-12.80\pm0.02$ & $2.16\pm0.02$ 
\enddata
%\tablecomments{}
%\tablenotetext{a}{}
\end{deluxetable}

\subsubsection{Nucleus Luminosities and Sizes}

In Figure~\ref{fig:virgo3}, histograms of nuclei luminosities
for both our Virgo and Fornax sample are compared. The parameters
of the weighted maximum-likelihood fit of a normalized Gaussian to each 
sample are given in Table~\ref{tab:lum_fun_virgo}, where the errors
on the fitted parameters are the standard errors. 
Although we find differences between $\bar{m}^0_n$ between the 
two surveys, there amounts are comparable to the errors estimated 
for the nuclei magnitudes.   

We plot a histogram of nucleus sizes in Figure~\ref{fig:virgo4} for 
both our Virgo and Fornax samples. Although there is a large range in size	
(the very large Virgo nucleus belongs to VCC~1178), 
most nuclei appear to have radii $<10$ pc. The typical sizes 
are in good agreement, with median values of $5.7$~pc in the 
$g$-band for both clusters, and $7.2$~pc and $7.0$~pc in the $z$-band for Virgo and Fornax 
respectively. 

\subsubsection{Other Properties} 

In addition to the above properties, we find the Virgo and Fornax nuclei 
to be remarkably similar in a number of other ways. First,  and most obviously,
both the ACSFCS and ACSVCS  galaxies  exhibit a trend along the luminosity 
function in which their central surface brightness profiles gradually change from having a luminosity 
``deficit'' to an ``excess'': see, e.g., Figures~3 and 4 in \citetalias{cote06}, Figure~1 of \citetalias{cote07}, Figures~\ref{fig:acs_images} 
and \ref{fig:sb_profiles} here, as well as a detailed discussion of this trend in \citetalias{glass11}.

Plotting surface brightness against magnitude, the nuclei are found to have different  
scaling relations than the GCs (see Figure~18 in \citetalias{cote06} and Figure~\ref{fig:scaling} here). 
Although \citetalias{cote06} used integrated nucleus colors in their study,
our use of aperture colors may be a more appropriate comparison to the King profiles
used to determine the integrated nucleus magnitudes. 
Indeed, the best-fit line parameters outlining the color-magnitude relation 
for the nuclei with $B_{T}\leq13.5$, given by Equation~13 in \citetalias{cote06} and Equation~\ref{eq:col} in 
this work, are in good agreement.

Overall, we find a striking similarity between the nuclei of Virgo and Fornax,
despite the clear environmental differences between the two clusters. 
This agreement suggests that the physical characteristics of
individual galaxy clusters (such as ICM density), or the processes that depend on them 
(such as ram pressure stripping 
efficiency), do not play a dominant role in the formation and evolution of nuclei
in early-type galaxies. Thus, it seems we can consider the nuclei examined 
here as being representative of those in early-type galaxies in general. 

% Local groups

\subsection{Extension to Low Luminosity: Comparison to the Local Group}

Although the ACSVCS and ACSFCS provide a reliable measurement of the nucleation frequency
for galaxies brighter than $M_{B}\lesssim-15$~mag, it is instructive to consider the frequency of nucleation in 
galaxies fainter than this magnitude limit. We can do so by examining the members of the Local Group, 
where the smallest observed dwarf galaxies reach magnitudes faint as $M_{V}=-1.5$~mag and can have 
effective radii on the order of $\sim30$~pc \citep[see, e.g.,][]{martin08}. As sample completeness is 
a concern for such faint, compact systems, we focus on the subset of early-type galaxies brighter
than $M_B \approx -8$.

At present, there are 25 known early-type galaxies in the Local Group brighter than this limit
\citep[compiled from][]{mateo98, mcconnachie05, brasseur11}. Of these, only
two (NGC~205 and M32) are brighter than the ACSVCS limiting magnitude of 
$M_{B}=-15$~mag \citep{mateo98}, both of which are known to be nucleated 
\citep[e.g.][]{kent87, lauer98, mateo98, butler05, derijcke06}.
Moving down the luminosity function, at most six other galaxies 
may contain either nuclei or kinematically/structurally distinct features near their core, listed
in order of decreasing luminosity: 
NGC 147 \citep{derijcke06}, Sagittarius \citep{mateo98, layden00, monaco05, bellazzini08},
Fornax \citep{coleman04, coleman05, coleman08}, Sextans \citep{kleyna04, walker06},
Andromeda~II \citep{mcconnachie06}, and 
Ursa~Minor \citep{kleyna03, palma03}.\footnote{Although it is traditionally classified as non-nucleated,
we include the Fornax dSph in this list since GC \#4 is located $\sim$ half a core radius from the
galaxy photocenter \citep[see Figure~1 of][]{coleman08} and 
might thus be classified as a dwarf with an offset nucleus if moved to the distance
of the Virgo or Fornax clusters.} 
Considerable caution is advisable here since, in some cases (e.g., in Ursa Minor and, 
especially, in Sextans), the ``nuclei"  are
rather subtle substructures (sometimes only apparent with the addition of kinematic data) 
that bear little resemblance to the prominent, compact nuclei seen in the faintest
ACSVCS and ACSFCS galaxies. Yet, even with this liberal definition of a ``nucleus",
only 8 out of the 27 Local Group early-type galaxies ($f_n = 30\%$) can be classified as nucleated. 
If we exclude Fornax, Ursa Minor and Sextans from the list of nucleated galaxies, then
$f_n$ falls to 19\%. While it is possible that some nuclei have 
yet to be discovered, it seems certain that many of the faint Local Group
galaxies do {\it not} contain a nucleus; e.g., \citet{mcconnachie06} 
studied of structural properties of six Andromeda satellites using deep, homogeneous imaging, 
and found a nucleus in only a single object (And~II).

We conclude that the frequency of nucleation along the Local Group
sample is clearly far lower than in either our ACS surveys of the Fornax or Virgo clusters.
Why is there such a large disparity in $f_n$?
We speculate that the lack of nuclei in very faint galaxies could be related to the general absence of GCs
in galaxies below $M_B \sim -12$ \citep[see, e.g.,][]{peng08}.
If nuclei in low-mass galaxies are assembled through GC infall and mergers
(see \S\ref{sec:gcinfall}), then the presence of GCs would obviously be a prerequisite for 
nucleus formation. The faintest galaxies in the Local Group known to contain GCs
are Sagittarius   and Fornax, with $M_{B}= -12.8$ and $-12.6$ respectively \citep{mateo98}.  
The former is unquestionably nucleated, while Fornax {\it may} meet the definition of a nucleated
galaxy (see above).
Because no Local Group dwarfs below this magnitude are known to contain GCs, such galaxies
might have been unable to form a nucleus if star cluster infall is the dominant mode of nucleus 
formation in low-mass systems.

It is also interesting to note that, assuming a constant nucleus-to-galaxy luminosity ratio of 0.4\%,
then the expected nucleus magnitude of a $M_B = -12.6$~mag  
host would be $M_B=-6.6$~mag. This corresponds closely to the mean turnover
magnitude of the globular cluster luminosity function, 
$M_V\approx-7.5$~mag \citep[e.g.,][]{jacoby92, harris01, brodie06}, suggesting that galaxies
may be unable to form nuclei at the point where the expected nucleus
luminosity would fall below the typical GC luminosity.  

However, as caveats we firstly note that the nucleus of Sagittarius \citep{monaco09} as well
as the very central region of the Andromeda satellite NGC~205 \citep{siegel07} have been 
observed to have undergone multiple star formation episodes, which indicates that other
processes in addition to GC accretion must have shaped their formation history. 
In addition, the nuclei late-type dwarfs  have been shown to \emph{not} form form
exclusively from GC infall \citep[e.g.,]{walcher06} or gas accretion \citep{hartmann11},
even though it has been observed that GC specific frequency is independent of morphology  
\citep{georgiev10} and thus should be the same for both early- and late-type dwarfs.

% Formation models

\subsection{Formation and Evolution Models}\label{sec:models}

The origin of nuclei remains an open theoretical
problem, with two main avenues of nucleus formation presently considered
most viable. The first proposes that a galaxy's star clusters 
will experience orbital decay due to dynamical friction and spiral inwards, 
eventually coalescing at the center of the galaxy. The second formation mode
focuses on gas accretion at the center of the galaxy, followed by star formation.
Some similarities in the scaling relations of nuclei and black holes  (see \S\ref{sec:introduction})
have also given rise to models that consider the formation of both types objects in a 
shared context. In this section, we shall examine theoretical studies of nuclei
formation in light of our new results, as well as models that explore the relationship between nuclei and black holes.

% Dissipationaless 

\subsubsection{Dissipationless Infall of Star Clusters} \label{sec:gcinfall}

\citet{tremaine75} first suggested that the nucleus of M31 was formed
from GCs that spiraled inward to the galaxy center due to dynamical friction,
and this mechanism continues to offer an attractive explanation for the
assembly of nuclei in at least some galaxies. Of course, not all clusters
that come close to the center of a galaxy will necessarily contribute to the formation,
or growth, of a stellar nucleus; as
\citet{capuzzo93} showed, dynamical friction and tidal
stripping are competitive processes, where GCs are more readily
destroyed by large nuclei, limiting nucleus growth. 

Nevertheless, some fraction of GCs are expected to avoid tidal disruption and 
could contribute to either nucleus formation, or the growth of pre-existing nuclei.
Evidence in favor of this process was described in 
\citet{capuzzo99}, who pointed out that the radial distribution of 
GCs in galaxies is less centrally concentrated than 
the halo stars (see also \citealt{mclaughlin95, mclaughlin99, cote01, cote03, peng08}). 
Such ``missing'' clusters could have contributed
to nucleus formation. Monte Carlo simulations based on this premise by \citet{lotz01}
predicted nuclei luminosities for dEs with $-17\lesssim M_B \lesssim -12$
that were consistent with observations for the brighter galaxies within
this range, although they were overestimated for less luminous ones.
The over-prediction of nuclear luminosities in their low-mass systems
resulted from their short dynamical times --- meaning that nuclei are able 
to grow very efficiently --- in spite of the fact that 
these galaxies have relatively few star clusters (see e.g., \citealt{peng08}). 

Numerical simulations by \citet{oh00} and similar, higher resolution N-body 
simulations by \citet{capuzzo08a, capuzzo08b} were able to successfully
reproduce the observed surface brightness profiles of known nucleated galaxies.  
A dependence on local tidal field was found in the \citet{oh00} model, where disruptive
tidal forces on the outskirts of galaxy clusters would alter GC
orbits, increasing dynamical friction timescales and decreasing nucleation frequency.
The \citet{capuzzo08a, capuzzo08b} models suggest that, if linear scaling is assumed, then the observed 
nuclei could have formed from the infall of tens, to hundreds, of GCs (see also \S4.9 and \S5.2.4 of \citetalias{cote06}). 
Both simulations found that nuclei may begin to coalesce away from 
the galaxy photocenter, although to quite different extents: i.e., up to 
$\sim0.3$~kpc and settling within $\sim1$ Gyr 
in \citet{oh00}, and $\sim4$ pc away in \citet{capuzzo08a, capuzzo08b}.

Other simulations by \citet{bekki04} observed that the scaling
relations of nuclei formed through mergers of GCs 
would be notably different than those of the GCs.  In \S\ref{sec:scaling},
we discussed that the predicted scaling relation for nuclei in these 
simulations, $R_e \propto {\cal M}_*^{0.38}$, was generally in good agreement
with observations (see Figure~\ref{fig:scaling}).
More recent work by \citet{bekki10a} focused on simulations of 
star cluster infall due to dynamical friction in {\it disk} galaxies. 
He found that the effectiveness of dynamical friction did not depend strongly on 
bulge mass, but increased with smaller disk mass, and with larger disk mass fraction, 
galaxy surface brightness, and star cluster mass. The ratio of nucleus mass to disk 
mass was found to decrease as a function of increasing disk mass, with a mass ratio 
of $\gtrsim0.4\%$ for smaller disks, and $\lesssim0.1\%$  for disks with masses 
$M\gtrsim10^{9}M_{\sun}$. 
However, star cluster mergers on to a disk may not be sufficient to explain
nuclei formed in $M_V \sim -19.5$ spirals. 
N-body simulations by \citet{hartmann11}, which aimed to reproduce the observed 
kinematics of the nuclei in M33 and NGC~4244, found that star cluster
accretion on to a disk did not produce the necessary line-of-sight velocity rise, and at
least half of the nucleus mass had to come from gas dissipation.

\begin{figure}
	\figurenum{19}
	\plotone{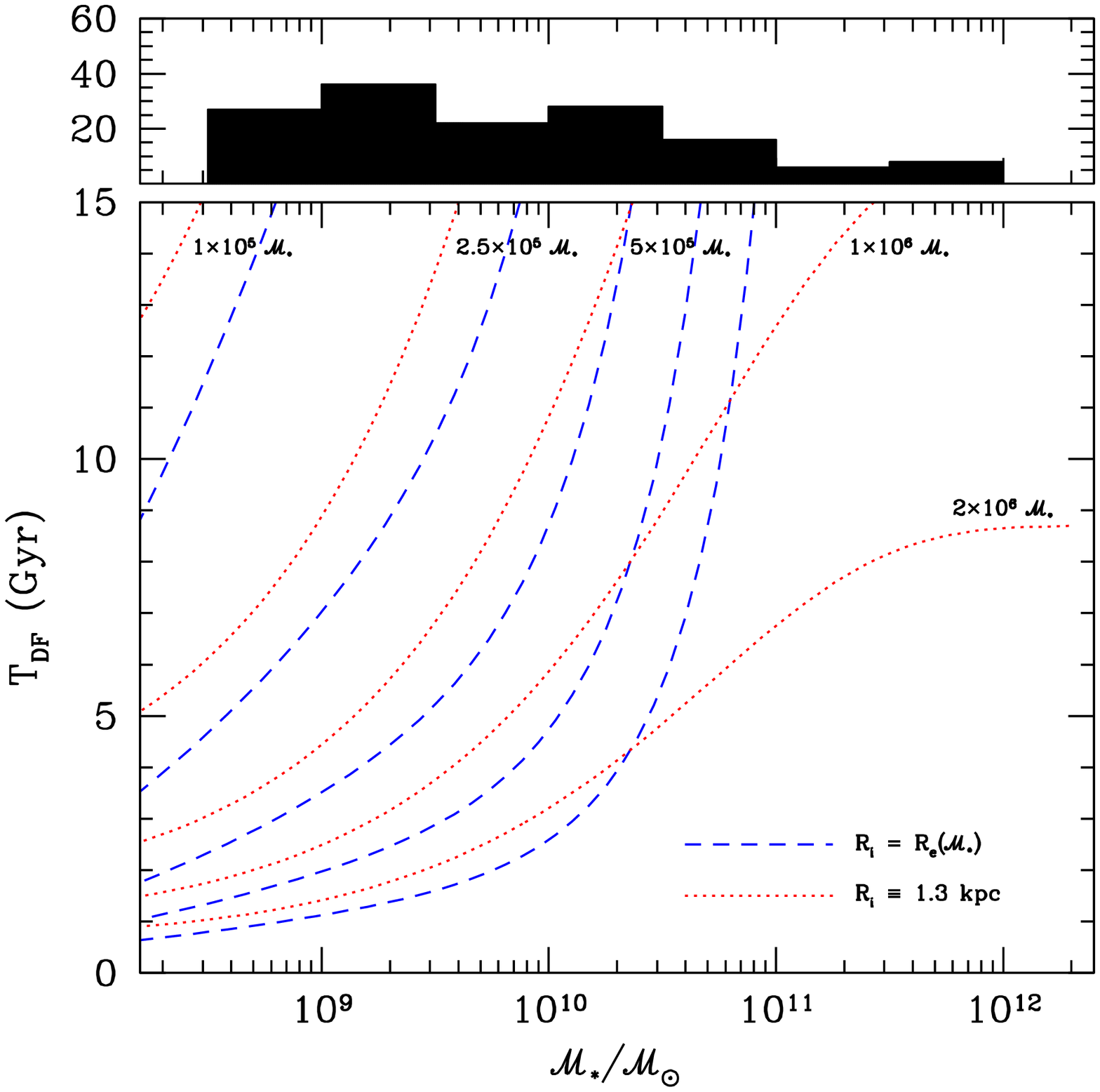}
	\caption{\emph{Top}: Histogram of masses for the 143 galaxies from the ACSVCS and ACSFCS surveys. 
	\emph{Bottom}: Dynamical friction timescales, T$_{\rm DF}$, plotted as a function of galaxy mass. Two sets 
	of curves are shown. The dashed blue curves show calculations for initial GC radii, $R_i$, equal to the galaxy
	effective radii (see the lower panel of Figure~\ref{fig:scaling}), while the dotted red curve shows $R_i$ fixed
	to 1.3~kpc, the median effective radius for ACSFCS galaxies. In both cases, T$_{\rm DF}$ is plotted for five GC masses:
	$0.1, 0.25, 0.5, 1.0$ and $2.0$ million solar masses. Note the sharp decline in T$_{\rm DF}$ for
	low-mass galaxies.}
	\label{fig:dyn}
\end{figure}

Some provisional evidence for dissipationless formation in at least some galaxies was presented in \citet{paudel2011}, 
who used optical spectroscopy for Virgo cluster dwarfs to study both their stellar populations and those
of their nuclei. Despite the small sample and the different environment (Virgo vs. Fornax), their
data present an interesting opportunity to speculate on possible formation mechanisms for the 
ACSFCS nuclei. \citet{paudel2011}
found that nuclei in a handful (5) of the faint ($-16 \lesssim M_B \lesssim -14$) galaxies in their sample
were older and more metal poor than their hosts, which is certainly suggestive of a connection 
to GCs. At higher luminosities, most of their nuclei were found to be {\it younger} than their hosts.  While
inconsistent with nucleus formation from old globular clusters, this observation may still be compatible with
cluster infall, as our observations and many others have shown that ongoing star cluster formation can be 
present throughout some galaxies \citep[e.g.,][]{anders04, kyeong10}. In the ACSFCS sample,
FCC~119, FCC~90 and FCC~26 are possible examples of $M_B > -19.5$ galaxies with young cluster systems.

Additional support for such a scenario may come from the 
GC luminosity functions in Virgo and Fornax galaxies.
The widths of GC luminosity functions 
are known to decrease significantly with galaxy luminosity, a trend that is accompanied by a slight decrease in 
turnover mass  (\citealt{jordan06, jordan07b}; \citetalias{villegas10}).
This truncation of the GC population on the bright end of the luminosity function 
may be caused, at least in part, by the shorter dynamical friction times
as galaxies become less massive, although other (external) processes 
could also play a role (see \S7.2 of \citealt{jordan07b}).

We revisit the question of star cluster infall efficiency by calculating the dynamical friction
timescale, T$_{\rm DF}$, for all galaxies in our ACS surveys of Fornax and Virgo.
The upper panel of Figure~\ref{fig:dyn} shows the distribution of galaxy masses from the 
combined surveys (filled histogram), while the lower panel show the dependence of T$_{\rm DF}$
on galaxy mass, ${\cal M}_*$, which is given by
$${\rm T}_{\rm DF}  =  {2.64\times10^2 \over {\ln{\Lambda}}}\biggl({R_i \over 2~{\rm kpc}}\biggr)^2\biggl({v_c \over 250~{\rm km~s^{-1}}}\biggr)\biggl({10^6{\cal M}_{\odot} \over {\cal M_{\rm GC}}}\biggr)~{\rm Gyr}.\eqno{(16)}$$
Here $R_i$ is the initial galactocentric radius of the star cluster, $v_c$ is the circular velocity of the (assumed isothermal) galaxy, and
${\cal M}_{\rm GC}$ is the mass of the star cluster \citep{binney08}. In this equation,
$\ln{\Lambda}$ is the coulomb logarithm, which is defined as 
$${\ln{\Lambda}}   =   \ln{\biggl[{b_{\rm max}v_c^2 \over G({\cal M_{\rm GC}} + \textsc{m})}\biggr]} \eqno{(17)}$$
where $b_{\rm max}$ is the maximum impact parameter between the cluster and the interacting particle (a star of mass $\textsc{m}$).
Following \citet{lotz01}, we assume $v_c \simeq \sqrt{2}\sigma$ where $\sigma$ is the integrated-light velocity dispersion measured within $R_e/4$ 
from McLaughlin et~al. (2012, in prep.). We also take $b_{\rm max} = R_e$ for all galaxies, with $R_e$ measured directly
from the ACS imaging (see \S\ref{sec:scaling} and Figure~\ref{fig:scaling}).

Calculations have been carried out for five different star cluster
masses (i.e., 0.1, 0.25, 0.5, 1 and 2 million solar masses)\footnote{Recall that in the Milky Way, the GC mass 
corresponding to the peak of the luminosity function is $2.4\times10^5{\cal M}_{\odot}$
\citep{mclaughlin99}.} and for two assumptions for $R_i$. In the first case, we take $R_i = R_e$ 
(see also \citealt{lotz01}) which is shown as the dashed blue curves in Figure~\ref{fig:dyn}. In
the second case, we simply fix $R_i$ at  the median effective radius, $1.3$~kpc, for all galaxies in
the ACSFCS sample. The results in this case are indicated by the dotted red curves in Figure~\ref{fig:dyn}.
Although T$_{\rm DF }$ clearly varies with the assumed cluster mass and the precise
choice of $R_i$, the strong mass dependence noted by previous investigators is clearly apparent in this figure.
In particular, the dynamical friction timescales are dramatically shorter in galaxies with ${\cal M}_* \lesssim10^{10}{\cal M}_{\odot}$
compared to higher-mass galaxies. We conclude that star cluster infall seems like a viable, indeed a likely,
candidate for the growth of nuclei in low- and intermediate-mass galaxies in our sample. For the highest-mass galaxies, the
mechanism appears much less viable given the fact that, in these systems, T$_{\rm DF}$ greatly exceeds the 
Hubble Time for all but the most massive and centrally concentrated star clusters.

Finally, we conclude this section with some final remarks on
Figure~\ref{fig:scaling}, which compared the structural parameters of nuclei to those of  GCs and their
host galaxies. While there is, as noted in \S\ref{sec:scaling}, good agreement with the nuclei size-mass relationship
found by \citet{bekki04} from simulations of GC mergers, there are reasons to believe
that a single relation cannot be appropriate for all nuclei which, in our sample, span more than four decades in 
mass.  For comparison, the simulated nuclei of \citet{bekki04} span a factor of just ten
in mass. It is to be expected that the precise form of the size-mass relation in the context of the GC merger 
model will be different in different mass regimes. For instance, when only a small number of mergers 
contribute to the nucleus, we expect from the virial theorem and conservation of energy that 
$R_e \propto {\cal M}_*^{0.5}$. At later times, when the mass of the nucleus greatly exceeds the mass of
an accreted GC, the relation should steepen to $R_e \propto {\cal M}_*$. These scaling relations,
shown in the lower panel of Figure~\ref{fig:scaling}, are in good agreement with the observed sizes 
and masses. 

All in all, based on the existing data, we believe that cluster infall must have played an important 
role in the formation of the nuclei the low- and intermediate-mass hosts within our sample. At the same time, the 
red colors of some of the largest and most massive nuclei (\S\ref{sec:colors}) present a strong challenge to this model, suggesting
that an additional process --- most likely the dissipational infall of metal-rich gas  --- likely begins to dominate
the formation of nuclei in galaxies of progressively larger masses (\citealt{mihos94}; 
\citetalias{cote06, cote07}; \citealt{hopkins08, hopkins09a}). 

% Dissipational

\subsubsection{Dissipational Infall of Gas} \label{sec:gas}

It has long been suspected that nuclei could form through star formation following the accretion of gas in
galaxy centers \citep{vandenbergh86}, although the exact origin of the gas, and the mechanism that triggers the inflow, 
are matters of debate.

In some models, the gas is assumed to originate from
outside the galaxy. \citet{davies88} proposed that dEs may be formed from 
fading stellar populations in dwarf irregulars, where the accretion of 
$\textsc{Hi}$ gas induced starbursts, the final one occurring 
in the center and forming the nucleus. \citet{silk87} predicted that
the intergalactic medium (IGM) could fall into dwarf galaxies when it is cooled and 
compressed during group formation. This model noted that dwarfs closer to large 
galaxies may not be able to form nuclei as efficiently, since the
large galaxy's tidal field makes it difficult for the dwarf to capture 
the gas. \citet{babul92} found an opposite trend with environment: they observe that nucleus evolution may depend on local IGM
density, because this determines whether supernova-driven gas outflows
are able to escape. Dwarfs in low-pressure regions would have their gas ejected 
and then fade away, while winds in dense environments would be restricted to the 
starburst region by the IGM. This confinement could cause gas to cool and recollapse,
creating two short or one prolonged starburst.

Gas might also be funneled to the centers of galaxies which have
disks and axisymmetric features.  \citet{milosavljevic04} suggested that 
in spiral galaxies, magneto-rotational instability in the disk transports gas to the 
center. \citet{bekki06} and \citet{bekki07} performed chemodynamic simulations of 
the inner $1$~kpc of dwarf galaxies with stellar masses  of $2.5\time10^7 \leq M_{\rm sph} \leq 1.0\times10^9$, 
to explore the remnant created through
dissipative merging of stellar and gaseous clumps formed from nuclear gaseous
spiral arms in a gas disk. The simulations produced nuclei which that rotating and flattened, 
consisting of stars with varying ages and metallicities. Although the initial clump was found to form
off-center (about $200$~pc by visual inspection of the 
simulation data), it would fall into the center within $100$~Myr. They found that overall,
the nuclei were characteristically younger and more metal rich than the host, with more 
massive hosts creating more metal-rich nuclei. Gas settling timescales increased 
with decreasing dwarf mass (due to feedback being more effective in smaller galaxies), 
so low mass dwarfs were found to have younger and bluer nuclei. 
More massive and dense nuclei were formed in more massive dwarfs with deeper central 
potentials, and both the mass and mass fraction of the nucleus were found to increase
with spheroid mass. Nuclei in high surface brightness galaxies
should also have higher surface brightness, owing to the increased 
dynamical friction due to higher stellar densities. 
The nucleus  surface brightness was strongly dependent on the gas 
fraction of the host, and  thus may be more likely to form in this manner 
in  late-type galaxies with relatively large amounts of gas. Finally, the addition of a central
black hole to the simulation had little effect on the properties of the remnant nucleus.

Another source of nuclear material, which was first proposed by \citet{bailey80},
could arise from stellar winds. It was found that only a small ($\sim10^6 M_{\sun}$) 
amount of gas was needed to cause an inflow for an elliptical galaxy with 
$M_{\rm gal}\sim10^{11} M_{\sun}$. \citet{seth10b} observed that such a 
mechanism could produce the age, abundance gradient, and rotation curve 
seen in the nucleus of M32.

The dissipative infall of gas to the galaxy center can also be induced by galaxy mergers.
\citet{mihos94} performed N-body simulations of disk galaxy mergers,
where they found that gas dissipation and the star formation that followed
created dense stellar cores in the remnant.  Similar higher resolution simulations
by \citet{hopkins08, hopkins09a}, showed that gravitational torques 
during gas-rich mergers removed the angular momentum of the gas, which would then undergo gravitational 
collapse. The amount of gas infall was found to largely depend on the progenitor galaxy
gas fraction, while the addition of a central black hole was not found to have a
significant effect on the the properties of the final remnant. 
Unfortunately, these models lacked the resolution to study typical nuclei, 
particularly those in the low-mass galaxies: i.e., 
apart from a small number of cE galaxies in the ACSVCS sample, which have likely been heavily tidally stripped
(e.g., \citealt{faber73, lauraa06, cote08, chilingarian09b, huxor11}; McLaughlin et~al. 2012, in prep.), 
the simulated galaxies of \citep{hopkins09a}
have masses $\gtrsim10^{10}{\cal M}_{\odot}$, more than ten times larger than the
masses of the faintest galaxies in the ACS surveys. However, in this restricted mass regime, the
properties of these simulated galaxies are in good agreement with our ACSFCS (and ACSVCS) observations.

Likewise, the simulations of \citet{bekki06} and \citet{bekki07}, which instead focused on the {\it low-mass} galaxies,
also appear to be consistent 
with observations, including those from our HST/ACS imaging and results from ground-based spectroscopy. 
First, the nuclei in these simulations were
found to be younger and more metal rich than their hosts, with 
nucleus metallicity increasing with host mass, a trend that was seen in \citet{paudel2011}. 
Second, their finding that low-mass dwarfs have younger and bluer nuclei is consistent
with some of the nuclei from \citet{paudel2011}, as well as with the nucleus colors 
observed in our study.  Finally, they also found that the mass fraction of the nucleus increased
with host spheroid mass, and that their simulated surface brightness profiles showed 
nuclei which become more prominent with increasing dwarf mass, whereas in low-mass
dwarfs the nuclei were barely distinguishable.
It is therefore possible, as discussed in \S\ref{sec:gcinfall}, that nucleus formation through gas infall may be most significant
for intermediate- and high-mass galaxies. In their analysis of the ACSVCS, \citetalias{cote06} noted that some of the reddest and brightest nuclei 
``may be candidates for the {\it dense stellar cores} that form in numerical simulations \citep{mihos94} when (chemically enriched) 
gas is driven inward, perhaps as a result of mergers."
Such a result can be reconciled with our nearly constant nucleus-to-galaxy luminosity ratio 
if star cluster infall accounts mainly for nucleus build up in lower-mass galaxies. At intermediate masses,
both processes could contribute significantly to the growth of nuclei; candidates for such {\it hybrid nuclei} in
the ACSFCS include FCC~43, FCC~249, FCC~310, FCC~148 and FCC~301, which may consist of both
compact and extended components.

\begin{figure}
	\figurenum{20}
	\plotone{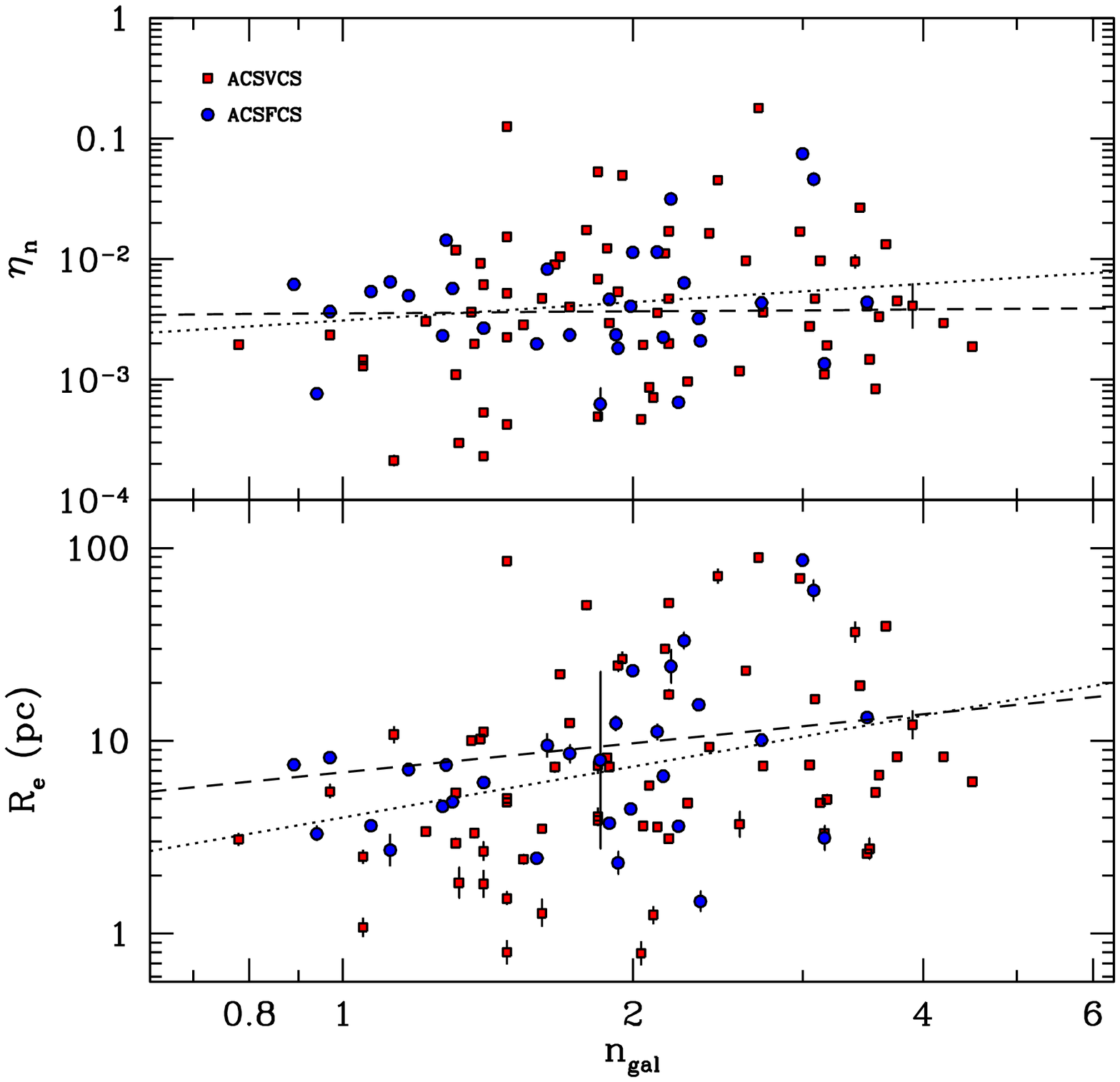}
	\caption{\emph{Top}: Nuclei luminosity fraction plotted against S\'ersic index of the host galaxy, $n_{\rm gal}$.
	\emph{Bottom}: Nuclei effective radius as a function of $n_{\rm gal}$. The dashed line in each panel shows the 
	weighted best-fit linear relation; unweighted fits are shown by the dotted lines. The nuclei in both
	clusters show weak trends with S\'ersic index (or, equivalently, galaxy mass) in the sense that the central 
	``excess" above the fitted S\'ersic model seem to be brightest and largest in galaxies with the largest $n_{\rm gal}$.
	These galaxies have likely undergone fewer mergers and accretions than those with $n_{\rm gal} \sim 1$.}
	\label{fig:sersicindex}
\end{figure}

This basic scenario is also consistent with the general view that 
mergers (which can drive gas to the central regions
of a galaxy) become increasingly important as galaxy luminosity increases, a consequence of the hierarchical merging
paradigm. The observation that galaxy concentration --- parameterized by S\'ersic index $n$ --- varies 
smoothly with galaxy luminosity (e.g., \citealt{jerjen97, graham03a, lauraa06}; McLaughlin et~al. 2012, in prep.; see also \S\ref{sec:sbp})
provides strong supporting evidence for this picture, as violent relaxation of merger remnants is thought to be responsible for 
the creation of de Vaucouleurs profiles \citep[e.g.][]{barnes88,barnes92}, while
S\'ersic index of both bulge and disks of spirals has been shown to increase after satellite 
infall \citep{eliche05}. Figure~\ref{fig:sersicindex} shows the dependence of two fundamental
parameters for nuclei --- luminosity fraction and effective radius --- against host galaxy S\'ersic index (Paper~III).
Those galaxies whose internal structure has likely been transformed most extensively
through mergers, accretions and harassment (i.e., those galaxies with high S\'ersic indices) tend to have the
most luminous and spatially extended nuclei (although the trend between $n$ and $\eta$ is statistically
significant only when an unweighted fit is used). These trends are generally consistent with an increasing importance for
gas dissipation as ones moves to higher and higher mass galaxies. 
Nuclei formed through merger-driven gas inflow could also be expected to follow a 
mass-radius scaling relation, as \citet{hopkins10b} found that stellar systems may 
have a maximum stellar surface density, due to feedback from massive stars. 

One complication with the gas inflow model is that it obviously requires the presence of gas, which is 
not consistent with the ``classical" picture of early-type galaxies. 
However,  both low-mass Es and high-mass ``dEs" are now recognized to be quite complex, having been found to contain
dust, spiral arms, embedded disks, and bars \citep{jerjen00, barazza02, derijcke03b, lisker06a, lauraa06},
as well as counter rotating and kinematically decoupled cores \citep{derijcke04,  thomas06, chilingarian08}, 
and ongoing star formation \citep[e.g.,][]{derijcke03a, lisker06b, cote06,michielsen07}. These features
suggest that a non-negligible fraction of intermediate-mass galaxies classified as ``early" types
have experienced some level of morphological transformation, likely through
mergers, accretions, or interactions with the cluster environment \citep{moore96, kazantzidis11}. 

It is, in fact, possible that the nuclei in some of our early-type galaxies formed in {\it late-type} progenitors. 
A recent finding by \citet{emsellem08} noted that galaxies with
S\'ersic indices of $n\lesssim3.5$ have compressive tidal
forces in their central regions, with the size of the compressive region
increasing with decreasing S\'ersic index. 
Assuming a constant S\'ersic index of $n = 1$, the amplitude of the tidal forces
was found to scale linearly with galaxy mass, and form a central massive object (CMO) with a constant host mass 
fraction of $\sim0.5\%$. A CMO growing through gas accretion
in this way would eventually reach a critical density and luminosity, altering
the galaxy profile such that it no longer has central compressive 
forces. Comparison of this theoretical threshold nucleus luminosity with \citetalias{cote06} 
reveals that many observed nuclei are much more luminous than would be predicted by this model,
which suggests that the nuclei in early-type galaxies may have formed in some low S\'ersic index, gas-rich 
progenitors that have since evolved morphologically.

High-resolution observations of molecular and neutral hydrogen
in these galaxies may be able to constrain the role of gas inflow and enrichment 
in nucleus formation, since H$_{2}$ will highlight regions of star formation, while
HI is a tracer of processes affected by the ICM and gravitational 
interactions. Subarcsecond-resolution  maps of molecular starburst 
gas --- by using ALMA to observe the CO transitions and EVLA
to detect HI through  1.4~GHz emission ---
would allow the relationship between galaxy nuclei and molecular gas 
to be examined in much greater detail than is currently possible.

% Black hole

\subsubsection{Possible Connections to Black Holes}

As discussed in \S~\ref{sec:introduction}, recent observations have uncovered
the coexistence of nuclei and black holes in intermediate-mass galaxies,
which may have implications for the evolution of the central regions of galaxies.
For instance, \citet{hopkins10b} performed simulations of gas accretion on to a 
black hole, which they find can form a lopsided, eccentric nuclear disk that
exerts a strong torque on and drives in the remainder of the gas, producing a 
system much like that found in M31. 

Another simulation by \citet{bekki10} examined the merging of 
two nuclei containing black holes, 
and found the dynamical heating of the cluster from the black hole 
binary expelled stars from the center, with the final stellar
density of the remnant decreased with increasing black hole mass fraction.
This type of merger could produce observed ``core'' galaxies with larger black 
holes (as originally noted by \citealt{ebisuzaki91}; see also \citealt{milosavljevic01}), and 
shape the inner regions of intermediate-luminosity galaxies in which a nucleus
is difficult to distinguish observationally (\S\ref{sec:freq}). Their simulations further showed that if
 only one nucleus had a black
hole, the decrease in stellar density of the nucleus was less pronounced, as
most of the heating comes from the black hole binary. In mergers where neither nucleus
had a black hole, the stellar density of the nucleus increased.

If black holes do become an increasingly dominant component of the CMO mass budget in high- and 
intermediate-luminosity galaxies, then
they could either hinder nucleus growth, or lower the density of the nucleus through mergers
until it is destroyed by black hole binary feedback. These effects could  
create the trends in intermediate-mass galaxy surface brightness profiles observed in this study, 
where the galaxies undergo a transition from central light ``excesses" to ``deficits" as they become 
more luminous (see also \citetalias{glass11}).

%
% SUMMARY
%

\section{Summary} \label{sec:summary}

This HST study examined 43 early-type galaxies in the Fornax cluster, imaged in the 
ACS F475W and F850LP bands. Our analysis --- performed in both one-
and two-dimensions --- extracted photometric and structural parameters for 31 compact stellar
nuclei in these early-type galaxies. The main results are summarized as follows:

\begin{enumerate}
	\item We have compared our 1D results to those obtained by using 2D image modeling
	 techniques, and found the extracted nucleus structural parameters to be 
in agreement for both methods. Although 2D fitting potentially allows for full structural
decomposition of a galaxy, 1D methods enable characterization the outer regions with a single 
surface brightness profile. We conclude that 1D fits are more appropriate for our
study, since they allow us to easily compare nucleus and galaxy parameters in an objective and
homogeneous way. 
	\item We find that $72\pm13$\% of the 43 galaxies in our sample are nucleated, which 
is a significant increase from ground-based studies. The nuclei --- defined as a central excess
relative to the inward extrapolation of a S\'ersic model \citepalias{cote06} ---
are found exclusively in galaxies with $M_{B}\gtrsim-19.5$ (${\cal M}_* \lesssim 10^{10.6}{\cal M}_{\odot}$),  and the frequency of 
nucleation for galaxies fainter than this magnitude is $89\pm16$\% (31/35). As was found  
previously in the Virgo cluster, nuclei are exceedingly
common in low-mass, early-type galaxies in the Fornax cluster (i.e., ${\cal M}_* \gtrsim 10^9{\cal M}_{\odot}$).
	\item Most nuclei are not significantly offset from their host 
photocenter --- only three are offset by more than $0\farcs5$. We do not find any trend between the magnitude 
of the offset and host galaxy luminosity. 
	\item We find a nearly constant nucleus-to-galaxy luminosity ratio of $\approx$0.4\%. 
The observed nucleus luminosity function can be understood therefore in terms of the
galaxy selection function (and the fact that galaxies brighter than $M_{B}\lesssim-19.5$ do not contain nuclei). 
If we parameterize the nucleus luminosity function as a normalized Gaussian, we find 
peaks at $\langle M_{g}\rangle = -11.5$ and 
$\langle M_{z}\rangle = -12.7$~mag, which is $\sim40$ times more
luminous than the peak of the GC luminosity function.
The nuclei are also found to have larger sizes and different effective
surface brightness scaling relations than the GCs. 
	\item The colors of the nuclei in hosts with $B_{T}<13.5$ are  found
to correlate with galaxy colors, as well as with galaxy and nucleus luminosities. 
In particular, both the galaxies and the nuclei were observed to become
increasingly red with increasing galaxy luminosity, with the trend being steeper 
for the nuclei. This leads to a relation between nucleus-and-host color difference 
and host magnitude, where nuclei that are more red than their hosts are found predominantly
in brighter galaxies, and vice versa. However, on average most of the nuclei are 
significantly bluer in $(g-z)$ color than their hosts by $0.28\pm0.04$ mag. 
	\item A comparison to \citetalias{cote06}, which examined the nuclei of early-type galaxies in Virgo,
reveals many similarities between the nuclei in the two environments. Both studies find similar frequencies of nucleation
(increasing sharply from 0 to $\gtrsim70$\% for galaxies with $M_B>-19.5$~mag), 
surface brightness selection effects, nucleus-to-galaxy luminosity ratios, 
nucleus luminosity functions, sizes, and color-magnitude relations. 
The trend along the luminosity function where the galaxy central surface 
brightness profiles gradually change from having a luminosity ``deficit'' 
to an ``excess'' is shared by both samples \citepalias[see also][]{cote07, glass11},
 which suggests that generic 
formation and evolution processes largely independent of the galaxy environment
are involved in shaping the central regions of galaxies. Rather, 
nucleus creation may be more contingent on local factors, especially 
host galaxy mass. 
\end{enumerate}

{\it Our conclusion is that, in low-mass galaxies, the dominant mechanism for nucleus
growth is probably infall of star clusters through dynamical friction, while at higher masses,
gas accretion resulting from mergers and torques becomes dominant.}
There is no reason to expect either of these
processes to be discontinuous, and we argue that the relative importance of
these processes vary smoothly as a function of galaxy mass. We examine the
efficiency of dynamical friction in our sample galaxies and confirm the finding of
many previous studies that  star cluster infall is most effective
in low-mass galaxies.  Based on simulations carried out by other researchers, we
argue that gas infall, followed by central star formation, becomes increasingly important
in high-mass galaxies having S\'ersic indices that may have been inflated by successive mergers and accretions. 
There is also some evidence for ``hybrid nuclei" in some of the
intermediate-mass galaxies in our sample: i.e., nuclear components with complex inner structures. 
Simulations that take into account {\it multiple} formation
mechanisms --- star cluster infall, gas accretion driven by
tidal torques and/or accretions and mergers, the influence of central black holes, etc --- are urgently needed to
elucidate the processes that drive nucleus formation in different mass regimes. 

Both dissipationless cluster infall and gas accretion models make predictions 
that nucleus formation would depend on local density \citep{oh00, babul92}. 
Although the fact that we do not find any
major differences between the nuclei of Virgo and Fornax suggests 
that local density may not be a dominant factor in their formation, observations
that examine the entire volume of a galaxy cluster (and that have the 
sensitivity necessary to detect the nuclei) may help determine the 
role environment plays in shaping the nuclei and their hosts. In this context, the
forthcoming {\it Next Generation Virgo Cluster Survey} (Ferrarese et~al. 2012),
which is imaging the entire Virgo cluster to a (10$\sigma$) depth of $g \approx 25.7$,
should provide important new constraints on formation models.

\acknowledgments

Support for programs GO-9401 and GO-10217 was provided through a grant from the Space Telescope Science 
Institute, which is operated by the Association of Universities for Research in Astronomy, Inc., under 
NASA contract NAS5-26555. 
MT thanks Chien Peng, Lisa Glass, and Kaushi Bandara for their kind assistance, and acknowledges 
support from the University of Victoria through their fellowship program, and from
the Marie Curie Initial Training Network CosmoComp (PITN-GA-2009-238356). 
AJ acknowledges support from BASAL CATA PFB-06, FONDAP CFA 15010003, Ministry of Economy ICM Nucleus
P07-021-F and Anillo ACT-086.
LI acknowledges support from the Chilean Center of Excellence in Astrophysics 
and Associated Technologies (PFB 06) and from the Chilean Center for Astrophysics FONDAP 15010003. 
EWP gratefully acknowledges partial support from the Peking University Hundred Talent Fund (985) 
and grants 10873001 and 11173003 from the National Natural Science Foundation of China (NSFC).
The authors thanks Lisa Glass for providing the surface brightness 
profiles in the bottom panels of Figure~\ref{fig:schematic}.  
This research has made use of the NASA/IPAC
Extragalactic Database (NED) which is operated by the Jet Propulsion Laboratory, California Institute of Technology, 
under contract with the National Aeronautics and Space Administration.

\textit{Facilities:} HST(ACS)

%\clearpage

%%%%%%%%%%%%%%%%%%%%%%%%%%%%%%%%%%

%
% APPENDIX
%

\appendix 

%
% 1D vs 2D
%

\section{Comparison with 2-Dimensional Surface Brightness Profile Fitting} \label{app:2D}

\S\ref{sec:observations} describes the determination of nuclei 
parameters through 1D fitting of surface brightness profiles from {\tt ellipse}. This appendix examines 
the pros and cons  of 1D and 2D methods when measuring parameters for the photometric and structural
parameters of nuclei in early-type galaxies.

In general, the decision to use a 1D or 2D
approach depends on the specific scientific goals. If a galaxy has multiple components (which even for early-type galaxies can
include, e.g., bulges, large-scale disks, embedded disks, outer/inner rings, bars, shells, dust filaments,
dusk disks,  and even faint spiral arms)\footnote{Although early-type galaxies are often considered structurally simple systems, 
{\it all} of these features are found in the sample of 143 early-type galaxies 
studied in the ACSVCS (\citealt{lauraa06}) and ACSFCS (Paper~III).}, then, by using 2D decomposition, individual structure can, in principle, 
be fitted with separate profiles and the  galaxy's composition examined 
in detail. The 2D fitting program GALFIT \citep{peng02, peng10} allows the implementation
of many surface brightness profile modifications, such as variability of their diskiness/boxiness, 
or the addition of spiral arms and non-axisymmetric bending modes --- an attractive feature of the
2D method. 
However,  full galaxy decomposition is, in practice, not always straightforward, particularly
for nearby galaxies observed at HST/ACS resolution.
In many situations, it is not clear how many components are needed to fully 
fit a galaxy, and the physical origin of each component may not be obvious.  
For example, sometimes multiple surface brightness profiles are required to 
fit what may be the same photometric component (see \citealt{peng02} for examples) 
due to the fact that  the models used in 2D methods have 
fixed center, ellipticity, 
and position angle, and have difficulty characterizing a galaxy profile in which 
these parameters are not intrinsically constant on all scales.

% Demo images for 1D treatment of FCC190 (a "typical" galaxy)

\begin{figure}
	\figurenum{21}
	\plotone{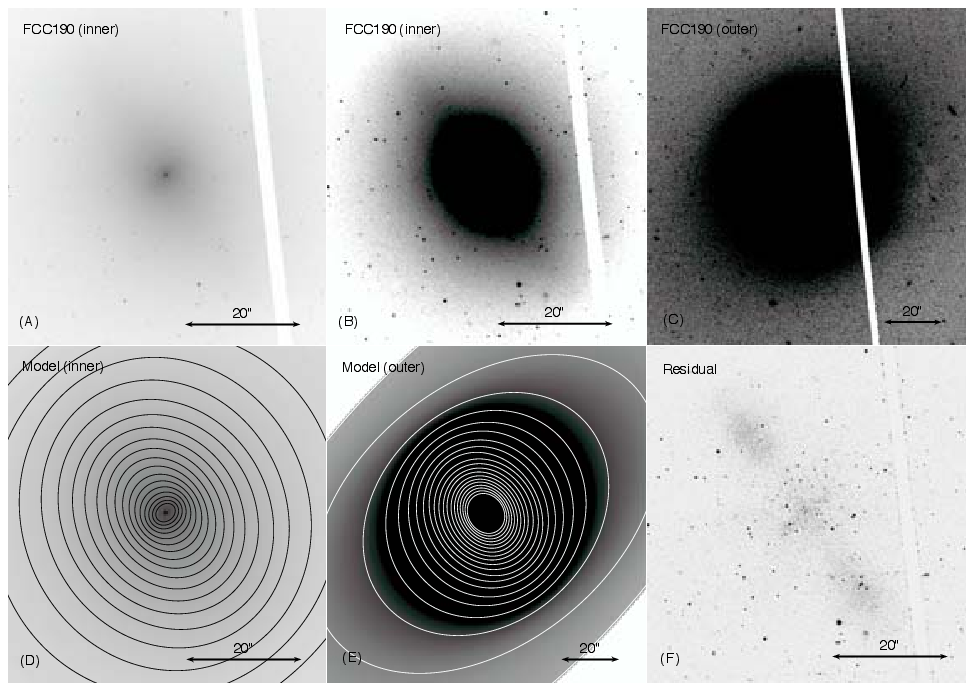}
	\caption[Demonstration of 1D fit treatment of FCC 190]
	{\emph{Upper row}: F475W image for FCC~190 displayed at three different intensity stretches ($A$, $B$ and $C$) and two different magnifications 
	($A/B$ vs. $C$). Note the prominent nucleus visible in {\it panel~(A)}, and the dramatic changes in ellipticity and position angle with radius.
		\emph{Panels (D) and (E)}: Galaxy model constructed using {\tt ellipse}, with contours overlaid to illustrate the gradual changes in 
		galaxy flattening and orientation.
		\emph{Panels (D) and (E)}: Residual image (observed $-$ model) showing a weak residual bar, with a peak intensity of $\sim$ 0.02 $e$ pixel$^{-1}$, corresponding to $\mu_g \sim  23.8$~mag~arcsec$^{-2}$.}
	\label{fig:fcc190}
\end{figure}

% 1D vs 2D Plot

\begin{figure*}
	\figurenum{22}
	\plotone{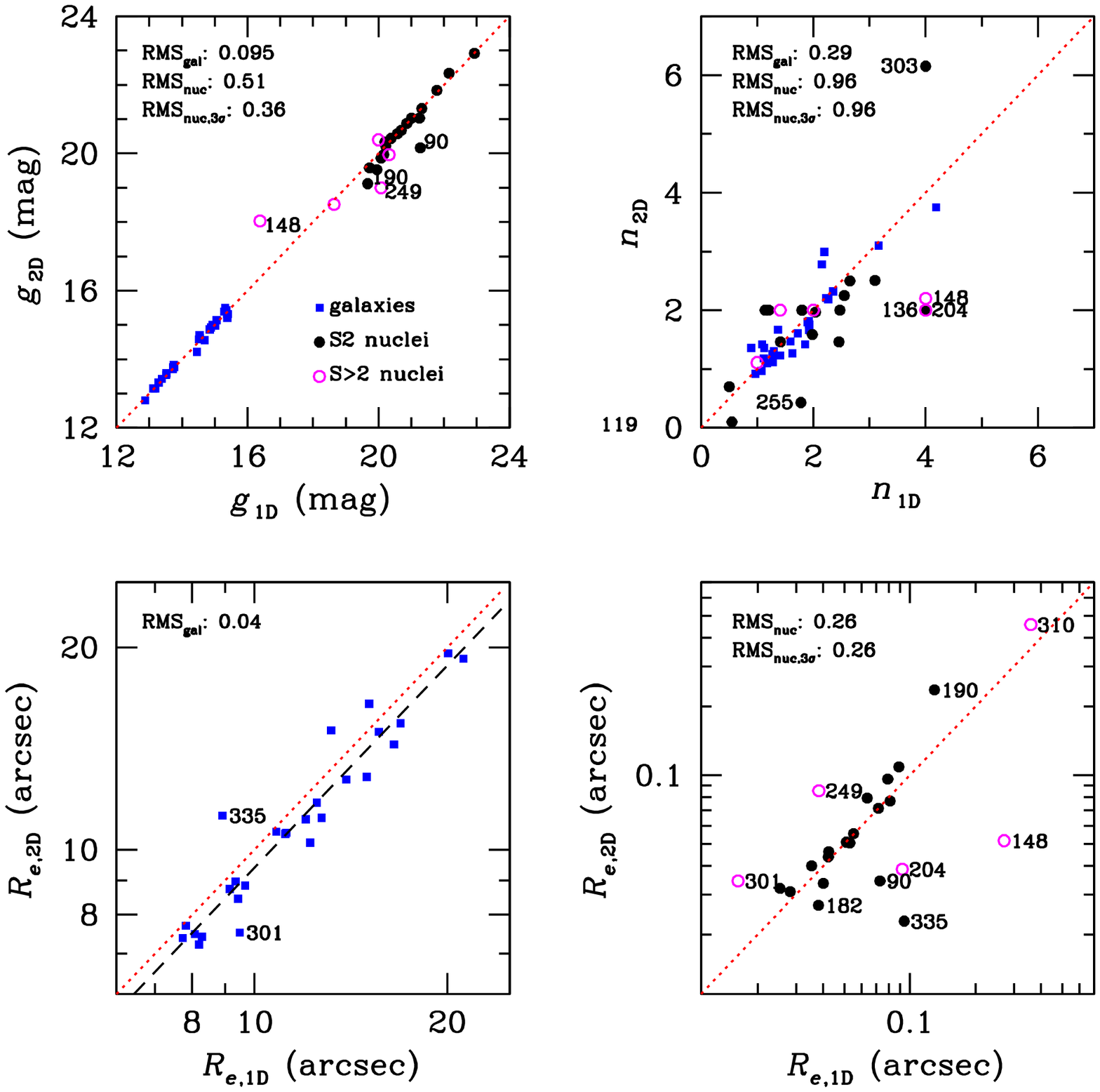}
	\caption{\emph{Top left}: Values for nucleus $g$-band luminosity
	obtained from the 1D and 2D fits (ordinate and abscissa, respectively). The filled blue squares
	show the host galaxies, the filled black circles indicate the nuclei
	from galaxies fit well by a S2 profile in 2D, and the filled magenta circles
	represent the nuclei belonging to galaxies in which more than two S\'ersic components (i.e., S$>$2) were
	required  in 2D.	The dotted red line is the relation where 
	the parameters obtained from both methods are equal. Galaxies and nuclei where the 
	measured magnitudes differ by more than 0.5 mag between methods are labeled.
	The root mean squared (RMS) error around the magnitude sample mean is shown for the 
	galaxies, nuclei, and nuclei again after applying $3\sigma$-clipping. 
	\emph{Top right}: Same as the top left, except for S\'ersic
	indices. Galaxies and nuclei where the measured S\'ersic indices
	differ by more than 1.0 between methods are labeled. We note that labels 
	for the blue filled squares are to the left of the points, while those for the
	magenta open squares are to the right.
	\emph{Bottom}: Same as top, except for galaxy (\emph{left}) and nucleus
	 (\emph{right}) geometric mean effective radii. Galaxies and nuclei where 
	 the measured effective radii in 1D and 2D differ by more than 0.1 in the log
	 are labeled. The black dashed line shows the best-fit line to the galaxy
	 effective radii, with a fixed slope of unity.}
	\label{fig:1Dvs2D}
\end{figure*}

The method of 1D profile fitting used in this work, however, allows the aforementioned parameters to vary, 
and we are therefore usually able to cleanly fit an entire galaxy with a 1D model. 
A demonstration of this is shown in 
Figure~\ref{fig:fcc190}, where we examine the structure of FCC 190 ({\it panels A--C}),
 and plot the {\tt ellipse} model derived from our fitted elliptical isophotes ({\it panels D--E}).
This figure illustrates the striking change in the shape of this galaxy's
isophotes when moving from small to large scales, and how this effect
is well-captured by the model. The residuals of the fit ({\it panel F}) are relatively clean,
and reveal a weak ($\mu_g \sim  23.8$~mag~arcsec$^{-2}$) central bar. To compare
to a 2D fit, the inner $10\times10\arcsec$ residuals from fitting 1S and 2S profiles 
to FCC 190 using GALFIT are shown in Figure~\ref{fig:2comps}a. Clearly, two S\'ersic
profiles with fixed ellipticity and position angle are unable to fully parameterize
this galaxy. However, the penalty in this approach is that the information 
about the shapes, sizes, and relative position angles of various galaxy components is lost, 
as their surface brightness profiles blend together into a single component which 
describes the galaxy on global scales.

Our study is concerned with the properties of the nuclei in comparison to their
host galaxies, and with the global trends in these properties as a function of galaxy 
luminosity or mass. Thus, we are not interested in a full decomposition
of any large-scale galaxy structure; rather, we are seeking to characterize 
the main galactic body component as a whole, so 1D techniques are 
most appropriate for our study. However, it is important to ensure that
the nucleus structural parameters extracted using 1D methods are robust. 
To test this assertion, we perform surface brightness profile fitting in 2D, and compare
the results obtained using the two techniques.

% Magnitude differences

\begin{figure}
	\figurenum{23}
	\plotone{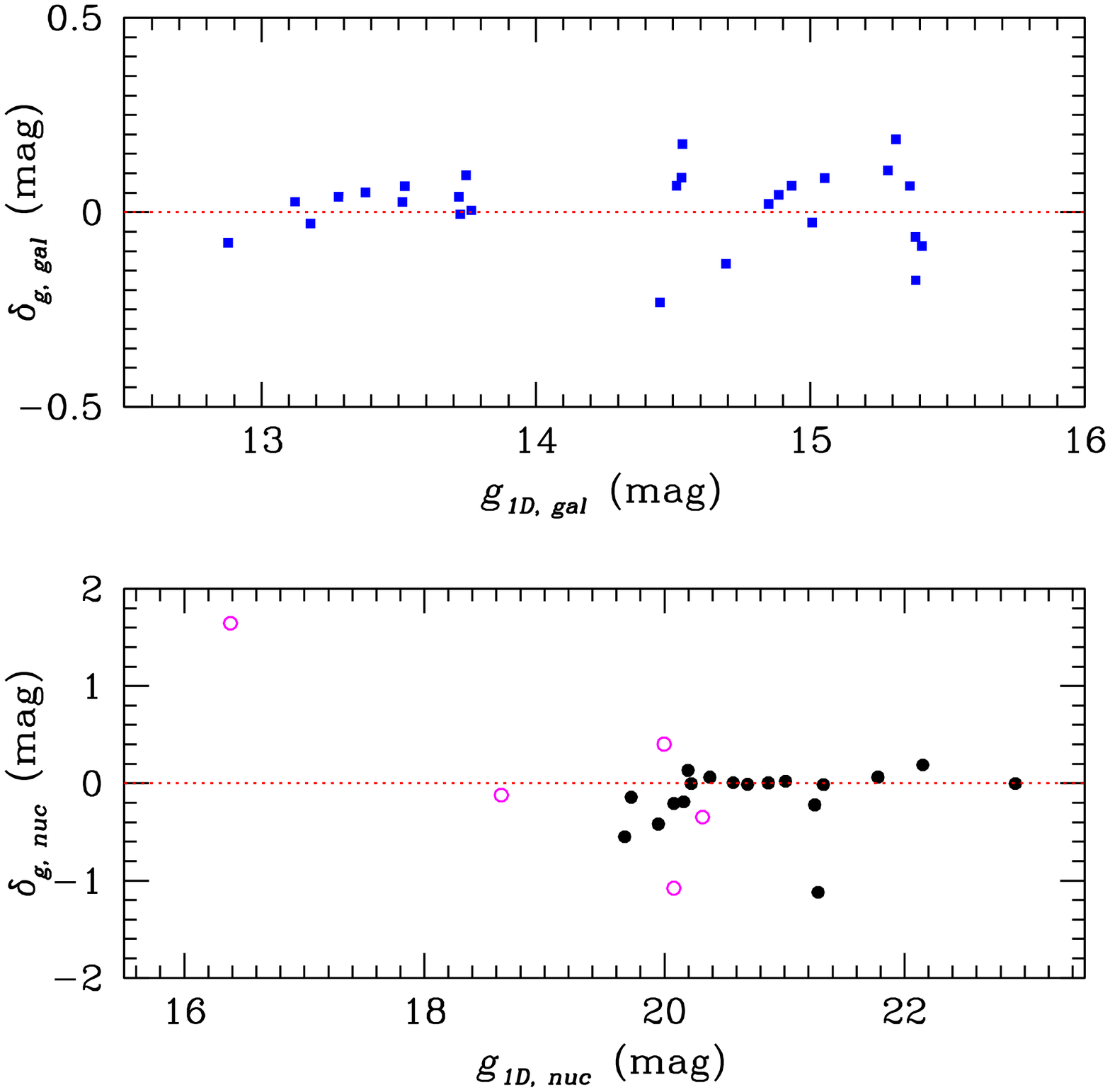}
	\caption{\emph{Top}: The difference between 2D and 1D galaxy magnitudes against
	1D galaxy magnitudes.
	 \emph{Bottom}: Same as the top, but for nuclei magnitudes.}
	\label{fig:1Dvs2Db}
\end{figure}

\subsection{Procedure}

To perform our 2D analysis, we use GALFIT \citep{peng02, peng10}, a program that fits 
galaxy images using multi-component 2D intensity profiles, using an iterative 
downhill gradient Levenberg-Marquardt algorithm.  This 2D analysis is performed on all 
galaxies in our sample with $B_T \geq 13.5$, a cutoff which was chosen to include most
of the nucleated galaxies, while avoiding those that are much 
more challenging to fit in either 1D or 2D. Galaxies brighter than this are known to often 
show a complex structure, regardless of their classification as Es, S0s, dEs or dS0s. 
For instance,  some of the brighter dEs are known to show  
substructures including disks, spiral arms, and bars \citep[e.g.,][]{lisker06a, lisker07}. Likewise,
more massive galaxies --- often classified as Es and S0s --- frequently show similar 
morphological complexities  
\citep[see, e.g.,][]{bender87, combes90, nieto92, scorza98, lauraa06, krajnovic11}. 
The substructures identified in these early-type galaxies could either be a sign that
they are intrinsically more complex objects, or a selection effect arising from their higher 
luminosities and surface
brightnesses, which aid in the detection of these distinct components. In any case, the 
sample of galaxies used in our 2D analysis consists of  27 galaxies, 24 of which are 
found to be nucleated in our 1D analysis. This sample includes roughly equal numbers 
of galaxies listed in Table~2 of \citetalias{jordan07a} as ``giants" (E/S0) or ``dwarfs" (dS0, dE, etc), although such 
classifications such be viewed with caution since there can be significant
discrepancies among classifiers: see, e.g., \citet{chen10} and Paper~III where
issues relating to the morphology of ACSVCS and ACSFCS galaxies are explored in more detail.

Our analysis proceeded by first measuring the background sky value. To do so, we used
 \textit{SExtractor} to mask out any background sources, 
and then convolved this mask with a Gaussian in order to thoroughly cover  
any diffuse outer edges. The galaxy was then masked with an ellipse of geometric
radius length between five and six effective radii (determined from the 1D analysis). 
We then used the biweight value of the remaining pixels as the sky value 
for each of the four ACS chips.
Although the sky value between different chips was found to vary up to $\sim1$ count,
we found that such a count difference resulted in no more than $5\%$ difference in any of 
the fitted parameters; we therefore adopted the average of the biweight estimates for
each of the four chips as the sky value.
          
We began by fitting each galaxy with a single S\'ersic ({\tt S1}) profile. We then
attempted to fit each of the 24 galaxies classified as nucleated in 1D by adding a second S\'ersic
component (for the central nucleus). In 13 cases, it was possible to fit the nucleus with a S\'ersic 
model with all fit parameters varying freely. Five more galaxies required 
a prior on the nucleus S\'ersic index which was fixed at $n = 2$ in analogy with Galactic GCs. 
For the six remaining galaxies,
GALFIT was not able to converge on a nucleus with only one S\'ersic component
fitted to the main body of the galaxy; at least one other large-scale component
needed to be added for before a fit to the nucleus could be achieved. 
However, in one case (FCC~43) the nucleus parameters were flagged as having caused
numerical convergence issues, and thus we do not include the {\tt S$>$2} fit in our results.	
In all cases, we did not impose any constraints on the relative position angles of 
of the fitted components. 

The above procedures are summarized in Fig~\ref{fig:1Dvs2D},
where we have plotted the 1D versus 2D magnitudes, S\'ersic indices, 
and effective radii for the galaxies and nuclei from our sample.
For the galaxies that require more than two S\'ersic components  to 
fit the nucleus, we use the parameters from our 2D {\tt S1} fit to plot
the galaxy portion. Although the galaxy main body (filled black circles) parameters
are generally in good agreement from both techniques, we note a slight offset
in effective radius, where those obtained from the 2D fits are usually somewhat smaller
 than in 1D (by a factor of $0.94\pm0.02$, derived from least-squares fit, with a fixed line slope of 1,
 to the galaxy main body effective radii in the log).
The {\tt S2} nuclei (filled blue squares) are also 
relatively consistent between techniques, although with some notable outliers 
that will be discussed below. Finally, the non-{\tt S2} nuclei
(magenta open squares) appear to show the most scatter. We note that the scatter 
in nucleus magnitudes appears to be the most significant for the brightest nuclei, 
probably due to the increased difficulty of extracting nucleus parameters from structurally
complex galaxies that often have high central surface brightness. This can be seen 
clearly in in Figure~\ref{fig:1Dvs2Db}, where we have plotted the magnitude differences as 
a function of 1D magnitudes. We now discuss findings for galaxies in these
different categories.

% S1

\begin{figure}
	\figurenum{24}
	\plotone{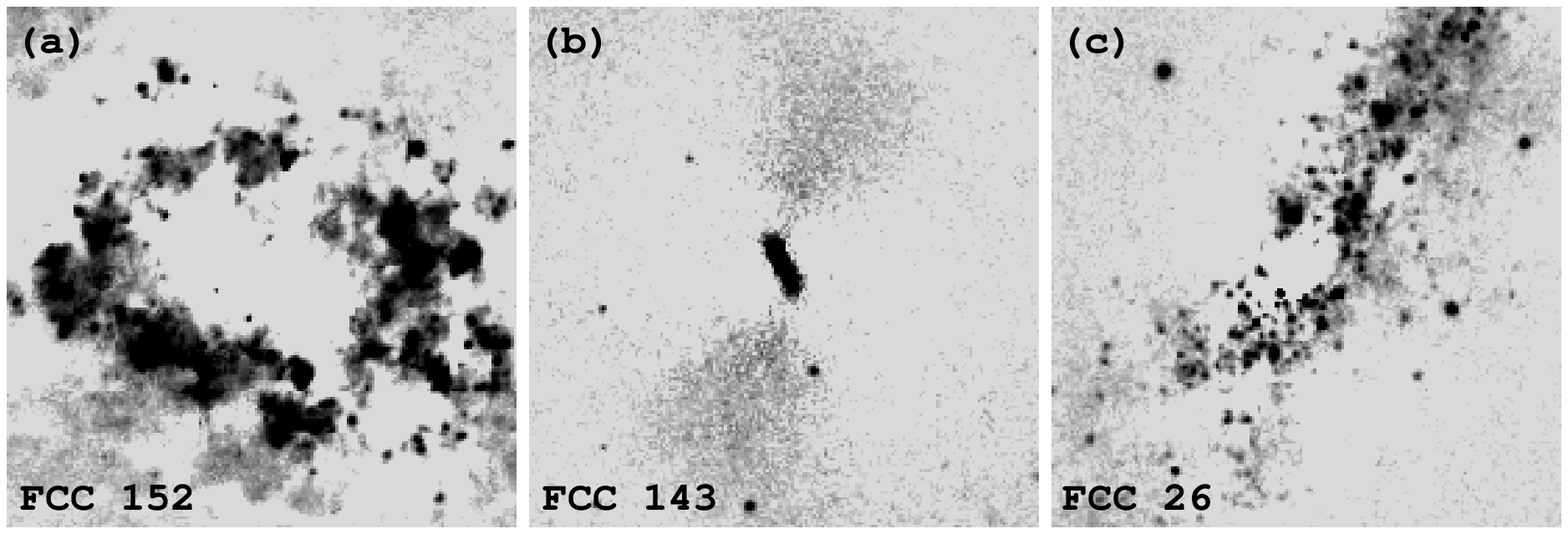}
	\caption{\emph{Panel (a)}: GALFIT residuals from an {\tt S1} model
	fitted to FCC 152, showing the inner $10\arcsec\times10\arcsec$ region.
	 \emph{Panels (b)--(c)}: Same as for {\it panel (a)} but for FCC~143 and FCC~26.}
	\label{fig:1comp}
\end{figure}

% S2

\begin{figure*}
	\figurenum{25}
	\plotone{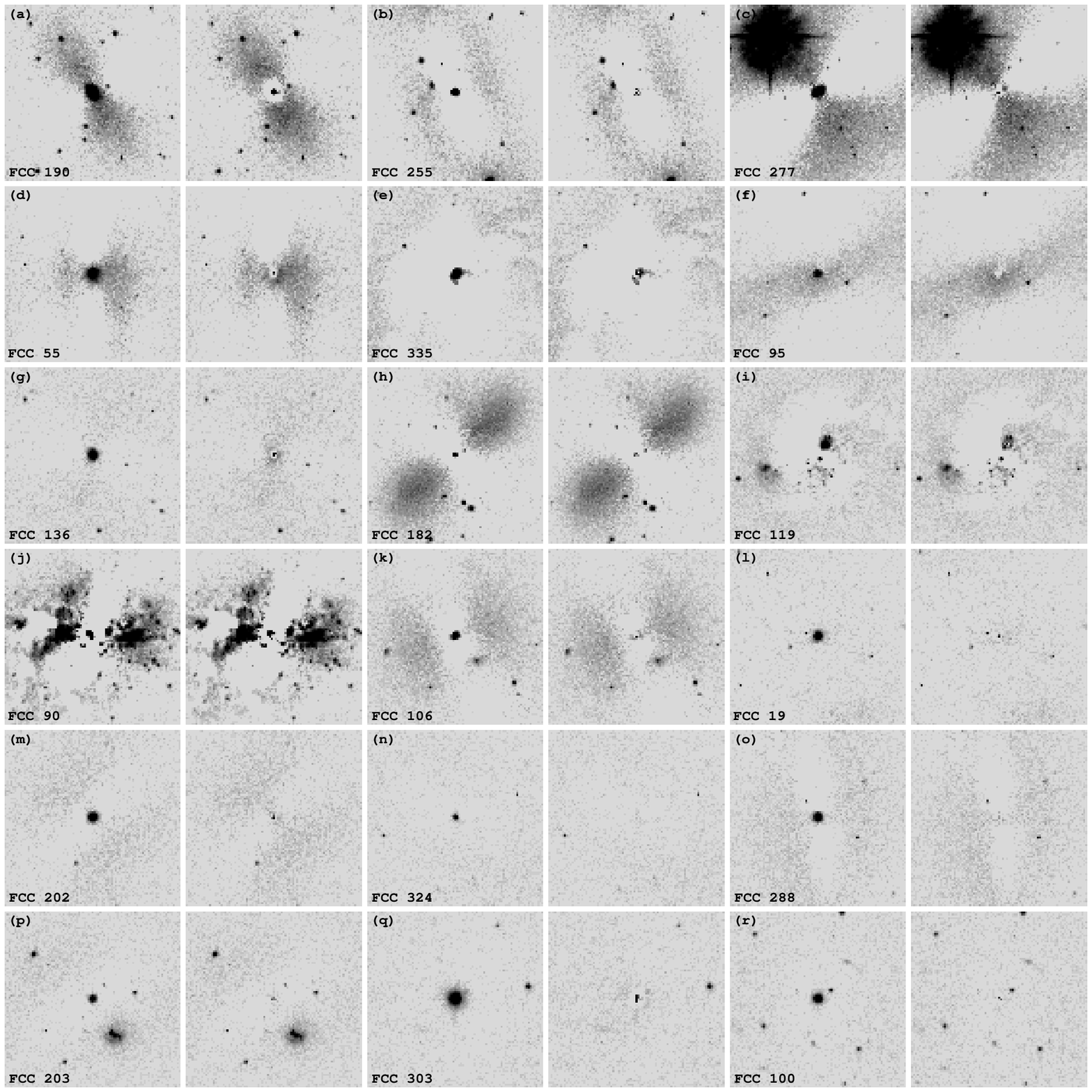}
	\caption{\emph{Panel (a)}: GALFIT residuals from {\tt S1} (\emph{left})
	and {\tt S2} (\emph{right}) models fitted to FCC 190, showing the inner 
	$10\arcsec\times10\arcsec$ region. These results should be compared to the {\tt ellipse} model (1D) results
	shown in Figure~\ref{fig:fcc190}.
	\emph{Panels (b)--(r)}: Same as for {\it panel (a)} but for the galaxy labeled. Galaxies have been
	ordered by increasing blue magnitude (decreasing luminosity) from the FCC.}
	\label{fig:2comps}
\end{figure*}

\subsection{Non-Nucleated Galaxies (S1)} 
    
There are three galaxies in our 2D sample where we do not find a nucleus in 
our 1D analysis, a result with which we find full
agreement in 2D. It is interesting to examine the 
residuals of a single-S\'ersic fit to these objects individually
to determine why they are not
found to be nucleated, since it is the {\it lack} of a nucleus that is unusual 
for galaxies in our sample.

The residuals of FCC 152 (Figure~\ref{fig:1comp}a) reveal large amounts
of dust, but with no nucleus-like object present in the central regions. 
FCC 143 (Figure~\ref{fig:1comp}b) shows a small bar in the residuals, 
which appears to have a bright excess in the center.\footnote{
Performing a 1D fit to the residuals, we find 
$R_e \sim 0\farcs02$ and $g\sim20.19$~mag for the inner 
and $R_e\sim0\farcs22$ and $g\sim20.63$~mag for the outer component.
However, due to the complexity of the inner structure, we consider
these results to be uncertain, and retain the non-nucleated 
classification. 
}
Nevertheless, even with the addition of one or two more large-scale S\'ersic
profile components, GALFIT is unable to fit a central nucleus.  
Finally, the low-mass galaxy FCC 26 (Figure~\ref{fig:1comp}c) has two bright compact objects located 
$0\farcs95$ and $1\farcs37$ away from the galaxy photocenter. However, it is unclear 
if either of these objects in this actively 
star-forming, ``dE/dIrr transition" galaxy can unambiguously called a true ``nucleus''.

\subsection{Nucleated Galaxies Fit With Double-S\'ersic Profiles (S2)}

Of the 23 nucleated galaxies in our 2D sample, we are able to fit the galaxy 
and nucleus using an {\tt S2} profile for 18 systems. The residuals from the
{\tt S1} and {\tt S2} fits to these galaxies are shown
in Figure~\ref{fig:2comps}, where the galaxies are displayed in order of 
increasing blue magnitude from the FCC. This figure illustrates how the structural complexity of 
the galaxies
seems to increase, and then diminish, as their luminosity decreases --- reaching an apparent maximum in the range 
$-19 \lesssim M_B\lesssim -17$ or $10^{10.4} \gtrsim {\cal M}_*/{\cal M}_{\odot} \gtrsim 10^{9.6}$
--- with the residuals for the faintest galaxies appearing much cleaner \citep[see][]{lauraa06, lisker06a}.
Of course, part of this 
apparent simplicity is likely related to the lower S/N of the available imaging for the faintest and
lowest surface brightness systems.

For five of the {\tt S2} galaxies, the S\'ersic index of the nucleus needed to be held fixed during the fit.
FCC 190 (Figure~\ref{fig:2comps}a), FCC 55 (Figure~\ref{fig:2comps}d), FCC 95 (Figure~\ref{fig:2comps}e), 
and FCC 136 (Figure~\ref{fig:2comps}g), all have substructure such as bars that overlap with the 
nuclear region which the second S\'ersic component attempts to fit. By fixing the  S\'ersic 
index of the nuclei at $n =2$ (appropriate for Galactic GCs), GALFIT is able to fit 
the nucleus, with the resulting magnitude and effective radius of the nucleus is in agreement 
with the 1D results in all cases except for FCC 190, which is discussed below. 
The other galaxy that requires a fixed nucleus S\'ersic index, FCC 335, contains a large
amount of dust in the central regions, and if the nucleus S\'ersic index is allowed to vary, 
then the nucleus effective radius and S\'ersic index converge to very small values that 
GALFIT warns may cause numerical convergance issues which cause the final solution to be unreliable.
The differences between the 1D and 2D results for this galaxy are also discussed below, where
we describe nuclei that are notable outliers in Fig~\ref{fig:1Dvs2D}.
Specifically, the nuclei of FCC 190, FCC 335 and FCC 90 have 1D and 2D magnitude 
differences of $>0.4$~mag and fractional differences in their effective radii of $>0.5$.  

\emph{FCC 190}: This nucleus is 0.55~mag brighter and twice as large in effective radius
in the 2D fit than in 1D. The galaxy exhibits distinct ``peanut-shaped" residuals, as shown
in Figure~\ref{fig:2comps}a. It should be noted that after 
fitting a both a bulge and a disk component along with the nucleus, 
the nucleus magnitude and radius %improve ($0.21\arcsec$ and 19.25 mag), but they 
are still notably disparate.    

\emph{FCC 335}: 
In 2D, the nucleus is 0.42~mag brighter, but four times smaller, than in the 1D fit.
The 2D residuals are shown in Figure~\ref{fig:2comps}e. This galaxy has a large amount of dust, 
and the center was held fixed during the ellipse fitting for the 1D analysis. 
However, the position  of the 1D fit ellipse center is actually $\sim 0.5$ pixels away from the 
nucleus (as determined by GALFIT and confirmed by eye). This could cause
the 1D analysis to overestimate the nucleus effective radius and underestimate
the magnitude, as the light from the nucleus effectively becomes smeared out.

\emph{FCC 90}: 
This nucleus is 1.12~mag brighter in 2D than in 1D. The residuals of a 
single-S\'ersic fit (Figure~\ref{fig:2comps}j) show a bright central nucleus as well as a secondary 
fainter object $\sim 0\farcs25$ away. This second object is the cause 
of a small secondary bump in the 1D surface brightness profile (see Figure~\ref{fig:sb_profiles}).
After simultaneously fitting this secondary object, the nucleus is still found to be 
1.08 mag brighter in 2D than in 1D.  Like FCC 90, there is large 
amounts of dust in the center of this galaxy, and the center was held fixed for the
1D ellipse fitting, at a point $\sim 1$ pixels away from the 2D nucleus center,
which may partly account for the smaller and brighter
nucleus found in 2D.   
     % S>2

\begin{figure*}
	\figurenum{26}
	\plotone{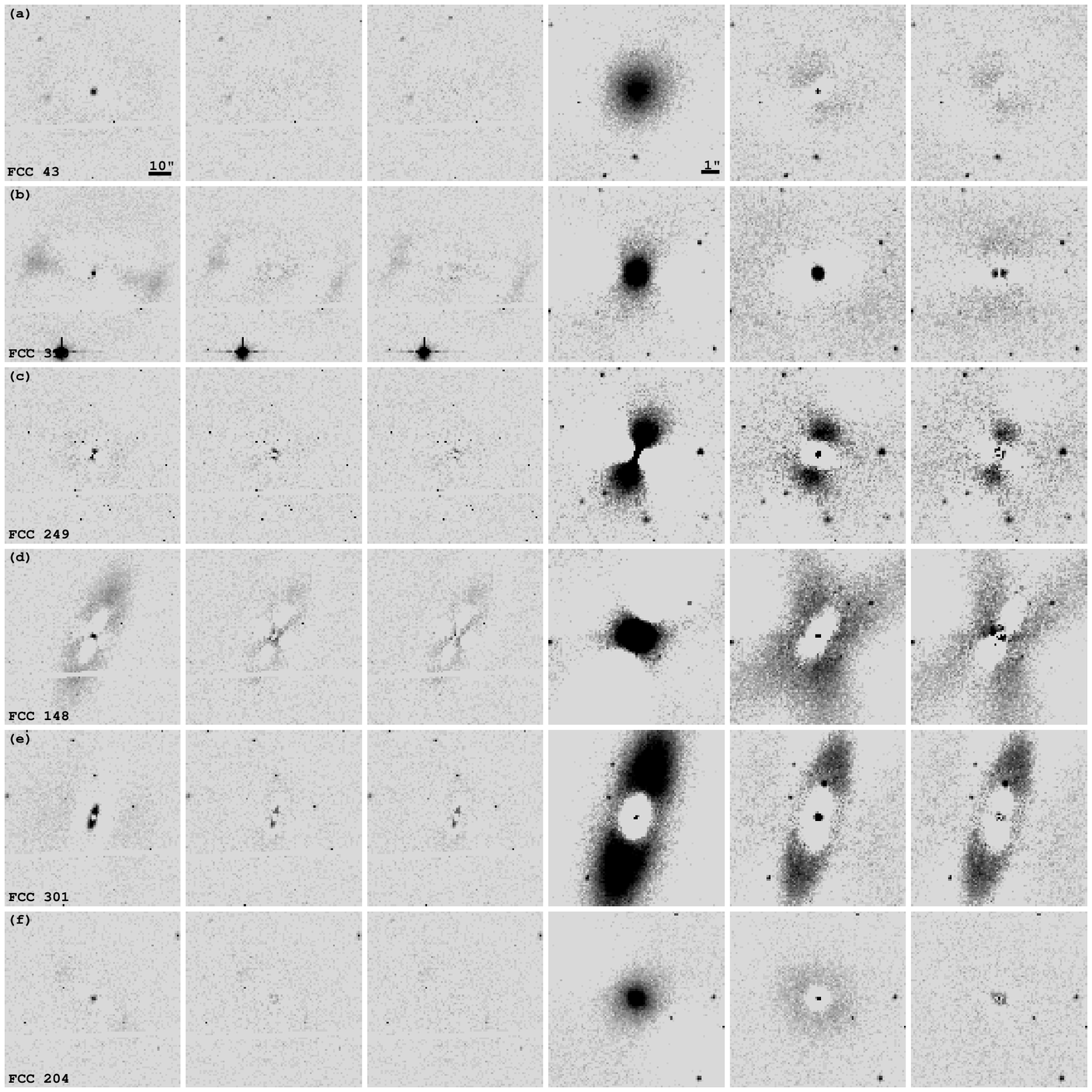}
	\caption{\emph{Row (a)}: The first three panels show, from left to right,
	$80\arcsec\times80\arcsec$ regions centered on FCC 43 with GALFIT
	residuals from an {\tt S1} fit, a two-S\'ersic component
	fit, and a two-S\'ersic component plus S\'ersic nucleus component fit.
	The last three panels show the same, but magnified to show the inner $10\arcsec\times10\arcsec$.
	\emph{Rows (b)--(f)}: The same as {\it panel (a)} but for the galaxies labeled. 
	In the case of FCC 310, the residuals show the results found using three, rather
	than two, S\'ersic components. }
	\label{fig:3comps}
\end{figure*}

% S>2 Parameter Table

\begin{deluxetable}{lcccccc}
%\tabletypesize{\scriptsize}
\tabletypesize{\small}
%\rotate
\tablecaption{1D and 2D Nucleus Parameters for Multi-Component (S$>$2) Galaxies \label{tab:3comps}}
\tablewidth{0pt}
\tablehead
{
\colhead{FCC} & 
\multicolumn{2}{c}{$g_{AB}$} & 
\multicolumn{2}{c}{$R_{e}$} & 
\multicolumn{2}{c}{$n$} \\
\colhead{} &
\multicolumn{2}{c}{(mag)} & 
\multicolumn{2}{c}{(arcsec)} & 
\multicolumn{2}{c}{} \\
\colhead{} &
\cline{1-6}
\colhead{} & 
\colhead{1D} & 
\colhead{2D} & 
\colhead{1D} & 
\colhead{2D} &
\colhead{1D} & 
\colhead{2D} 
}
\startdata
 310  &  $18.6$  &  $18.5$  &  $0.36$  &  $0.46$  &  $1.4$  &  $2.0$  \\
 249  &  $20.1$  &  $19.0$  &  $0.04$  &  $0.09$  &  $2.0$  &  $2.0$  \\
 148  &  $16.4$  &  $18.0$  &  $0.27$  &  $0.05$  &  $4.0$  &  $2.2$  \\
 301  &  $20.3$  &  $20.0$  &  $0.02$  &  $0.03$  &  $1.0$  &  $1.1$  \\
 204  &  $20.0$  &  $20.4$  &  $0.09$  &  $0.04$  &  $4.0$  &  $2.0$ 
\enddata
%\tablecomments{}
%\tablenotetext{a}{}
\end{deluxetable}

\subsection{Nucleated Galaxies with Multiple Large-Scale Components (S$>$2)}

There are six nucleated galaxies in our sample which we were unable to model in 2D 
using an {\tt S2} model, as a second S\'ersic component in GALFIT will,
even with its S\'ersic index held fixed, attempt to 
fit a different component of the underlying galaxy.
Thus, we need to add a second, or even third, S\'ersic component  to the main body of the galaxy in order 
to fit the nucleus (i.e., three or four components
in total).  A comparison of the 1D and 2D nucleus parameters for these six galaxies is 
given in Table~\ref{tab:3comps}. Note that the nuclei in several of these galaxies
are candidates for ``hybrid nuclei" with a complex structure (e.g., a compact, high surface 
brightness component  embedded in an small-scale disk-like feature) that are suggestive of 
multiple, parallel formation processes (see \S\ref{sec:gcinfall} and \S\ref{sec:gas}).

\emph{FCC 43}: 
A small ($\lesssim10\arcsec$ in diameter) round central component, seen in 
Figure~\ref{fig:2comps}a needs to be fitted before it is possible to 
fit to the nucleus. However, because of the small size
of the fitted nucleus (i.e., an effective semi-major axis of 0.39 pixels), 
the output of GALFIT indicated that this parameter may have caused numerical 
convergence issues, making all parameters from this solution unreliable. 
We therefore did not include these results in Fig~\ref{fig:1Dvs2D}.

\emph{FCC 310}: 
After fitting with a single S\'ersic profile, the bar and 
and outer envelope-like structure of this galaxy become apparent from the residuals, 
seen in Figure~\ref{fig:2comps}b.
To fit the nucleus, we must first fit an $n=2.38$ bulge-like component, an 
$n=0.26$ bar, and an $n=0.20$ outer envelope. After these three 
components are fit, the nucleus appears quite clearly in the residuals, and it 
can be fitted by adding a fourth S\'ersic profile with its S\'ersic index held fixed at $n = 2$. 

\emph{FCC 249}: 
A single-S\'ersic fit reveals a peanut-shaped residual in the 
center (Figure~\ref{fig:2comps}c), with a possible nucleus. After a second 
small component is added, a nucleus becomes apparent in the residuals. The nucleus
can then be fitted with its S\'ersic index held fixed at $n = 2$.

\emph{FCC 148}: 
This galaxy shows a very boxy inner bulge, with X-shaped
isophotes in intermediate regions (Figure~\ref{fig:2comps}d).
Since we are unable to fit the host and nucleus with a double-S\'ersic model,
a second large-scale component with disk-like properties
($n=1.04$, and an axis ratio of 0.36) is added, after which
 GALFIT will converge on the nucleus. Although a nucleus is not very prominent
in the two-component fit residual, the S\'ersic
index of the bulge-like component grows to $n = 9.35$ if a nucleus is not included in the fit.
After a nucleus is included, the fitted bulge S\'ersic index is $n = 5.1$. 
The disk-like component does not change significantly with the addition of the nucleus.

\emph{FCC 301}: 
The complex structure of this galaxy, seen in the single S\'ersic component 
fit residuals in Figure~\ref{fig:2comps}e, can be appreciated from the 1D surface
brightness profile (Figure~\ref{fig:sb_profiles}), where the intensity is slightly over-subtracted at $1\arcsec$,
and then under-subtracted out to $\sim 5\arcsec$. There are also
bright outer wings, at $>10\arcsec$ scales. After a single-S\'ersic 
component is fitted, a second component will converge on the larger bright central excess,
and a third S\'ersic profile will then fit the nucleus. However, the bright residuals show that the 
the main body of this galaxy is not well described in 2D, even by two S\'ersic profiles.   

\emph{FCC 204}: 
As is the case for FCC 43, there appears to be an embedded disk
in this galaxy, which can be seen in the residuals of a single S\'ersic
fit, Figure~\ref{fig:2comps}f. The nucleus is found to be  
slightly fainter in the 2D fit than in 1D.
It is possible that in 1D, the central disk might be contributing to nucleus 
component and causing the nucleus luminosity to be over-estimated. 

Overall, the 2D nuclei parameters from these complex galaxies
are in reasonable agreement with those found in the 1D analysis. In terms of magnitude, only FCC 148
and FCC 249 show differences of $>0.5$ mag, although all except for FCC 310
show discrepancies of $>50\%$ in effective radius and S\'ersic index. 
However, the differences do not appear to be systematic, in the sense that 
there does not seem to be consistent under- or over-estimation of a specific
parameter in 1D or 2D. The nuclei parameters for these cases are 
likely to be more uncertain overall, and thus larger differences between the extracted 
parameters in these structurally complex galaxies is to be expected.

\begin{figure}
	\figurenum{27}
	\plotone{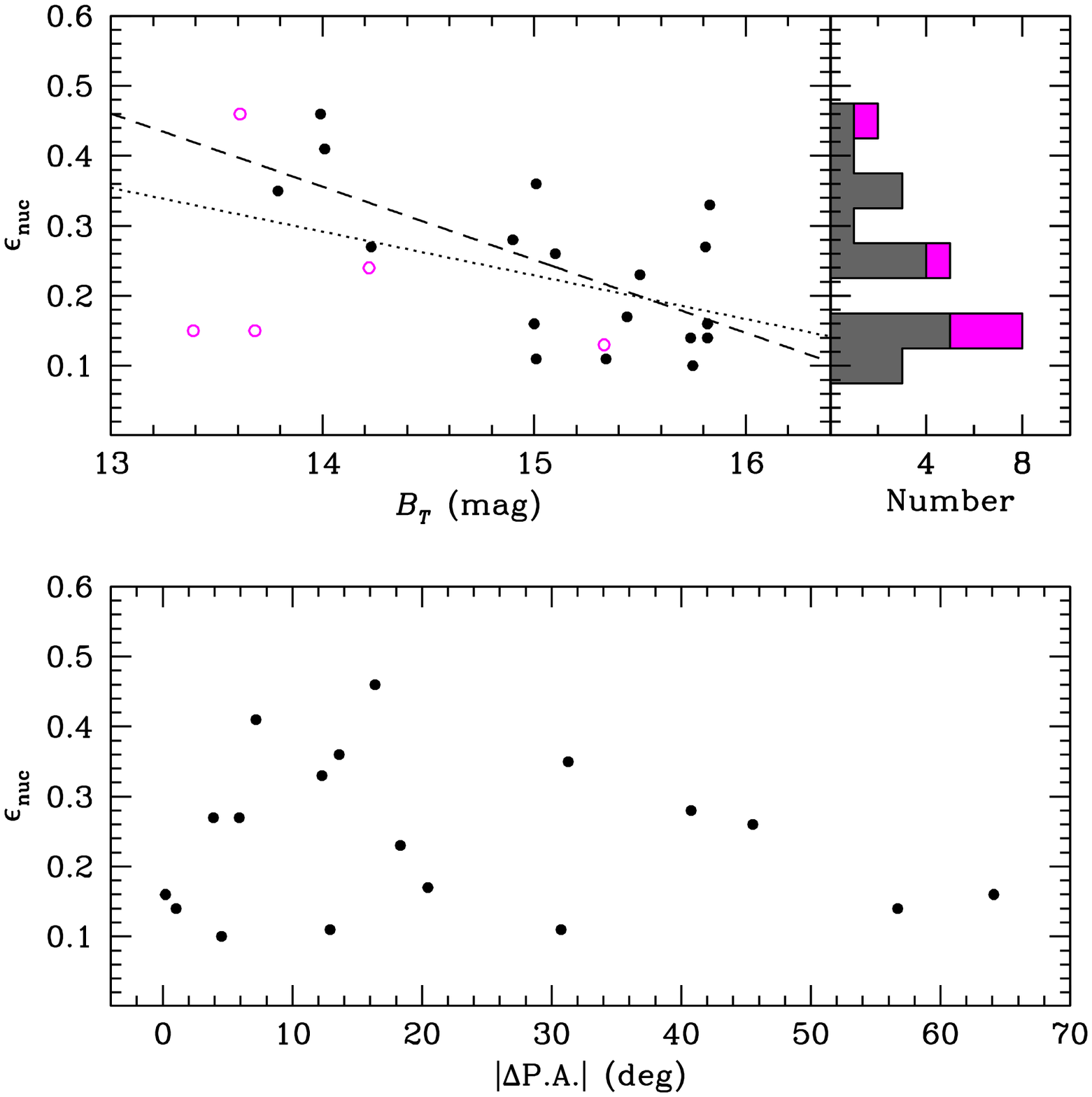}
	\caption{\emph{Top left}: Nuclei ellipticity vs. host galaxy magnitude, for 
	the S2 (closed black circles) and  S$>$2 (open magenta circles) nuclei. 
	The dotted and dashed lines show fits to the full sample and 
	only the S2 nuclei, respectively.
	 \emph{Top right}: Histogram of the nuclei ellipticities. The gray regions
	  represents the S2 nuclei, while the magenta regions indicate S$>$2.
	  \emph{Bottom}: Nuclei ellipticity plotted against the absolute 
	  difference between nuclei and host galaxy position angles.}
	\label{fig:nuce}
\end{figure}

\subsection{Nuclei Ellipticities}

One advantage of performing the 2D analysis is that we are able to measure 
the nuclei ellipticities. In general, we find the nuclei to be flattened,
with median ellipticities of 0.20 and 0.25 for the full and only S2
samples, respectively. The top right panel of Figure~\ref{fig:nuce} shows
a histogram of the ellipticity distribution, while 
in the top left panel, we plot 
ellipticity against host galaxy magnitude. 
A least-squares fit to all of the nuclei hints at a trend of
increasing nuclei ellipticities with galaxy luminosity,
\begin{equation}
\epsilon_{\rm nuc} = -(0.062\pm0.027)\, B_T + (1.2\pm0.40),
\end{equation}
and after removing the more uncertain S$>$2 nuclei, we obtain
a significant relation:
\begin{equation}
\epsilon_{\rm nuc,S2} = -(0.10\pm0.03)\, B_T + (1.8\pm0.45).
\end{equation}
This result suggests that nuclei in brighter (and higher mass) galaxies
are more flattened, and may be more likely to contain edge-on disk-like 
components.  

In the bottom panel of Figure~\ref{fig:nuce}, we show nuclei
ellipticities versus the difference in position angle between the 
nuclei and their host galaxies. We find that the most highly flattened nuclei
are aligned with their hosts, although over one third (7/18) of
our sample are significantly ($\Delta{\rm P.A.} > 20$~deg)
misaligned. 

\subsection{2D Analysis Conclusion}

Figure~\ref{fig:1Dvs2D} shows that there is reasonable agreement between the nuclei parameters measured in 1D and 2D. We
conclude that the 1D nucleus parameters are for the most part robust, and note that the brightest and most structurally complex
galaxies --- which typically have $ \mu_g(1\arcsec) \lesssim 19$ mag~arcsec$^{-2}$ --- present a challenge for measuring
nuclei parameters using {\it either} approach. 

Indeed, even in cases where
adding a second or third profile to the main body is required to fit the nucleus in 2D, 
it is unclear how many components must be added until a ``best" fit is actually achieved, 
and it is usually difficult to say whether one method yields parameters closer to those of the true nucleus. 
In our study of Fornax nuclei, we are primarily interested in extracting the nuclei parameters relative to the  {\it average}
outer profile. Although much of the power of 2D techniques
lies in their ability to fit multiple large-scale components, in galaxies that require more than
one outer S\'ersic profile, it becomes more difficult to perform a fully objective and homogeneous comparison 
between the nuclei and galaxy parameters. Our analysis therefore uses the results from 
our 1D fits, which meets the dual criteria of {\it objectivity and homogeneity}. 
The general consensus between methods indicates that the main conclusions in this
work  are independent of the specific
approach used to model the galaxies and their nuclei.

%%%%%%%%%%%%%%%%%%%

\bibliographystyle{apj_8}
\bibliography{nuclei}

%%%%%%%%%%%%%%%%%%

\end{document}